\documentclass[sn-standardnature]{sn-jnl}

\usepackage[dvipsnames]{xcolor}
\usepackage{subfig} 
\usepackage{doi}
\usepackage{pdflscape}	
\usepackage{lineno}
\usepackage[none]{hyphenat}
\usepackage{hyperref}

\usepackage[separate-uncertainty=true,multi-part-units=single]{siunitx}
\DeclareSIUnit\jansky{Jy}
\DeclareSIUnit\gauss{G}
\DeclareSIUnit\percent{\%}
\DeclareSIUnit\beam{beam}
\DeclareSIUnit\mjpb{\milli\jansky\per\beam}
\DeclareSIUnit\parsec{pc}
\DeclareSIUnit\rsol{R_\odot}
\DeclareSIUnit\angstrom{\text {Å}}

\newcommand{\hide}[1]{}
\newcommand{\askap}{ASKAP~J174508.9\ensuremath{-}505149}
\newcommand{\askapshort}{ASKAP~J1745\ensuremath{-}5051}
\newcommand{\gaia}{{\textit{Gaia}}~$5946454415417964032$}
\newcommand{\gaiashort}{{\textit{Gaia}}~$4032$}
\newcommand{\gaiaother}{{\textit{Gaia}}~$5946454411127231488$}
\newcommand{\gaiaothershort}{{\textit{Gaia}}~$1488$}

\def\arcsec{\hbox{$^{\prime\prime}$}}

\newcommand\arcmin{\hbox{$\,\!\!^{\prime}$}}
\def\Msun{\ifmmode{{\rm M}_\odot}\else${\rm M}_\odot$\fi}
\newcommand{\tj}{{\texttt{The Joker}}}

\begin{document}

\title[Periodic Radio and X-ray Emission from an Accreting White Dwarf Binary]{Periodic Radio and X-ray Emission from an Accreting White Dwarf Binary}

\author*[1,2]{\fnm{Kovi} \sur{Rose}}\email{kovi.rose@sydney.edu.au}
\author[2]{\fnm{Joshua} \sur{Pritchard}}
\author[1,3]{\fnm{Tara} \sur{Murphy}}
\author[1]{\fnm{L. N.} \sur{Driessen}}
\author[4]{\fnm{D. L.} \sur{Kaplan}}
\author[1,3]{\fnm{M.} \sur{Caleb}}
\author[5]{\fnm{Ziteng} \sur{Wang}}
\author[2,3]{\fnm{A.} \sur{Zic}}
\author[6]{\fnm{I.} \sur{Andreoni}}
\author[6]{\fnm{J.} \sur{Carney}}
\author[6]{\fnm{B. N.} \sur{Barlow}}
\author[1,3]{\fnm{D.} \sur{Dobie}}
\author[7,8]{\fnm{M.} \sur{Gu}}
\author[9,10]{\fnm{G.} \sur{Heald}}
\author[]{\fnm{D.} \sur{Huber$^{\textnormal{11}}$}}
\author[2]{\fnm{E.} \sur{Lenc}}
\author[12,13,14]{\fnm{J. K.} \sur{Leung}}
\author[15]{\fnm{W.} \sur{Lu}}
\author[16,17]{\fnm{R.} \sur{Momose}}
\author[1]{\fnm{M. G.} \sur{Pedersen}}
\author[18,19]{\fnm{Y.} \sur{Qu}}
\author[]{\fnm{N.} \sur{Rea$^{\textnormal{20,21}}$}} 
\author[1,3]{\fnm{I.} \sur{de Ruiter}}
\author[1,2,3]{\fnm{K.} \sur{Shaji}}
\author[]{\fnm{G. R.} \sur{Sivakoff$^{\textnormal{22}}$}} 
\author[9,10]{\fnm{A. J. M.} \sur{Thomson}}
\author[]{\fnm{Y. L.} \sur{Wang$^{\textnormal{20,21,23,24}}$}}
\author[23,24]{\fnm{G. J.} \sur{Yang}}
\author[25]{\fnm{F.} \sur{Zahedy$^\dagger$}}

\affil[1]{\orgdiv{Sydney Institute for Astronomy, School of Physics}, \orgname{The University of Sydney}, \orgaddress{\city{Sydney}, \postcode{2006}, \state{NSW}, \country{Australia}}}

\affil[2]{\orgname{Australia Telescope National Facility, CSIRO, Space \& Astronomy}, \orgaddress{\street{PO Box 76}, \city{Epping}, \postcode{1710}, \state{NSW}, \country{Australia}}}

\affil[3]{\orgname{ARC Centre of Excellence for Gravitational Wave Discovery (OzGrav)}, 
\orgaddress{\country{Australia}}}

\affil[4]{\orgname{Center for Gravitation, Cosmology, and Astrophysics, Department of Physics \& Astronomy, University of Wisconsin-Milwaukee}, \orgaddress{\street{P.O. Box 413}, \city{Milwaukee}, \postcode{53201}, \state{WI}, \country{USA}}}

\affil[5]{International Centre for Radio Astronomy Research, Curtin University,
Kent Street, Bentley WA, 6102, Australia}
\affil[6]{Department of Physics and Astronomy, University of North Carolina, Chapel Hill, NC 27599, USA}
\affil[7]{Department of Astronomy, Tsinghua University, Beijing 100084, People’s Republic of China}
\affil[8]{The Hong Kong Institute for Astronomy and Astrophysics, The University of Hong Kong, Hong Kong, People’s Republic of China}

\affil[9]{SKA Observatory, SKA-Low Science Operations Centre, 26 Dick Perry Ave, Kensington WA 6151, Australia}

\affil[10]{Australia Telescope National Facility, CSIRO, Space \& Astronomy, 26 Dick Perry Ave, Kensington WA 6151, Australia}

\affil[11]{Institute for Astronomy, University of Hawai`i, Honolulu, HI, USA}
\affil[12]{David A. Dunlap Department of Astronomy and Astrophysics, University of Toronto, 50 St. George Street, Toronto, ON M5S 3H4, Canada}
\affil[13]{Dunlap Institute for Astronomy and Astrophysics, University of Toronto, 50 St. George Street, Toronto, ON M5S 3H4, Canada}
\affil[14]{Racah Institute of Physics, The Hebrew University of Jerusalem, Jerusalem 91904, Israel}
\affil[15]{Department of Astronomy and Theoretical Astrophysics Center, University of California at Berkeley, Berkeley, CA 94720, USA}
\affil[16]{Carnegie Science Observatories, 813 Santa Barbara Street, Pasadena,
CA 91101, USA}
\affil[17]{Department of Astronomy, School of Science, The University of Tokyo, 7-3-1 Hongo, Bunkyo-ku, Tokyo 113-0033, Japan}

\affil[18]{Nevada Center for Astrophysics, University of Nevada, Las Vegas, NV 89154 USA}

\affil[19]{Department of Physics and Astronomy, University of Nevada Las Vegas, Las Vegas, NV 89154, USA}

\affil[20]{Institute of Space Sciences (ICE-CSIC), Campus UAB, C/ de Can Magrans s/n, Cerdanyola del Vallès (Barcelona) 08193, Spain}

\affil[21] {Institut d’Estudis Espacials de Catalunya (IEEC), Esteve Terradas 1, RDIT Building, Of. 212 Mediterranean Technology Park
(PMT), 08860, Castelldefels, Spain}

\affil[22]{Department of Physics, University of Alberta, CCIS 4-181, Edmonton, AB T6G 2E1, Canada}

\affil[23]{National Astronomical Observatories, Chinese Academy of Sciences, 20A Datun Road, Beijing 100101, China}

\affil[24]{School of Astronomy and Space Science, University of Chinese Academy of Sciences, 19A Yuquan Road, Beijing 100049, China}

\affil[25]{Department of Physics, University of North Texas, 210 Avenue A, Denton, TX 76201, USA}

\abstract{Long period radio transients (LPTs) are coherent bursts of polarised radio emission that repeat periodically on timescales of minutes to hours. Little is known about the physical origins of these systems. Astronomers have proposed magnetars that rotate slowly and white dwarfs that rapidly orbit with a companion star as potential explanations. While several recent examples appear to support the latter hypothesis, the mechanism generating these bright radio pulses remains poorly understood. Here we report our discovery and classification of the LPT \askap\, as an accreting white dwarf binary. This object has a \SI{\sim1.3}{\hour} spectroscopic orbital period and exhibits orbitally-modulated X-ray emission and radio bursts.
These elliptically polarised radio bursts drift in emission frequency, potentially due to a longer beat period, and turn off for several hours at a time.
Some long period radio transients have been associated with non-interacting white dwarf binaries. We have spectroscopically confirmed this system as an accreting cataclysmic variable, identified through characteristic optical emission lines and an ongoing X-ray outburst. Our results strengthen the link between at least some long period radio transients and white dwarf binaries.}

\maketitle

\section*{Main}



We discovered \askap\ (hence \askapshort) with the Australian SKA Pathfinder radio telescope \citep[ASKAP;][]{hotan_21} in an untargeted search for circularly polarised sources in the \SI{1.365}{\giga\hertz} Rapid ASKAP Continuum Survey \citep[RACS-mid;][]{racs_mid_duchesne_2024} --- see Methods. In follow-up observations with the MeerKAT
\citep{2016mks..confE...1J.GRS} 
radio telescope (see Methods) we refined the initial RACS-mid J2000 position to $\rm{RA} = 17\rm{h} 45\rm{m }08\rm{s}.929\pm0.06\arcsec$  and 
$\rm{Dec}=-50^{\circ}51\arcmin49\arcsec.86 \pm0.03\arcsec$. 

\begin{figure}
\centering
\hspace{-1.0 cm}
\includegraphics[width=13.0cm]{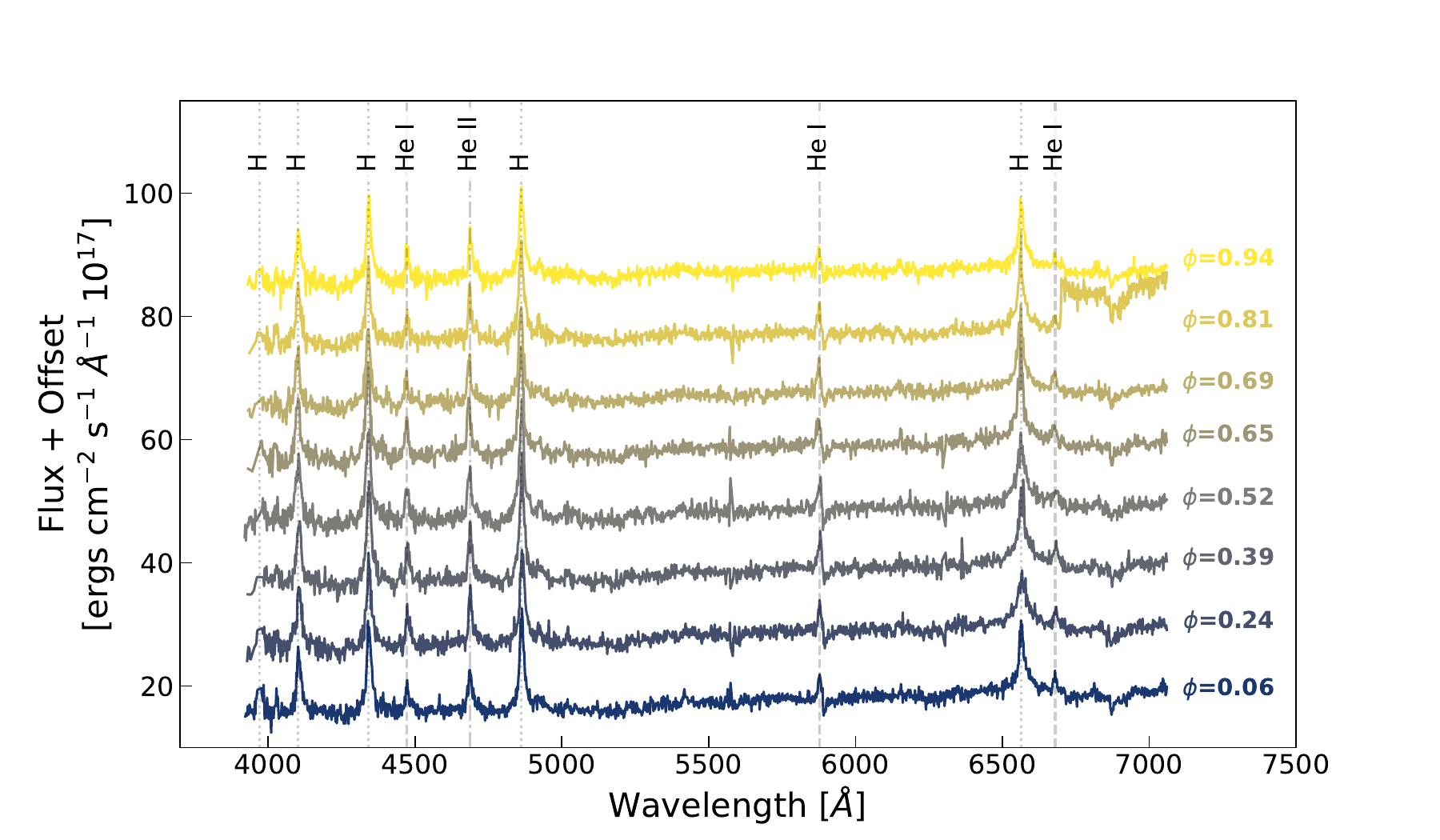}
\caption{\textit{SOAR} spectra of \gaiashort. We show each of the consecutive $10$\,min spectra with an offset and plot the rest wavelengths for the Hydrogen Balmer series (dotted), Helium I (dashed), and Helium II emission lines (dot-dashed). The orbital phases $\phi$ are shown next to each corresponding spectrum. The red excess in the $\phi=0.81$ spectrum (second from the top) is likely due to a calibration error.}
\label{fig: soar}
\end{figure}

We identified an optical counterpart in \textit{Gaia} Data Release 3 \citep[\textit{Gaia} DR3;][]{gaia_dr3}, with the apparent magnitude $m_{\rm{G}} = 19.45 \pm 0.04$ mag (see Methods). In follow-up spectroscopy with the \textit{SOAR/Goodman} \citep{clemens_goodman_2004} and \textit{LDSS-3/Magellan} \citep{2008SPIE.7014E..0AO} telescopes (see Methods) we found \askapshort\ to have a flat spectrum with a blue excess and strong, narrow emission features in Hydrogen (Balmer) and Helium (HeI, HeII) --- see Fig. \ref{fig: soar} and Extended Data Fig. 1. 
The combination of strong HeII lines and the flat spectra with narrow Balmer lines are characteristic of magnetic cataclysmic variables (CVs) 
\citep[e.g.,][]{2024AJ....167..186S,2025MNRAS.540..821P}. 
Magnetic CVs are compact binary systems composed of a strongly magnetised white dwarf and a main-sequence companion (usually of spectral type K to M) \citep{1995cvs..book.....W}, with polars and intermediate polars being the two main subtypes.
Polars have close orbits ($P_{\rm{orb}}$\SIrange{\sim1.3}{4}{\hour}), and strong magnetic fields ($B\gtrsim10^7$\,G) which synchronise the white dwarf spin to the orbital period \citep{1995cvs..book.....W}. Intermediate polars tend to have weaker magnetic fields ($10^6\lesssim B\lesssim10^7$\,G) such that the white dwarf spin and orbital periods ($P_{\rm{orb}}$\SIrange{\sim1.3}{12}{\hour}) are not synchronised \citep{1995cvs..book.....W,2023MNRAS.524.4867I}. Polars may also deviate from synchronisation for periods of $\sim$\SIrange{100}{1000}{} years following a nova outburst \citep{2004ApJ...614..349N}. White dwarfs in these asynchronous polars usually have spin periods a few percent faster than $P_{\rm{orb}}$ but some systems, like Paloma (RX J0524$+$42), have been observed with spin periods up to $\sim$\SI{20}{\percent} faster than their orbital periods \citep[][and references therein]{2007A&A...473..511S, 2023ApJ...943L..24L}.

Another unique radio-emitting CV is AR Scorpii \citep[AR Sco;][]{ar_sco_marsh}, which is more radio-luminous than most CVs \citep{ar_sco_marsh} and has been suggested as an evolutionary progenitor to intermediate polars and LPTs \citep{ar_sco_marsh,rodriguez2025}; though some have argued against this interpretation \citep{castrosegura2025}. Like \askapshort, AR Sco also has flat optical spectra with narrow Hydrogen and Helium lines.
We note that similar features are also seen in the two other known AR~Sco-like systems J191213.72$-$441045.1 \citep[J1912;][]{j1912_pelisoli} and SDSS~J230641.47$+$244055.8 \citep[SDSS~J2306;][]{castrosegura2025}. All three of these systems have orbital periods between \SIrange{3.4}{4.1}{\hour}. Measuring Balmer line radial velocities (see Methods), we found that \askapshort\ has a far shorter orbital period of
$P_{\rm{orb}}=$ \SI{1.368\pm0.053}{\hour}. This period is also shorter than ILT~J1101$+$5521 \citep[ILT~J1101, $P_{\rm{orb}}=$ \SI{2.1}{\hour};][]{iris_ilt_j1101} and GLEAM-X~J0704$-$37 \citep[GLEAM-X~J0740, $P_{\rm{orb}}=$ \SI{2.9}{\hour};][]{2024ApJ...976L..21H}, LPTs thought to be associated with white dwarf-M dwarf binaries, though lacking the characteristic spectra of a magnetic CV \citep{2025A&A...695L...8R}. 
\askapshort\ has properties broadly consistent with LPTs, namely: coherent and highly polarised radio bursts which repeat periodically.
The observed LPT-like radio emission and magnetic CV-like spectral features of \askapshort\ confirm this relationship and suggest that magnetic CVs may be the progenitor for a subset of LPTs.\\



\begin{figure}
\centering
  \includegraphics[width=5.0 in]{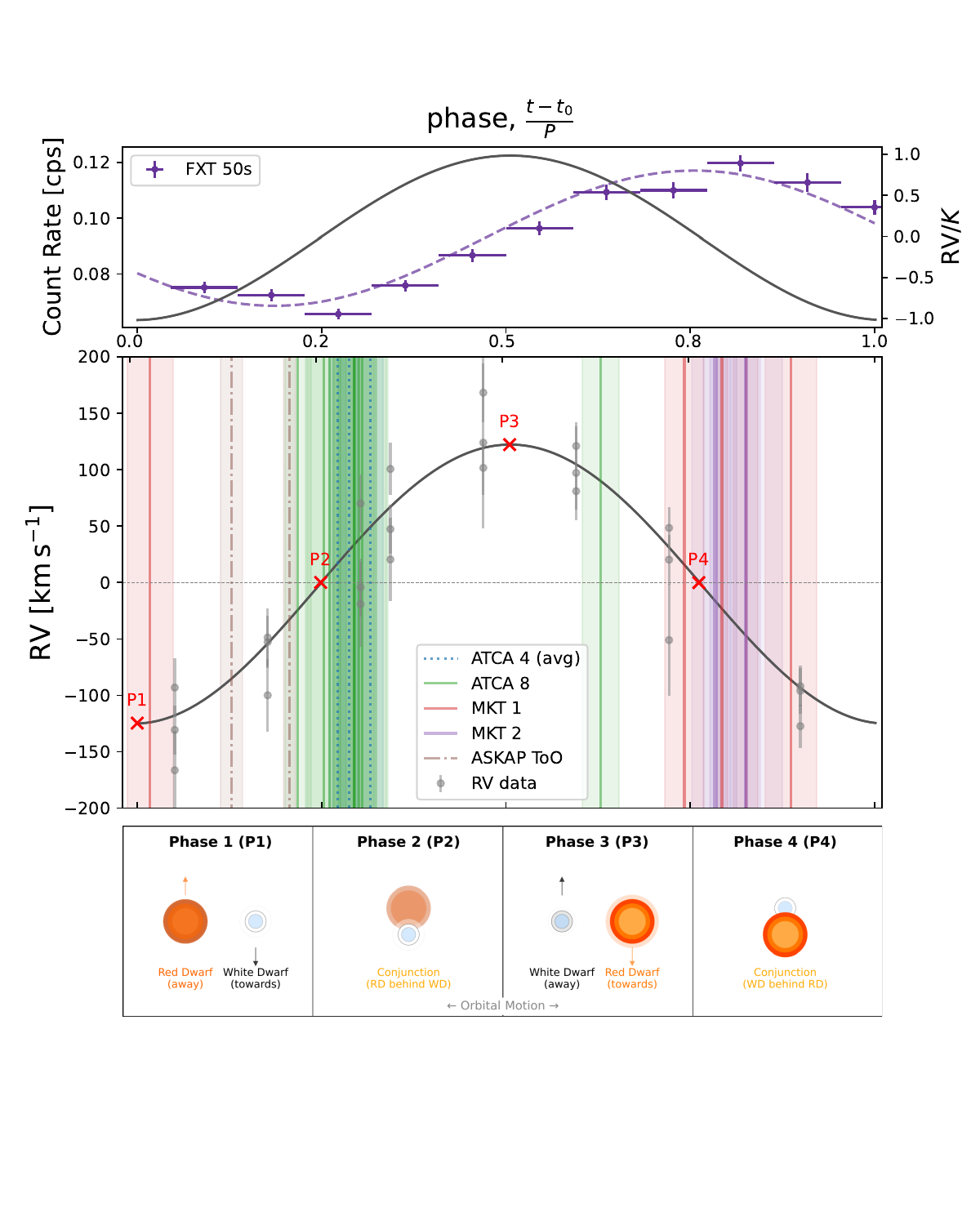}
  \vspace{-2.5cm}
    \caption{Phase-folded pulse timing. \textit{Upper}: 
    \textit{Einstein Probe}-FXT X-ray data compared to the normalised median two-body radial velocity posterior from \tj\ (black curve), with the standard error of the binned count rates shown as vertical error bars and the width of the phase bins shown as horizontal error bars. We also show the sinusoid fitted to the X-ray data (purple dashed curve) peaking at an orbital phase of $\phi_{\rm{X}}=0.89 \pm0.19$.
    \textit{Middle}: 
    arrival times of radio pulses compared to \textit{SOAR} radial velocity measurements (gray markers with $1\sigma$  error bars) and median two-body radial velocity posterior from \tj\ (black curve). We show pulses from MKT Epoch~1, MKT Epoch~2, and ATCA Epoch~8, in red, purple, and green. The ASKAP ToO pulses are denoted as dot-dashed brown lines. For the double-peaked pulses of ATCA Epoch~4 we show, in blue, the average pulse arrival times (dotted lines) taken halfway between the two peaks of each pulse. The light shaded regions denote the pulse width. The red crosses denote the binary phases shown below. \textit{Bottom}: Sketch of orbital phases with a face-on inclination. Phases 1 and 3 correspond to the binary quadratures -- with the two stars side-by-side -- where the Doppler shift maxima/minima occur. Phases 2 and 4 correspond to the binary conjunctions, when the radial velocity is zero.}
\label{fig: folded pulses}
\end{figure}

Roughly $50$ CVs have been seen to produce radio emission, including non-magnetic CVs ($B\lesssim10^6$\,G) \citep{coppejans2015novalike, coppejans2016dwarf, barrett2020radio, ridder_cv_2023}. Of these, none have been reported to exhibit periodic radio emission, and the observed radio emission in these CVs is far less luminous than that seen in LPTs; by a factor of at least \SIrange{100}{1000}{}. There have, however, been detections of coherent and highly circularly polarised radio emission from several magnetic CVs \citep{barrett2020radio} and one nova-like CV \citep{coppejans2015novalike}, supporting a possible CV-origin for LPTs.
It has been shown that there is a canonical $P_{\rm{orb}}$\SI{\sim1.3}{\hour} lower limit on CV orbital periods \citep{gansicke2009}, at which the white dwarf and its low-mass companion detach and begin to drift apart \citep{2006MNRAS.373..484K}. 
\askapshort\ falls near this boundary, with an orbital period of $P_{\rm{orb}}=$\SI{1.368\pm0.053}{\hour}. This spectroscopic period is consistent with the radio pulse period $P_{\rm{radio}}=1.34497^{+0.00003}_{-0.00004}$ \SI{}{\hour}, obtained from observations with the Australia Telescope Compact Array \citep[ATCA;][]{atca_wilson_2011} and ASKAP radio telescopes spanning nearly two years; see Extended Table \ref{tab: radio_summary} and Methods. Moreover, phase-folding the arrival times of the radio bursts from separate observations revealed that these bursts occur around the same orbital phase near conjunctions, which occur at phases 
$\phi=0.25,0.75$, with a median phase $\phi_{\rm{median}}=0.31 \pm 0.03$ for the ATCA and ASKAP bursts and $\phi_{\rm{median}}=0.8 \pm 0.1$ for the MeerKAT bursts; see Fig. \ref{fig: folded pulses}. Similar behaviour was observed from both AR~Sco and ILT~J1101, with radio lightcurves that peak around orbital conjunction \citep{marcote_2017_arsco,circ_pol_ar_sco_stanway, iris_ilt_j1101}. It is noteworthy that, in Fig. \ref{fig: folded pulses}, we see the MeerKAT radio bursts are half an orbit out of phase with respect to the ASKAP and ATCA bursts, despite observing the complete orbital phase, indicating that there may be emission at both orbital conjunctions.
We find no evidence for a seconds-long white dwarf spin period (see Methods) similar to the seconds-long radio pulse structure seen in both AR~Sco and ILT~J1101 \citep{ar_sco_marsh,iris_ilt_j1101}, and cannot directly constrain a white dwarf spin period on longer timescales.

\begin{figure}
\centering
\includegraphics[width=10.0cm]{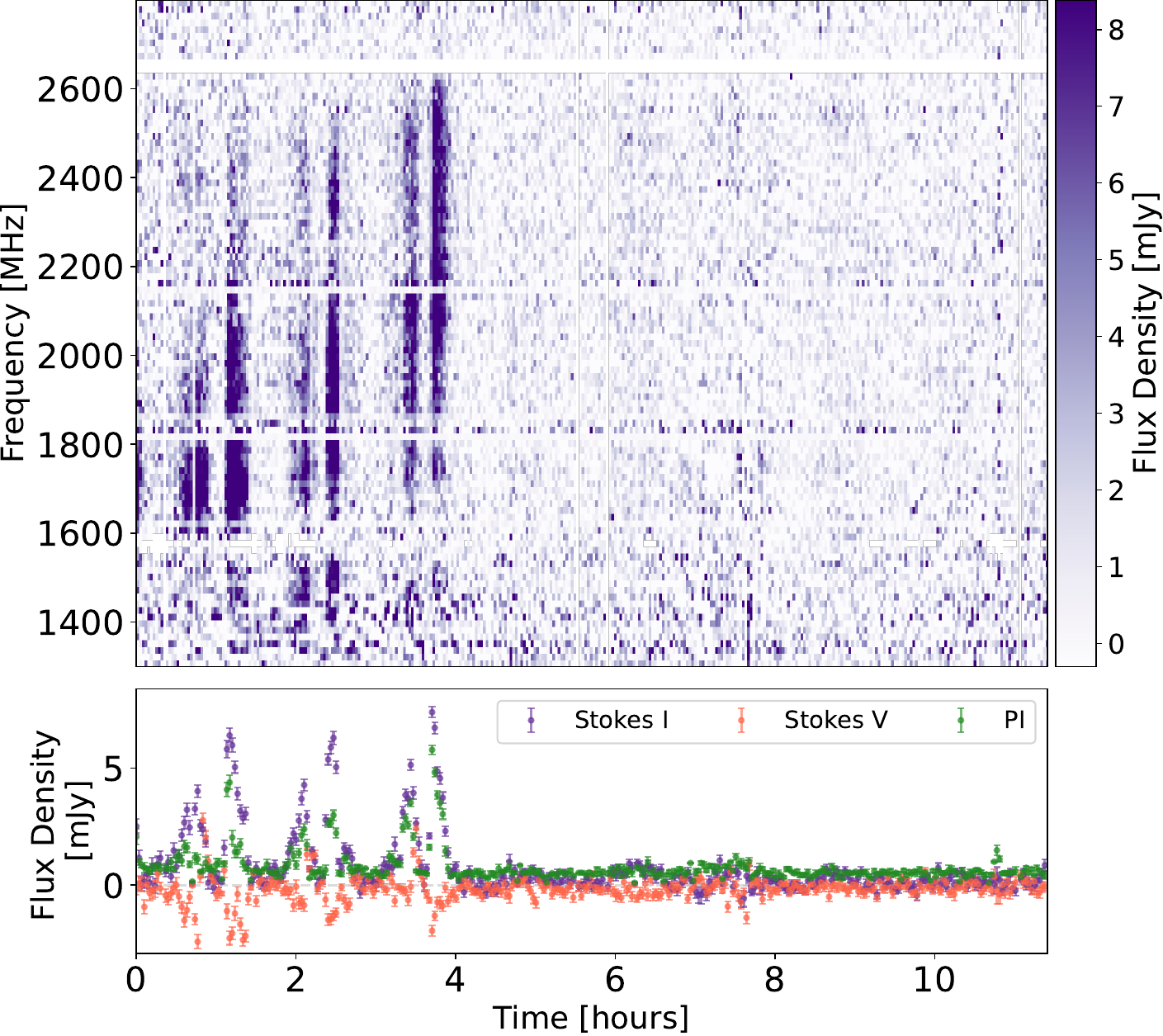}
\includegraphics[width=10.cm]{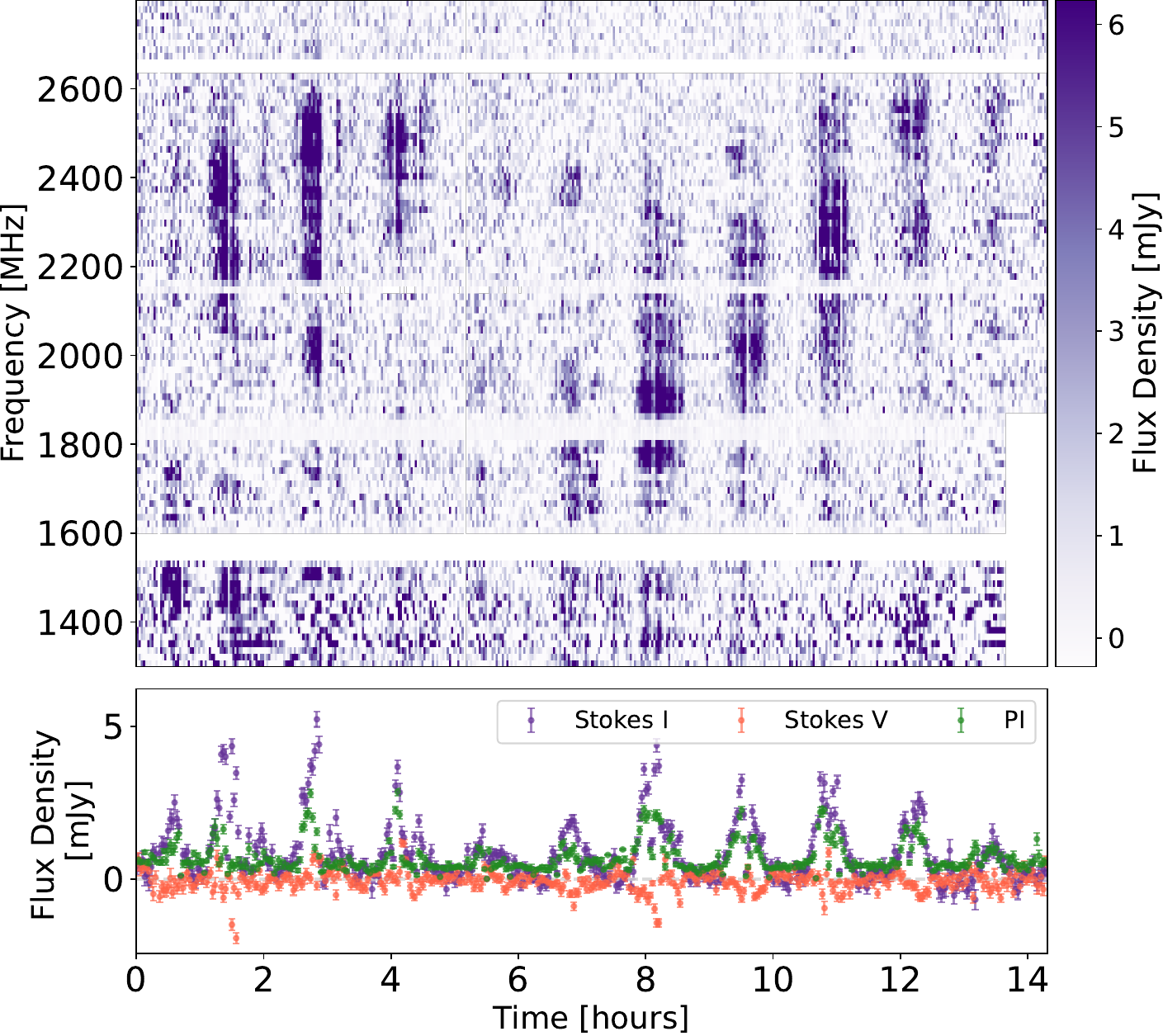}
        \caption{Total intensity (Stokes \textit{I}) dynamic spectra of \askapshort\ from ATCA. Corresponding Stokes \textit{I},\textit{V}, and $PI=\sqrt{Q^2+U^2}$ lightcurves are shown with $1\sigma$ standard error of the mean  error bars. Both dynamic spectra, ATCA Epoch~4 (Top) and ATCA Epoch~8 (Bottom), use \SI{15}{\mega\hertz} frequency averaging and \SI{120}{\second} time averaging. Empty regions of white space denote data that was flagged for radio frequency interference.}
        \label{fig: ATCA DS}
\end{figure}
\begin{figure}
\centering
\includegraphics[width=12.5cm]{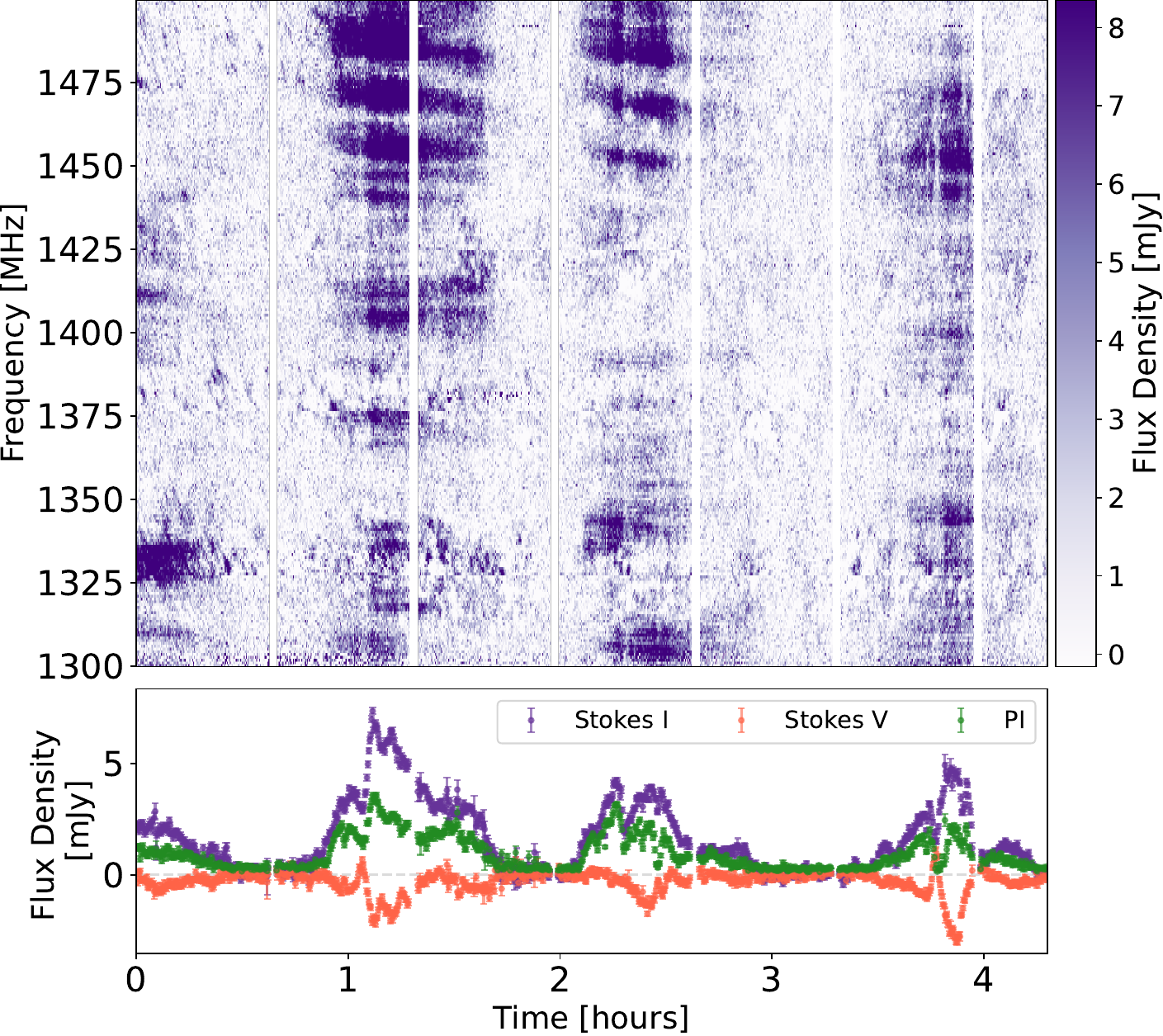}
        \caption{Total intensity (Stokes \textit{I}) dynamic spectra of \askapshort\ pulses from MKT Epoch~1. Corresponding Stokes \textit{I},\textit{V}, and $PI=\sqrt{Q^2+U^2}$ lightcurves are shown with $1\sigma$ standard error of the mean error bars. These dynamic spectra (top) use \SI{0.84}{\mega\hertz} frequency averaging and \SI{16}{\second} time averaging. 
        We show the first half of the observation in the top of the observing band to highlight the modulation lane effect. Vertical white space corresponds to calibration scans.}
        \label{fig: MKT DS}
\end{figure}

The radio pulses from \askapshort\ are elliptically polarised and display variability in their polarisation properties; see Extended Data Fig. 2 and Supplementary Data 1. \askapshort\ also exhibits complex pulse morphology, narrowband emission structure, and intermittency, including switching off for several hours at time (Fig. \ref{fig: ATCA DS} and \ref{fig: MKT DS}).

\askapshort\ exhibits pulse properties not previously observed in LPTs, providing valuable insights into the progenitor system. The pulses are seen to drift up and down in frequency over a longer beat period, with a\hide{potential \SI{\sim8}{\hour}} modulation of the \SIrange{2}{3}{\giga\hertz} upper cut-off frequency --- see Fig. \ref{fig: ATCA DS}.
\askapshort\ also exhibits narrow ($\sim$\SI{10}{\mega\hertz}) frequency structure within the pulses, shown in the MeerKAT dynamic spectra in Fig. \ref{fig: MKT DS}. This sort of intensity modulation -- commonly observed in the decametric emission from Jupiter \citep{2002JGRA..107.1081I} -- is absent from all LPTs except for ASKAP J144834$-$685644 \citep[ASKAP J1448;][]{2025MNRAS.tmp.1214A}. Such variability cannot be explained by interstellar propagation effects, with typical refractive interstellar scintillation having longer timescales ($\sim$months) and lower relative intensity variations (\SIrange{\sim10}{30}{\percent}), while diffractive interstellar scintillation would occur at much shorter timescales (\SI{\sim10}{\second}) than we observe.
This is the only time that these intensity patterns (also known as ``modulation lanes") have been detected in any binary system other than the Jupiter-Io system. The intensity modulation suggests the presence of local plasma acting as an interference screen to the beamed radio emission. The highly elliptical polarisation and the radio frequency modulation indicate that the radio emission is coming from a strongly magnetised plasma.


This plasma in the \askapshort\ system may be the result of accretion onto the white dwarf. This is supported by the detection of coincident UV and X-ray emission in both archival observations and target-of-opportunity observations we conducted with the Neil Gehrels Swift Observatory \citep[\textit{Swift;}][]{gehrels_swift_2004} and the \textit{Einstein Probe} X-ray Telescope \citep{ep_paper} --- see Methods. 
We note that \askapshort\ is only the third LPT detected at X-ray wavelengths, after the recent discoveries of ASKAP~J1448 and ASKAP~J1832$-$0911 \citep[ASKAP~J1832;][]{2025Natur.642..583W}. AR~Sco and J1912 also show pulsed X-ray emission \citep{2021JApA...42...83S, j1912_pelisoli}, the exact origin of which is still debated though for J1912 some residual accretion has been claimed \citep{2023A&A...674L...9S}. Accretion in LPTs has been suggested only with the discovery of an X-ray outburst in ASKAP~J1832 \citep{2025Natur.642..583W} but never proven unambiguously.
The flux across the X-ray observations of \askapshort\ varies by more than an order of magnitude, providing further evidence of variable accretion in the system.
As seen with ASKAP~J1832 we found that X-ray emission in \askapshort\ varies periodically, at the same period as the radio pulsations,  $P_{\rm{X}}=$\SI{1.32\pm0.13}{\hour}; see Fig. \ref{fig: folded pulses} and Methods. For \askapshort\ this demonstrates that the X-ray periodicity is modulated by the orbital period and suggests that the same may be true for ASKAP~J1832; with possible implications for the isolated neutron star or isolated white dwarf interpretations for ASKAP~J1832. The X-ray emission is anti-phase with respect to ASKAP and ATCA but in phase with the MeerKAT radio bursts. Specifically, we find the \textit{Einstein Probe} data peaks at an orbital phase of $\phi_{\rm{X}}=0.89\pm0.19$. This is consistent with the MeerKAT burst median phase and radial velocity posterior but, with respect to the ATCA and ASKAP bursts, there is a phase delay of $\Delta\phi = 0.58\pm0.19$; see Supplementary Information.
The distance to \askapshort\ is poorly constrained between \SIrange{0.4}{9.1}{\kilo\parsec} (see Methods). We therefore calculate a limiting range of X-ray luminosities. We find that detections in the \SIrange{0.2}{10}{\kilo\electronvolt} band with luminosities $L_{\rm X}\sim 10^{30}$--$10^{33} \ {\rm erg \ s^{-1}}$ are a good match for the typical range of accretion-generated X-ray emission in CVs \citep{2017PASP..129f2001M}. Similarly, we constrain the RACS-mid radio luminosity $L_{\rm{R}}\sim10^{18}$--$10^{21} \ {\rm erg \ s^{-1}\ Hz^{-1}}$ at \SI{1.365}{\giga\hertz} with a bandwidth of \SI{288}{\mega\hertz}. This is
more luminous than $\sim99\%$ of all known radio stars \citep{srsc}, making it unlikely that the radio emission originates from the stellar companion. We find that \askapshort\ is also over-luminous in the radio by a factor of $\sim100$ (even at the lower distance limit) compared to all known CVs and most LPTs with both radio and X-ray detections; see Extended Data Fig. 3. 
The notable LPT exception is ASKAP J1832, which has an estimated maximum radio luminosity of $L_{\rm{R}}\approx4\times10^{23}\ {\rm erg \ s^{-1} \ Hz}^{-1}$ \citep{2025Natur.642..583W}. \\



As evidenced by the optical spectra, \askapshort\ appears to be a polar or asynchronous polar, though without a constraint on the white dwarf spin period, we leave definitive classification to a future publication.
Optical photometry and spectroscopy suggest a low-mass red or brown dwarf companion for \askapshort. Specifically, the apparent \textit{Gaia} DR3 magnitude ($m_{\rm{G}}=19.45\pm0.04$) is faint and the spectra lack any obvious absorption lines or other spectral features. A white dwarf companion may be possible but we consider this less likely as \askapshort\ is redder and more luminous than most white dwarfs in the \textit{Gaia} DR3 colour-magnitude diagram; see Extended Data Fig. 4. Blackbody fits to the available photometry also suggest a low-mass spectral type M6.5$\pm0.5$ companion (see Methods and Extended Data Fig. 5), though these estimates may be contaminated by an unrelated nearby star and possibly by the accretion structure itself.

Assuming the companion has filled its Roche lobe, which is the case for accreting CVs \citep{1995cvs..book.....W}, we can use the orbital period to estimate a companion mass and radius of the M dwarf (MD): $M_{\rm{MD}}=0.0963\pm0.0047$\,$M_{\odot}$ and $R_{\rm{MD}}=0.1321\pm0.0055$\,$R_{\odot}$ \citep{1995cvs..book.....W}; see Methods. These values fall on the lower end of M dwarf values, corresponding to an $\sim$M6 companion --- in line with the blackbody spectral type. Taking the empirical mean mass for white dwarfs in a CV: $M_{\rm{WD}}=0.83\pm0.23$\,$M_{\odot}$ \citep{2011A&A...536A..42z}, we obtain an orbital separation of $a=4.2\pm0.4\times10^{10}$\,cm $=0.61\pm0.05$\,$R_{\odot}$. Using this white dwarf mass with the orbital period and radial velocity amplitude, the binary mass function for the estimated $M_{\rm{MD}}\approx0.10$\,$M_{\odot}$ companion constrains the system inclination to $i = 14\pm3$\,deg; see Extended Data Fig. 6 and Methods. We find that the system is highly inclined (face-on), regardless of the exact companion mass.\\

Low-mass M dwarfs and cooler, fully-convective brown dwarfs can produce detectable radio emission \citep[e.g.,][]{pritchard_2021,rose_2023}. These dwarfs possess surface magnetic fields up to $\sim$kG, which are understood to be involved in the generation of this radio emission \citep[e.g.,][]{2017ApJ...846...75P,2018ApJS..237...25K} --- with typical luminosities four orders of magnitude lower than \askapshort\
at $\sim$\,GHz frequencies \citep{pritchard_2024}.
While white dwarfs in CVs typically have much stronger surface fields ($\sim$\,MG), the magnetic field strength at any emission site would depend on its location relative to the two objects in the binary. The detected emission from \askapshort\ -- with a brightness temperature lower limit of $T_{\rm{B}}>10^{12}$\,K (see Methods) -- is necessarily produced by a coherent process, likely arising in the combined magnetic field interaction between the white dwarf and its companion.

For example, it has been suggested that the orbital motion of a weakly magnetised M dwarf within a strong white dwarf magnetosphere can produce a unipolar inductor effect \citep{Qu&Zhang2025}. As electrons from the accreted plasma are accelerated along the interacting magnetic field lines, both the background white dwarf field strength and the electron Lorentz factor grow. This can produce the observed $L_{\rm{X}}\sim10^{30}$--$10^{33}$\,erg s$^{-1}$ X-ray emission from relativistically boosted cyclotron radiation. We note that accretion and inverse Compton scattering could also produce similar levels of X-ray emission \citep{2017PASP..129f2001M}.
Electron cyclotron maser emission (ECME) \citep[e.g.,][]{2006A&ARv..13..229T} can plausibly be generated in low density regions of the same accreting plasma, for example at higher altitudes of the accretion column between the two stars.
The coherent, circularly polarised emission from low-mass stars is widely thought to be generated by ECME \citep{2008ApJ...684..644H}.

However, the degree of linear polarisation and high radio luminosity in the \askapshort\ pulses are not typical of standard ECME or other emission mechanisms operating in typical stellar atmospheres \citep[e.g.,][]{srsc}, making it unlikely that the emission originates solely from stellar magnetic activity of the M dwarf companion.

We suggest a contribution from relativistic ECME -- possibly due to the magnetospheric interaction \citep{Qu&Zhang2025,2026ApJ...997..124Y} --  may account for the high linear polarisation and boost the radio luminosity. \askapshort\ also exhibits rapid changes in polarisation and swings in polarisation position angle; see  Extended Data Fig. 2. This may be due to the precession of the emitting region relative to our line-of-sight \citep{RVM_1969} and the interaction of the ECME beam with surrounding magnetospheric plasma. In the Jupiter-Io system, hollow-cone ECME is generated as the moon Io energises particles along the field lines in Jupiter's magnetosphere. This beamed decametric emission produces a thin-film interference pattern when it passes through local plasma \citep{2002JGRA..107.1081I}. In the case of \askapshort, we propose that similar plasma enhancements from accreted material may be responsible for the observed intensity modulations. 


We see evidence of this plasma environment from Balmer emission lines, with an equivalent width ratio of $\rm{H}_{\alpha}/\rm{H}_{\beta}\le1$ indicative of electron densities $n_{\rm{e}}\gtrsim10^{13}$\,cm$^{-3}$ \citep[][and references therein]{2023MNRAS.521.4190K}. This strength ratio varies over time but is, on average, consistent between observations, with median values of $0.66 \pm 0.03$ in \textit{SOAR} (Fig. \ref{fig: soar}) and $0.68 \pm 0.07$ in \textit{LDSS-3} (Extended Data Fig. 1). 
We also see short-timescale variability in the $\rm{He II}/\rm{H}_{\beta}$ ratio, indicative of channelised accretion \citep{2025MNRAS.540..821P}. As with the Balmer lines, this equivalent width ratio is also consistent across observations, with
median values of $0.415 \pm 0.018
$ in \textit{SOAR} and $0.39 \pm 0.08$ in \textit{LDSS-3}. We tabulate the line strengths and ratios in Supplementary Tables 3 and 4. The variability in relative $\rm{H}_{\alpha}$ emission indicates changes in the local electron density and may be suggestive of variable accretion. It may also be indicative of instabilities in an accretion disk. 

The intermittency and frequency cutoff modulation in \askapshort\ could also be explained by asynchronous rotation of the white dwarf and an inclined magnetic axis, which may be the result of a past nova outburst in the system \citep{2004ApJ...614..349N}. We find that a simple geometric model of dipolar magnetic fields in an asynchronous orbit, with strengths of $\sim$MG and $\sim$kG for the white dwarf and M dwarf, respectively, can reproduce both the variability and frequency evolution \citep{2026ApJ...997..124Y}; see Methods for details. In this model the radio emission is produced in an interaction region. This is required for a magnetic CV as emission from closer to the white dwarf surface would require a gyrofrequency an order of magnitude larger than the observed $\sim$\,GHz emission. In Extended Data Fig. 7 we show simulated dynamic spectra generated with this approach. While our model does not include the plasma physics and gravitational interaction relevant to accreting binaries, our model can reproduce the observed intermittency, radio frequency cutoff modulation, and variable gap width between pulse pairs (Fig. \ref{fig: ATCA DS}). This interaction model may also be applicable to other LPTs with white dwarf binary progenitors.

Varying conditions in the local plasma density and magnetic field interaction may explain the intermittency and unique pulse morphologies in the observed radio pulsations from \askapshort. The fact that these pulsations from \askapshort\ are mostly phase-aligned around conjunction shows similarity to AR Sco, which was found to produce orbitally modulated radio bursts around the same orbital phase --- at or near conjunction \citep{marcote_2017_arsco}. Evidence of similar behaviour was found in ILT J1101, with spectroscopic analysis suggesting the LPT was associated with an M dwarf in a binary system; along with a blue photometric excess hinting at a white dwarf companion \cite{iris_ilt_j1101}. We note that this is not the case for GLEAM-X~J0740, however recent work suggests that this  may be a geometry-dependent effect \citep{2026NatAs.tmp...27H}.  \\


Our observations of \askapshort\ demonstrate that magnetically-driven accretion likely plays a key role in the generation of emission across the electromagnetic spectrum in magnetic cataclysmic variables, including coherent radio pulses and variable X-ray emission. The discovery of \askapshort, and its modulated emission in radio and X-ray that are associated with the spectroscopic orbital period, clearly establishes that accreting cataclysmic variables make up at least part of the population of LPTs. 
Future long-duration optical photometry and spectropolarimetry observations will help to constrain the properties of the low-mass companion. 
Coordinating these observations with simultaneous radio and X-ray observations will further establish the role magnetically-driven accretion plays in generating periodically pulsed emission in these systems. 
Determining if these processes can explain the properties of the entire emerging class of LPTs will require detailed simulations and modelling, as well as the discovery and investigation of new LPTs.


\clearpage


\backmatter

\clearpage



\clearpage


\clearpage


\clearpage


\clearpage



\setcounter{table}{0}
\captionsetup[table]{name={\bf Extended Data Table}}
\begin{table*}
\caption{Radio detections of \askap. We include the barycentric observation start time, duration, central frequency $\nu$, and bandwidth $\Delta \nu$. $S_{\nu,\rm peak}$ is the fitted peak time- and frequency-averaged flux density from the total intensity continuum image, and N$_{\rm{p}}$ is the number of pulses detected during the observation above $3\sigma$. Flux density errors  are the quadrature addition of the fitted error, RMS, and brightness uncertainty scaling --- $6$\% for ASKAP and $10$\% for ATCA and MeerKAT. We also include shorthand names used throughout the text for reference.} 
\vspace{0.45\baselineskip}
    \centering
    \setlength{\tabcolsep}{2.0pt}
    \small
    \begin{tabular}{lcccccll}
    \toprule
    Obs. Start & Telescope & $\nu$ & $\Delta \nu$ &  
     Duration & $S_{\nu,\rm peak}$ & 
    N$_{\rm{p}}$ & Obs. Name \\
    
    {[TDB]} &  &[GHz] & [MHz]  &  
     [hr] &  [mJy beam$^{-1}$]  &
    \\
    \hline
    2020-12-27 03:25\hide{:29.666}   &   ASKAP   &   $1.365$  &   $288$   &
    $0.25$  &   $22.9\pm2.3$  &   $1$ & RACS-mid \\
    2023-08-05 06:58\hide{:32.948}   &   ATCA    &   $2.11$  &   $2048$  &
    $4$     &   $0.376\pm0.047$  & 
     $3$  & ATCA Epoch 1\\
    2023-08-18 03:23\hide{:17.386}   &   ATCA    &   $2.11$  &   $2048$   &
    $6$     &   $0.148\pm0.027$ &  
    $1$ & ATCA Epoch 2\\
    2024-02-10 15:09\hide{:56.805}   &   ATCA    &   $2.11$  &   $2048$   &
    $15$    &   $0.299\pm0.035$  &   
    $3$ & ATCA Epoch 3\\
    2024-02-11 15:14\hide{:13.041}   &   ATCA    &   $2.11$  &   $2048$   &
    $12$    &   $0.643\pm0.086$  &   
    $5$ & ATCA Epoch 4\\
    2024-03-22 12:28\hide{:51.730}   &   ATCA    &   $2.11$  &   $2048$   &
    $15$    &   $0.381\pm0.046$  &   $   7$ & ATCA Epoch 5\\
    2024-07-16 08:06\hide{:10.549}   &   ASKAP   &   $0.943$  &   $288$   &
    $10$  &   $1.053\pm0.085$      &   
    $2$ & EMU \\
    2024-09-08 01:21\hide{:27.383}   &   ATCA    &   $2.11$  &   $2048$   &
    $15$    &   $0.282\pm0.032$  &   
    $8$ & ATCA Epoch 6\\
    2024-09-14 01:06\hide{:02.538}   &   ATCA    &   $2.11$  &   $2048$   &
    $15$    &   $0.209\pm0.026$  &    $5$ & ATCA Epoch 7\\
    2024-09-15 00:58\hide{:55.031}   &   ATCA    &   $2.11$  &   $2048$   &
    $15$    &   $0.831\pm0.087$  &   
    $9$  & ATCA Epoch 8\\
    2024-12-22 18:41\hide{:21.701}  &   ATCA    &   $5.5/9.0$  &   $2048$   &
    $6$    &   $<0.027 \,(3\sigma)$   & - & ATCA Epoch 9\\
    2024-12-27 03:25\hide{:52.987}   &   MeerKAT    &   $1.284$  &   $856$   &
    $10$    &   $0.832\pm0.084$  &   $5$  & MKT Epoch 1 \\
    2024-12-31 03:35\hide{:50.137}    &   MeerKAT    &   $1.284$  &   $856$   &
    $10$    &   $0.305\pm0.037$  &     $4$ & MKT Epoch 2\\
   2025-01-03 03:05\hide{:55.047}   &   MeerKAT    &   $1.284$  &   $856$   &
    $10$    &   $0.330\pm0.040$  &     $8$ & MKT Epoch 3\\
    2025-09-19 11:33\hide{:12.048}  & ASKAP
        &   $1.365$  &   $288$  &
        $3$     & $2.51\pm0.26$ &     $2$  & ASKAP ToO \\  
    \bottomrule
    \end{tabular}
    \label{tab: radio_summary}
 \end{table*}

\clearpage

\section*{Methods}
\label{sec: methods}

\section*{Observations}
\label{sec: observations}

\subsection*{ASKAP}
\label{subsec:askap}

Of the $\sim3\times10^6$ sources detected in the RACS-mid survey, only $\sim100$ are highly circularly polarised (polarisation fraction $>10$\%). Following the approach used for polarisation searches of RACS \citep{pritchard_2021,rose_2023}, we selected \askapshort\ for further study as it was the only one of these $100$ circularly polarised sources without a known astronomical identification within \SI{10}{\arcsec}.

\askapshort\ was detected with a time- and frequency-averaged Stokes \textit{I} flux density of \SI{1.053\pm0.085}{\mjpb} at \SI{943.5}{\mega\hertz} (SBID $63789$), and was detected at \SI{1.365}{\giga\hertz} with flux densities of \SI{22.9\pm2.3}{\mjpb} (SBID $20398$) and \SI{2.51\pm0.26}{\mjpb} (SBID $76988$). We obtain $5\sigma$ flux density limits of \SI{1.22}{\mjpb} at \SI{887.5}{\mega\hertz} (SBID $8646$) and \SI{0.53}{\mjpb} at \SI{855.5}{\mega\hertz} (SBID $34553$). The continuum detections and limits from ASKAP are summarised in Extended Table \ref{tab: radio_summary}. 

\subsection*{ATCA}
\label{subsec: atca}

We conducted follow-up radio observations of \askapshort\ with the Australia Telescope Compact Array \citep[ATCA;][]{atca_wilson_2011}; project codes C3363, CX553, and C3587. We obtained \SI{97}{\hour} of L-band (\SIrange{1.1}{3.1}{\giga\hertz}) and \SI{6}{\hour} of simultaneous C- (\SIrange{4.5}{6.5}{\giga\hertz}) and X-band (\SIrange{8.0}{10.0}{\giga\hertz}) observations, using the extended \SI{6}{\kilo\metre} array configuration. The details of these observations -- including shorthand names for each observation -- are summarised in Extended Table~\ref{tab: radio_summary}. 

We used standard continuum data reduction routines with \textsc{miriad} \citep{Sault1995} to flag and calibrate the data. We used the ATCA primary calibrator source PKS~B1934$-$638 to calibrate the bandpass response and flux scale for all observations except ATCA Epoch~6 and ATCA Epoch~7, for which we used PKS~B0823$-$500. For all observations we corrected time-varying gains using interleaved scans on the calibrator PKS~1740$-$517.

\subsection*{Murriyang}
\label{subsec: Murriyang}

We conducted a \SI{2}{\hour} Director's Discretionary Time observation with Murriyang, the CSIRO Parkes Radio Telescope. We used \SI{32}{\micro\second} time sampling and \SI{1}{\mega\hertz} channel frequency resolution across the $26\times$\SI{128}{\mega\hertz} sub-bands in the \SIrange{704}{4032}{\mega\hertz} range of the ultra-wide-bandwidth, low-frequency receiver \citep{parkes_uwl_hobbs_2020}. 

\subsection*{MeerKAT}
\label{sec:meerkat}

We conducted follow-up radio observations of \askapshort\ with the MeerKAT radio telescope \citep{2016mks..confE...1J.GRS}; project ID: SCI$-$20241101$-$KR$-$02. We obtained three \SI{10}{\hour} observations with the L-band (\SIrange{856}{1712}{\mega\hertz}) receiver, using the c856M1k correlator configuration with an \SI{8}{\second} integration time. We used the SARAO Science Data Processor pipeline to flag and calibrate the data, using PKS~J1939$-$6342 to calibrate the delays, bandpass response, flux scale, and polarisation leakage, PKS~J1744$-$5144 to calibrate the time-varying gains, and PKS~J1331$+$3030 to set the absolute polarisation position angle. We additionally corrected the visibilities for mislabelling of the \rm{X} and \rm{Y} feeds, which produce a sign inversion in Stokes $Q$ and Stokes $V$ and rotate the polarisation angle by \SI{90}{\degree} if left uncorrected \citep{Perley2022}.

\subsection*{Archival radio searches}
\label{subsec:archivalradio}
We searched the archives of the most sensitive radio, millimeter, and sub-millimeter arrays that can observe a source at $-50$ declination. We did not identify archival radio observations covering the position \askapshort\ from the Atacama Large Millimeter/submillimeter Array \citep[ALMA;][]{alma_2009}. Nor were there any  observations made with ATCA, prior to this work, available on the Australia Telescope Online Archive (\href{ATOA}{https://atoa.atnf.csiro.au/query.jsp}). 
We identified \askapshort\ in the MeerKAT Absorption Line Survey \citep[MALS;][]{2025A&A...698A.120G}, conducted at \SI{1.28}{\giga\hertz} with a bandwidth of \SI{856}{\mega\hertz}. \askapshort\ is active at $3$--$5\sigma$ levels in an observation starting on 2020-09-20 13:50 UTC. We provide some details on these nominal detections in the Supplementary Information.

\subsection*{\textit{Gaia} (including distance estimate)}
\label{subsec: Gaia}
In the third data release of the \textit{Gaia} \citep{gaia_mission_2016} space-based optical telescope \citep[\textit{Gaia} DR3;][]{gaia_dr3} we identify two faint sources within $\sim0.6$\arcsec\ of \askapshort, located at distances between $\sim$\SIrange{0.4}{2.0}{\kilo\parsec}; accounting for large parallax uncertainties. Both the original RACS-mid position and the average position across the ATCA observations are within $\sim0.3$\arcsec\ of \gaia\ (henceforth \gaiashort), with the MKT Epoch~1 position separated by $0.16$\arcsec\ from \gaiashort. By comparison the offsets from the other \textit{Gaia} source -- \gaiaother\ (henceforth \gaiaothershort) -- are $0.61$\arcsec\ (RACS-mid),  $0.90$\arcsec\ (MKT Epoch~1) and $0.85$\arcsec\ (ATCA median separation). This supports an association between \askapshort\ and \gaiashort.

\gaiashort\ is located in the Main Sequence of the \textit{Gaia} colour-magnitude diagram, see Extended Data Fig. 4, with an apparent magnitude of $m_{\rm G}=19.40\pm0.04$ mag and colour index of $B_{\rm P}-R_{\rm P}=1.079$.
\gaiashort\ has a DR3 parallax of $\varpi=1.75\pm0.91$\,mas, corresponding to a parallax distance of $d_{\varpi}=0.57^{+0.62}_{-0.19}$\,kpc and an absolute \textit{Gaia} magnitude $M_{\rm G,parallax}=10.67^{+1.39}_{-0.88}$\,mag, without correcting for extinction.
This absolute magnitude is typical of an M dwarf \citep{2013ApJS..208....9P} and the space velocity, calculated with the \textit{Gaia} parallax and proper motions, is consistent with the Galactic average \citep{blandhawthorn2016}. 
However, as the uncertainties on the \textit{Gaia} parallax distance and proper motion are poorly constrained (see Supplementary Information), and since our conclusions are not sensitive to the exact distance adopted, we choose not to assume a single preferred value.

Instead, we select a larger plausible distance range, including the ``Bailer-Jones" photogeometric distance estimate \citep{bailer-jones_21}, to ensure a conservative interpretation.
The photogeometric distance is determined from \textit{Gaia} parallax and photometry using a probabilistic approach. This method is considered more reliable for \textit{Gaia} sources with fractional parallax uncertainties in the range $0.1\leq\sigma_{\varpi}/\varpi\leq1$ \cite{bailer-jones_21}, as is the case for \gaiashort. This catalogue provides the median photogeometric distance of 
$d_{\rm{photogeo}}=6.5^{+2.6}_{-1.1}$\,kpc for \gaiashort. We note that the photogeometric distance may be skewed by emission from the companion and accretion, as it relies on photometric priors.
The Bailer-Jones \citep{bailer-jones_21} photogeometric distance corresponds to the absolute magnitude $M_{\rm G,photogeo}=5.39^{+0.40}_{-0.73}$\,mag, without correcting for extinction.
Hence, combining the parallax and photogeometric distances, we adopt the distance range \SIrange{0.4}{9.1}{\kilo\parsec}. We note that this range is extremely conservative, and that considerations of typical absolute magnitudes for M dwarfs suggest that the true distance is closer to the parallax distance. Regardless of the precise value, our interpretations of the system and emission mechanisms remain unaffected. 


\subsection*{\textit{LDSS-3}}
\label{subsec: LDSS-3}
We conducted optical spectroscopic observations with the \textit{LDSS-3} spectrograph on the \SI{6.5}{\metre} Magellan Clay telescope at the Las Campanas Observatory (LCO). We used the volume phase holographic (VPH)-All grism (\SIrange{4250}{10000}{\angstrom}) with a \SI{1}{\arcsec} slit. 

In the first observation, starting at 2024-03-07 07:36 UTC, the slit was aligned in an E-W configuration and we obtained $2\times$\SI{600}{\second} of data. With this slit alignment we obtained spectra for \gaiashort\ and \gaiaothershort\ simultaneously. We conducted a second observation starting at 2024-03-08 07:53 UTC, aligning the slit in an N-S configuration to observe each of the \textit{Gaia} sources separately. We obtained $2\times$\SI{600}{\second} exposures of \gaiashort\ and  $4\times$\SI{600}{\second} of \gaiaothershort.
We used a version of the reduction pipeline developed for the Magellan IMACS spectrograph \cite{2017ApJ...845..157N}, with updated wavelength
calibration and sky subtraction \cite{2022ApJ...932..103G}. See the Supplementary Information for further details.

\subsection*{\textit{SOAR}}
\label{subsec: SOAR}

We took optical spectroscopic observations of \askapshort\ with the Goodman High Throughput Spectrograph \citep[GHTS;][]{clemens_goodman_2004} on the SOuthern Astrophysical Research (\textit{SOAR}) \SI{4.1}{\metre} telescope. We obtained $8\times$\SI{600}{\second} consecutive exposures (PI: Andreoni) starting between 2024-09-25 00:34 and 2024-09-25 01:52 UTC. Each observation covered a wavelength range of \SIrange{3800}{7050}{\angstrom} and was conducted using a single \SI{1}{\arcsec} slit mask (aligned with the parallactic angle) and a VPH grating with $400$\,lines mm$^{-1}$. 

The data were reduced with \texttt{PypeIt} \citep{2020JOSS....5.2308P} using standard procedures for bias subtraction, flat-fielding, wavelength calibration, and fluxing using observations of a bright calibration star. 

\subsection*{\textit{GALEX}}
\label{subsec: GALEX}
We identified a nearby UV source, \textit{GALEX}~J174508.8$-$505149, separated from \askapshort\ by \SI{0.42}{\arcsec}. \textit{GALEX} is the Galaxy Evolution Explorer \citep{galex_bianchi} space telescope. This source has a far-UV (\SI{1528}{\angstrom}) magnitude of $M_{\rm{FUV}}=19.84\pm0.15$\,mag and a near-UV (\SI{2310}{\angstrom}) magnitude of $M_{\rm{NUV}}=19.67\pm0.11$\,mag, with the zero-point calibrated on the AB magnitude scale. These magnitudes correspond to the calibrated flux densities $F_{\rm{FUV}}=42.1\pm5.8$\,$\mu$Jy and $F_{\rm{UV}}=49.3\pm4.8$\,$\mu$Jy.

\subsection*{\textit{eROSITA}}
\label{subsec: eROSITA}

We identified a nearby X-ray source, \textit{1eRASS}~J174508.8$-$505151, detected by the Russian-German Spektr-RG (SRG) space-based telescope in the \textit{eROSITA} (extended ROentgen Survey with an Imaging Telescope Array) all-sky survey \citep[\textit{SRG/eRASS};][]{erosita_merloni_24}. In the \SIrange{0.2}{2.3}{\kilo\electronvolt} band, \textit{1eRASS}~J174508.8$-$505151 is offset from \askapshort\ by \SI{1.2}{\arcsec} in both MKT Epoch~1 and \textit{Gaia} DR3. 
In the \SIrange{2.3}{5.0}{\kilo\electronvolt} band, \textit{1eRASS}~J174508.8$-$505151 is offset from \askapshort\ by \SI{0.2}{\arcsec} in both MKT Epoch~1 and \textit{Gaia} DR3. 
The \textit{eROSITA} flux is $F_{\rm{X,soft}}=3.1 \pm 0.6\times10^{-13}$\,erg s$^{-1}$ cm$^{-2}$ in the soft \SIrange{0.2}{2.3}{\kilo\electronvolt}, and $F_{\rm{X,hard}}=1.7 \pm 0.4\times10^{-13}$\,erg s$^{-1}$ cm$^{-2}$ in the hard \SIrange{2.3}{5.0}{\kilo\electronvolt} band.

\subsection*{\textit{Swift}}
\label{subsec: Swift }
We did not find any archival detections of \askapshort\ with the \textit{Neil Gehrels Swift Observatory} \citep[\textit{Swift;}][]{gehrels_swift_2004}, a space telescope which performs simultaneous UV and X-ray photometry; within a \SI{10}{\arcmin} cone search.
We obtained a \SI{1.1}{\kilo\second} \textit{Swift} target-of-opportunity (ToO) observation (\href{ID 00016563005}{https://heasarc.gsfc.nasa.gov/FTP/swift/data/obs/2024_05//00016563005/}), starting at 2024-05-16 17:43 UTC. We split this observation into three \SI{0.36}{\kilo\second} exposures, with one exposure for each \textit{Swift} UVOT band used here -- UVW2, UVM2, and UVW1 -- which correspond to \SI{1928}{\angstrom}, \SI{2246}{\angstrom}, and \SI{2600}{\angstrom}. The \textit{Swift} X-Ray Telescope (XRT) instrument observed simultaneously in \textsc{PhotonCounting} mode for the duration of the observation.


We reduced and analysed the \textit{Swift} data on the \textit{SciServer} online compute platform \citep{Taghizadeh-Popp_sciserver_2020} using NASA's High Energy Astrophysics Science Archive Research Center (\textit{HEASARC}) (\texttt{\href{HEASoft} {https://heasarc.gsfc.nasa.gov/docs/software.html}}) software package. See Supplementary Information for additional details. 
In observation ID 00016563005 we found a single source in all three UV bands, each within \SI{<0.8}{\arcsec} of the \askapshort\ radio position --- see  Supplementary Table 2. 

We also inspected simultaneous data from \textit{Swift} XRT to obtain a count of \SIrange{0.3}{10.0}{\kilo\electronvolt} photons in a \SI{15}{\arcsec} aperture. We find an XRT count of $14.0\pm3.7$ photons in the observation (ID 00016563005), corresponding to a rate of $1.3^{+1.6}_{-0.9} \times10^{-2}$\, counts s$^{-1}$. We used the \textit{HEASARC} Portable Interactive Multi-Mission Simulator (\href{PIMMS}{https://heasarc.gsfc.nasa.gov/cgi-bin/Tools/w3pimms/w3pimms.pl}), with a power law photon index of $2$, to predict \SIrange{0.3}{10.0}{\kilo\electronvolt} flux of 
$F_{\rm{PIMSS}}=4.8^{+6.1}_{-3.6} \times10^{-13}$\,erg s$^{-1}$ cm$^{-2}$.

\subsection*{\textit{Einstein Probe}}
\label{subsec: EP }

We triggered ToO observations of \askapshort\ with the \textit{Einstein Probe}'s Follow-up X-ray Telescope (FXT). Three observations were respectively performed at 2025-09-13 18:04:13 UTC (\SI{9.2}{\kilo\second} exposure), 2025-09-19 13:27:05 UTC (\SI{10.5}{\kilo\second}) and 2025-09-23 19:42:47 UTC (\SI{10.0}{\kilo\second}). During the observations, both modules of FXT were set in Full Frame mode, which has a timing resolution of \SI{50}{\milli\second}. Spectral analysis was performed with \textsc{Xspec} v12.14.1. We applied a $\texttt{tbabs} \times \texttt{powerlaw}$ model to fit the X-ray spectra. The source flux in the \SIrange{0.5}{10}{\kilo\electronvolt} band showed a slow decreasing trend, from $2.75^{+0.25}_{-0.23} \times10^{-12}$\,erg s$^{-1}$ cm$^{-2}$ in the first FXT observation to $1.67^{+0.14}_{-0.15} \times10^{-12}$\,erg s$^{-1}$ cm$^{-2}$ in the last one. Within errors, the fitted values of the photon index are consistent in the three observations, varying from $1.22^{+0.19}_{-0.17}$ to $1.28^{+0.19}_{-0.16}$. The fitted values of n$_{\rm H}$ are also roughly consistent with the Galactic value. Further analysis of the X-ray emission from \askapshort\ will be published in a follow-up paper.

\section*{Analysis}
\label{sec: Analysis}

\subsection*{Chance Coincidence}
\label{subsec: Chance Coincidence}

We conducted chance coincidence trials to quantify the likelihood of spatially matching \askapshort\ to an unrelated multi-wavelength source. For this we used our best astrometric measurements, from MKT Epoch~1. 

We produced catalogues of all sources within a search radius from \askapshort\ of \SI{0.2}{\deg} for \textit{Gaia} DR3, \SI{1}{\deg} for \textit{GALEX}, and \SI{3}{\deg} for \textit{eROSITA}; respectively containing \SI{21982}{}, \SI{8871}{}, and \SI{736}{}, sources.
We then ran $n=10^5$, trials shifting the position of \askapshort\ by an angle selected randomly from a uniform distribution between \SIrange{0}{360}{\degree} and a radial separation generated from a random uniform distribution ranging from \SI{0.25}{\arcmin} to the maximum extent of the respective search radius. For each trial we identify a match as being within \SI{0.3}{\arcsec} (\textit{Gaia}), \SI{0.5}{\arcsec} (\textit{GALEX}), and \SI{11.25}{\arcsec}  (\textit{eROSITA}). These are based on $5\sigma$ regions defined by the respective astrometric uncertainties of MeerKAT, \textit{GALEX} \citep{2007ApJS..173..682M}, and \textit{eROSITA} \citep{erosita_merloni_24}, respectively. 

From these random trials we defined the chance coincidence probability as ${(m+1)}/({n+2)}$,
where $m$ is the total number of matches to these catalogues within the given crossmatch radius.
We found that the probability of finding a random source within the respective crossmatch radius is $0.061\%$ for \textit{eROSITA} (with $m=61$ matches), $0.022\%$ for  \textit{GALEX} (with $m=21$ matches), and $0.37\%$ for \textit{Gaia} (with $m=361$ matches).

\subsection*{Line Fitting}
\label{subsec: Line Fitting}

We observed strong $\rm H_{\beta}$, $\rm H_{\gamma}$, and $\rm H_{\delta}$ emission lines from \gaiashort\ in all \textit{LDSS-3} exposures --- see Extended Data Fig. 1. We do not observe any obvious emission features from \gaiaothershort\ in any observations. The narrow spectral feature around  \SI{\sim 7300}{\angstrom} is likely the result of cosmic rays.

The \textit{SOAR} spectra -- see Fig. \ref{fig: soar} -- also show strong Balmer lines and He lines in emission, with no obvious absorption features. 
The narrow spectral features around \SI{\sim 5600}{\angstrom} and \SI{\sim 6300}{\angstrom} are likely residuals from the sky lines at \SI{5578.5}{\angstrom} and \SI{6301.7}{\angstrom}. The red excess in the penultimate spectrum (second from the top in Fig. \ref{fig: soar}) is likely due to a calibration error.

We fit each of the high signal-to-noise emission lines in the \textit{SOAR} and \textit{LDSS-3} spectra and calculated the radial velocity using $\lambda_{\rm{fit}}$. 
We calculated the equivalent width for each of these high signal-to-noise emission lines in both datasets. Supplementary Tables 4 and 4 contain values obtained for the equivalent widths, radial velocities, and full width at half maximum (FWHM) for H$_{\alpha}$, H$_{\beta}$, H$_{\gamma}$, H$_{\delta}$ (only \textit{LDSS-3}), and HeII.

We obtained the radial velocities for the three lines with best signal-to-noise ratios in the \textit{SOAR} spectra (H$_{\alpha}$, H$_{\beta}$, H$_{\gamma}$) and input these data into \tj\ Monte Carlo sampler of radial velocity curves for two-body systems \citep{thejoker_2017} to extract the binary orbital parameters. We generated $10^6$ samples and used the default prior distributions. In Supplementary Figure 1 we show the posterior sample lightcurves with the RV values over-plotted. A clear sinusoidal trend can be seen in the radial velocity over the \SI{\sim1.3}{\hour} in which the spectra were taken. Specifically, we obtained a median period of $P_{\rm{orb}}=1.369\pm0.053$ \SI{}{\hour} as well as median velocities of $K=114.2\pm7.5$\,km s$^{-1}$
for the semi-amplitude and $v_0=15.8\pm5.0$\,km s$^{-1}$ for the barycentre velocity of the binary system. 

These narrow emission lines with wide bases are typical of polar and intermediate polar systems \citep{2024AJ....167..186S}.  Strong HeII lines are common in polars as well as longer-period CVs, with the \SI{4686}{\angstrom} HeII line indicating a magnetic white dwarf in most CVs where it is detected \citep{2007A&A...474..951S}. When such magnetic systems have highly channelled accretion, the strength of the HeII emission tends to be comparable to that of the Balmer lines \citep{2025MNRAS.540..821P}. 
The combination of strong, narrow Balmer and HeII lines -- usually seen in polars undergoing active accretion \citep{2024AJ....167..186S} -- and the shorter \SI{\sim 1.3}{\hour} orbital period support the polar interpretation. 
While the optical spectra of polar and intermediate polars are often quite similar, the HeII in intermediate polars is usually weaker than the H$_{\beta}$ \citep{1995cvs..book.....W}. The HeII / H$_{\beta}$ line ratios from \textit{SOAR} and \textit{LDSS-3} -- see Supplementary Tables 3 and 4 -- might therefore suggest that \askapshort\ is an intermediate polar.

\subsection*{Physical Properties}
\label{subsec: Physical Properties}

We use two independent approaches to estimate the mass of the companion $M_{\rm{MD}}$. 
The first assumes that the companion has filled its Roche lobe and relies on the relationship between the mean density of the companion and the orbital period of the system \citep{1995cvs..book.....W}. These provide the mean empirical mass-period and radius-period relationships

\begin{equation}
    \label{eq: mass-period}    
    M_{\rm{MD}}=0.065 P_{\rm{orb}}^{5/4} \left[M_{\odot} \right];\,\, R_{\rm{MD}}=0.094 P_{\rm{orb}}^{13/12} \left[R_{\odot} \right],\
\end{equation}

\noindent where $P_{\rm{orb}}$ is the orbital period in hours within the range $1.3\leq P_{\rm{orb}}\leq 9$; see Equation 2.100 and its derivation in ``Cataclysmic Variable Stars" \cite{1995cvs..book.....W}. For the orbital period of \askapshort\ these corresponds to

\begin{equation}
    \label{eq: mass-period_result}    
    M_{\rm{MD}}=0.096 \pm 0.008 \left[M_{\odot} \right];\,\, R_{\rm{MD}}= 0.13 \pm 0.01\left[R_{\odot} \right].\
\end{equation}

We use the ``A Modern Mean Dwarf Stellar Color and Effective Temperature Sequence" \href{website}{http://www.pas.rochester.edu/~emamajek/EEM_dwarf_UBVIJHK_colors_Teff.txt} \citep{2013ApJS..208....9P} 
to approximate the corresponding M6--M6.5 spectral type and $\sim2800$\,K temperature.

The second approach relies on the expected mass range from the spectral types corresponding to the fitted temperatures. We fit a blackbody function to the available photometric data from our observations and archival sources -- see Supplementary Information and Supplementary Table 1 -- separately for wavelengths below and above $\lambda=500$\,nm; see Extended Data Fig. 5. Archival measurements were obtained from VizieR \citep{vizier} for \textit{GALEX}, \textit{Gaia}, and the SkyMapper Southern Survey \citep{2018PASA...35...10W}. 

We found the short wavelength data corresponds to a blackbody temperature of $26\,641\pm4139$\,K, with a reduced $\chi^2$ of $4.27$ --- see Supplementary Information. This temperature is unusually high for a white dwarf in a CV below the period gap, where the mean temperature is of order $15\,000$\,K \citep{pala2017}. For a similar outlier CV SDSS J153817.35$+$512338.0, with an effective temperature of $35\,284^{+ 600}_{-688}$\,K and a period of \SI{\sim1.5}{\hour}, it is thought that the high temperature may be the result of a recent nova outburst or ongoing accretion onto the white dwarf \citep{pala2017,pala2022}. \askapshort\ exhibits features of ongoing accretion and hints, in the asynchronous polar scenario, of a possible post-nova outburst disequilibrium. The temperature of \askapshort\ may therefore indicate a heightened period of accretion onto the white dwarf and provide evidence for a possible evolutionary link between LPTs and AR Sco-like systems \citep{rodriguez2025,castrosegura2025}. From the long wavelength data we obtained a temperature of $2781\pm59$\,K with a reduced $\chi^2$ of $130.53$. This fitted temperature is consistent with the effective temperature, and therefore companion spectral type, estimated from the mass-period relationship. We note that the blackbody fitting may be impacted by contamination from an accretion disk/stream, the nearby star \gaiaothershort, or possibly both. This contamination, along with the poorly constrained blackbody fit, may imply higher than reasonable brightness at longer wavelengths. \\

With these mass estimates we calculate the orbital centre of mass separation from Kepler's third law \citep{1995cvs..book.....W}

\begin{equation}
    \label{eq: orbital separation}    
    a = 3.53\times10^{10}M_{\rm{WD}}^{1/3}(1+M_{\rm{MD}}/M_{\rm{WD}})^{1/3}P_{\rm{orb}}^{2/3},
\end{equation}

\noindent where the masses are normalised to solar mass units, $P_{\rm{orb}}$ is in hours, and the separation $a$ is in cm. Using the orbital period $P_{\rm{orb}}=1.369$, the median white dwarf mass $M_{\rm{WD}}=0.83$ $M_{\odot}$, and the companion mass calculated above, we obtained:
$$ a = 4.24\pm0.37\times10^{10} \,\rm{cm} = 0.61\pm0.05\,R_{\odot}.$$

We similarly used the mass estimates and orbital period to calculate the binary mass function:

\begin{equation}
    \label{eq: binary mass function}    
    f = \frac{M_{\rm{WD}}^3 \sin^3i}{(M_{\rm{WD}}+M_{\rm{MD}})^2} = \frac{P_{\rm{orb}}K^3}{2\pi G},
\end{equation}

\noindent where $i$ is the inclination of the orbit and $G$ is the gravitational constant. For this calculation we also used the radial velocity semi-amplitude  $K=114.2\pm7.5$\,km s$^{-1}$ obtained from \texttt{The Joker}. In Extended Data Fig. 6 we show how this constrains the inclination of the system to be near face-on.

\subsection*{Dynamic Spectra}
\label{subsec: DynamicSpectra}

To explore the time and frequency structure of \askapshort\ radio emission, we formed dynamic spectra from the observations with ASKAP, ATCA, and MeerKAT using \textsc{\href{DStools}{https://github.com/askap-vast/dstools/}} \citep{dstools}. For each observation, we first used \textsc{WSclean} \citep{Offringa2014} and \textsc{CASA} \citep{casa_2022} to image and self-calibrate the data and produce a model of all sources of emission within the primary beam, masking the position of \askapshort\ from the model. We then formed model-subtracted visibilities and baseline-averaged the data to produce dynamic spectra, discarding visibilities on baselines shorter than \SI{500}{\meter} to reduce the impact of poorly modelled diffuse emission and radio frequency interference (RFI).

We used the \textsc{\href{rm-lite}{https://github.com/AlecThomson/rm-lite/}} implementation of \texttt{RM-Tools} \citep{rmtools_purcell_2020} to improve the signal-to-noise of linearly polarised emission via rotation measure (RM) synthesis \citep{Brentjens2005} and to correct for Ricean bias \citep{2012PASA...29..214G} of the linearly polarised intensity. Finally, we re-binned the data to an optimal time- and frequency-resolution for each observation, and generated frequency-averaged visibility lightcurves in all Stokes parameters. 

We did not detect quiescent or bursting emission in the single \SI{6}{\hour} ATCA C/X-band observation above a $3\sigma$ limit of \SI{268}{\micro\jansky\per\beam} (ATCA Epoch~9). In L-band observation ATCA Epoch~2 we detected a quiescent source showing no significant variability or bursting emission on shorter timescales. In all other observations (ASKAP, ATCA Epoch~1, ATCA Epochs~3--8, MKT Epochs~1--3) we detected multiple highly elliptically-polarised bursts, showing varying degrees of fractional polarisation, intermittency, and burst sub-structure. The properties of each burst are summarised in Supplementary Table 1.


In Fig. \ref{fig: ATCA DS} we show Stokes $I$ dynamic spectra from the ATCA Epoch~4 and ATCA Epoch~8 observations, along with frequency-averaged lightcurves in Stokes $I$, Stokes $V$, and linearly polarised intensity $PI = \sqrt{Q^2 + U^2}$. In the ATCA Epoch~4 observation we detected three sets of bursts appearing as double-peaked pulses. The burst sets are separated by \SI{\sim1.3}{\hour} while the separation between each component narrows from \SI{\sim30}{\minute} to \SI{\sim10}{\minute}. The pulses have a lower frequency cutoff near \SIrange{1500}{1700}{\mega\hertz}, and upper cutoff around \SIrange{2500}{2600}{\mega\hertz}, and drift upwards in frequency with time at a rate of $\sim$\SIrange{1}{2}{\mega\hertz\per\minute}. We do not detect any pulsed emission from \askapshort\ for the remaining \SI{\sim7}{\hour} following the three detected burst sets.

In ATCA Epoch~8 we detected bursts occurring throughout the \SI{15}{\hour} observation. While less intermittent than in ATCA Epoch~4, the pulses in this observation show a similar modulation of lower and upper frequency cutoff and strong pulse-to-pulse variability. The upper frequency cutoff appears to drift between \SIrange{2500}{2700}{\mega\hertz} sinusoidally over a period of \SI{\sim8}{\hour}.

In Fig. \ref{fig: MKT DS} we present the Stokes~$I$ dynamic spectrum of four pulses detected in the \SIrange{1300}{1500}{\mega\hertz} sub-band of the MKT Epoch~1 observation. The pulses exhibit multi-peaked profiles that vary significantly in both total intensity and polarisation over the observation. The pulses have narrow-band frequency structure, with 
peaks spaced apart by $\sim$\SIrange{15}{35}{\mega\hertz}, decreasing in spacing towards the top of the band. The peaks increase in width from $\sim$\SIrange{1}{10}{\mega\hertz}, and drift downwards in frequency at a rate of \SI{1}{\mega\hertz\per\minute} within each pulse. In some cases, the drift continues between pulses with peaks appearing 
to smoothly connect from one pulse to the next. \\

\subsection*{Pulse Periodicity}
\label{subsec: Periodicity}

We used the dynamic spectra extracted with \texttt{DStools} from ATCA, MeerKAT, and ASKAP observations to produce Stokes \textit{I} lightcurves. For each observation with multiple pulses detected we extracted the pulse periodicity using a Lomb-Scargle periodogram. We used the \texttt{nifty-ls} \citep{nifty_garrison_2024} implementation of the \texttt{astropy} Lomb-Scargle method \citep{astropy:2022}. For each frequency peak in the periodogram, we take the HWHM as the nominal frequency uncertainty.

In this analysis we found a radio period of $P_{\rm{radio}}=$\SI{1.345\pm0.084}{\hour} combining the lightcurves from ATCA Epochs~6--8 -- see Supplementary Figure 2. For the combined MKT Epochs~1--3 we determined an initial radio period of $P_{\rm{radio}}=$\SI{1.31\pm0.13}{\hour}. Similarly for the \textit{Einstein Probe} data, binned to \SI{200}{\second} resolution, we find an X-ray period of $P_{\rm{X}}=$\SI{1.32\pm0.13}{\hour}. We used same Lomb-Scargle approach with the \texttt{scipy.signal} \citep{2020SciPy-NMeth} implementation of a Savitzky–Golay filter, using a $1^{\rm{st}}$ degree polynomial and a filter size of $30$ to smooth the noisy power spectrum; see Supplementary Figure 3.

Using the initial period to fold the radio lightcurves, we then measured pulse times of arrival (ToAs) to determine a more accurate period using pulsar timing techniques (see Supplementary Information for further detail). Fitting the times of arrival (ToAs) across the ATCA and ASKAP detections with \texttt{PINT} \citep{2021ApJ...911...45L} we determined a period of $P_{\rm{radio}}=1.34497^{+0.00003}_{-0.00004}$ \SI{}{\hour} ($4841.89^{+0.12}_{-0.14}$ \SI{}{\second}). The ToAs have a phase-connected solution across a $\sim2$ year baseline. We also obtained an upper limit on the period derivative of $\dot{P}$ \SI{<1.5e-8}{\second\per\second}, taking the $95$\% absolute value. 

We found that the radio periods, as well as the \textit{Einstein Probe} X-ray period, are all consistent with both the ToA period, and the spectroscopic orbital period for \askapshort. Further, we found that the X-ray emission appears to be phase-aligned with the MeerKAT pulses, but anti-phase with the ATCA and ASKAP pulses; see Supplementary Information for phase delay calculations.

\subsection*{Pulse Polarisation}
\label{subsec: pulse polarisation}

All \askapshort\ pulses feature a high degree of elliptical polarisation, with a linear fraction of \SIrange{23}{97}{\percent} and an absolute circular fraction of \SIrange{0}{56}{\percent}. None of our detected pulses show a significant RM. The polarisation position angle (PA) is typically \SIrange{+20}{40}{\degree} and remains flat or slowly wanders over the pulse profile, though on short timescales the polarisation state often undergoes significant variability. 

In Extended Data Fig. 2 we show the polarisation characteristics of a multi-component pulse detected in the MKT Epoch~1 observation, beginning at 2024-12-27 03:00 UTC. In the initial stages of this pulse the polarisation state is $\sim$\SI{80}{\percent} linear and $\sim$\SI{0}{\percent} circular, with a PA of \SI{+35}{\degree}. This state is retained during a sharp drop in total intensity at \SI{13}{\min}, and emission then transitions to \SI{50}{\percent} linear and \SI{50}{\percent} circular as the next pulse component builds in intensity. From \SIrange{22}{24}{\minute} the total intensity dips and the emission becomes significantly depolarised. As the emission returns to a highly polarised state, the PA swings through two complete revolutions of the Poincar{\'e} sphere with a constant fractional circular polarisation. The emission then returns to the initial polarisation state at \SI{26}{\minute} and stays constant as the pulse intensity decays. Further analysis of the polarisation properties from \askapshort\ will be published in a follow-up paper.

\subsection*{Pulse Sub-Structure}
\label{subsec: pulse sub-structure}


We conducted a single-pulse search of the \SI{2}{\hour} Murriyang data with \texttt{\href{PRESTO}{https://github.com/scottransom/presto}}. Using a dispersion measure (DM) range of up to \SI{350}{\parsec\,\cm^{-3}} -- double the maximum Galactic DM for a distance of \SI{6}{\kilo\parsec} from the NE2001 electron density model \citep{ne2001_dm}. Searching in the \SIrange{1100}{2600}{\mega\hertz} frequency range we found no convincing pulsar-like candidates. We provide further details on a periodicity search in the Supplementary Information.


We used MeerKAT's PTUSE backend to record and search the full Stokes data for sub-burst structure on timescales of \SI{300}{\micro\second}--\SI{300}{\milli\second}. We conducted a single-pulse search of the PTUSE data from MKT Epochs~1--3 using the \texttt{\href{HEIMDALL}{https://sourceforge.net/projects/heimdall-astro/}} package \citep{2012MNRAS.422..379B}. The search covered a DM range of \SIrange{0}{1000}{pc\,\cm^{-3}} with a DM tolerance factor of $1.1$, and a detection threshold of S/N $\geq7$. We did not find any evidence of sub-burst structure on timescales of \SI{100}{\micro\second}--\SI{7}{\sec}. See the Supplementary Information for information about the candidate inspection.

\section*{Modelling}
\label{sec: modelling}

\subsection*{Emission Mechanism}
\label{subsec: Emission Mechanism}

Assuming a conservatively large $r=1$\,$R_{\odot}$ upper limit on the emission
region, a minimum distance of \SI{0.4}{\kilo\parsec}, and a typical
flux density of $F_{\nu}=\SI{1}{\milli\jansky}$ at \SI{1}{\giga\hertz}, pulses from
\askapshort\ have a brightness temperature of at least

\begin{equation}
    \label{eq: bright temp}    
    T_{\rm{B}} = \frac{F_{\nu}c^2}{2k_{\rm{B}}\nu^2 \Omega} > \SI{e12}{\kelvin},
\end{equation}

\noindent where $\Omega = \pi \arctan^2(r/d)$ and $\nu$ is the central
observing frequency. This lower limit necessitates a coherent emission
mechanism. 
We note that for all  \textit{Gaia} parallax and Bailer-Jones photogeometric distances, $T_{\rm{B}}$ exceeds the \SI{1e12}{\kelvin} Compton limit. The observed pulsed emission reaches high degrees of fractional polarisation nearing \SI{100}{\percent} and features a slowly varying position angle. This implies the emission must originate from a region with highly ordered magnetic fields; otherwise the superposition of emission from randomly oriented field structures would lead to substantial depolarisation \citep[e.g.,][]{sokoloff1998}. This further constrains the emission region of \askapshort\ pulses to a much smaller region than $\SI{1}{R_{\odot}}$ and a correspondingly larger lower limit on the minimum $T_{\rm{B}}$.
Therefore, we do not consider incoherent mechanisms like gyrosynchrotron emission as they cannot produce the observed brightness temperatures in excess of \SI{1e12}{\kelvin}. Nevertheless, we note that some radio emission from AR Sco-like systems and CVs may be produced by gyrosynchrotron processes \citep[e.g.,][]{arsco_gyro2025}.

We instead propose ECME generated from a relativistic electron population \citep[e.g.,][]{Qu&Zhang2025} as a likely candidate to explain the observed radio properties of \askapshort. ECME can produce high brightness temperatures with large circular or elliptical polarisation when electrons
develop a population inversion in the velocity distribution in the
presence of a strong magnetic field \citep[e.g.,][]{1982ApJ...259..844M}. High degrees of linear polarisation, as observed here, can arise when the emitting electrons have relativistic energies, shifting the emission from purely circular to elliptical polarisation. 

Our optical spectra and X-ray observations show evidence for ongoing
accretion, and the presence of strong HeII emission lines is
suggestive of accretion via magnetically channelled streams. These
accretion streams can provide the necessary ingredients for the production 
of relativistic ECME: a source of energetic electrons, a means to
develop a population inversion in the converging magnetic field of 
the accretion stream, and a mechanism to accelerate electrons to relativistic energies. We note that the electron density lower limit from the accretion region is three orders of magnitude higher than the $n_{\rm{e}}<9\times10^{10}$\,cm$^{-3}$ upper limit, inferred from the plasma frequency condition; see Supplementary Information. This provides additional evidence that the ECME is produced in a lower density plasma region, not co-spatial with the origin of the Balmer line emission.

\subsection*{Simulated Dynamic Spectra}
\label{subsec: simulated dynamic spectra}

We modelled the observed radio pulse behaviour using a geometric
simulation of interacting magnetic dipoles representing a magnetic white dwarf and M dwarf in an asynchronous binary system --- implemented with \textsc{PyVista} \citep{sullivan2019pyvista}, see Supplementary Information for simulation details. By tracing the combined magnetic field topology and identifying regions satisfying the conditions for ECME, we generated synthetic dynamic spectra that reproduce multiple key features of the observations; see Extended Data Fig. \ref{fig: DS sim}. The model produces double-peaked radio pulses recurring at the same orbital phases immediately before and after inferior conjunction of the white dwarf, variation in the spacing of each pulse pair, modulation of the upper and lower frequency cutoffs, and intermittency  persisting over a timescale spanning multiple orbits. The intermittent  phases occur when the white dwarf magnetic pole rotates away from the M dwarf, causing the flux tube connection to become disrupted.

These results show that several of the observed radio emission features from \askapshort\ can arise purely from the evolving geometry of the interacting magnetic fields; although additional unmodelled effects due to gravitational influence, plasma flows,and variable particle acceleration are expected to further shape the emission. 
The same changes in magnetic connectivity that determine radio pulse visibility may also modulate the accretion rate, providing an explanation for the observed differences in X-ray brightness between our \textit{Einstein Probe} and \textit{Swift} observations. Continued coordinated monitoring in both bands will test whether the radio and X-ray variability share this common origin.



\setcounter{figure}{0}
\captionsetup[figure]{name={\bf Extended Data Figure}}

\begin{figure}
\centering
\includegraphics[width=11.5cm]{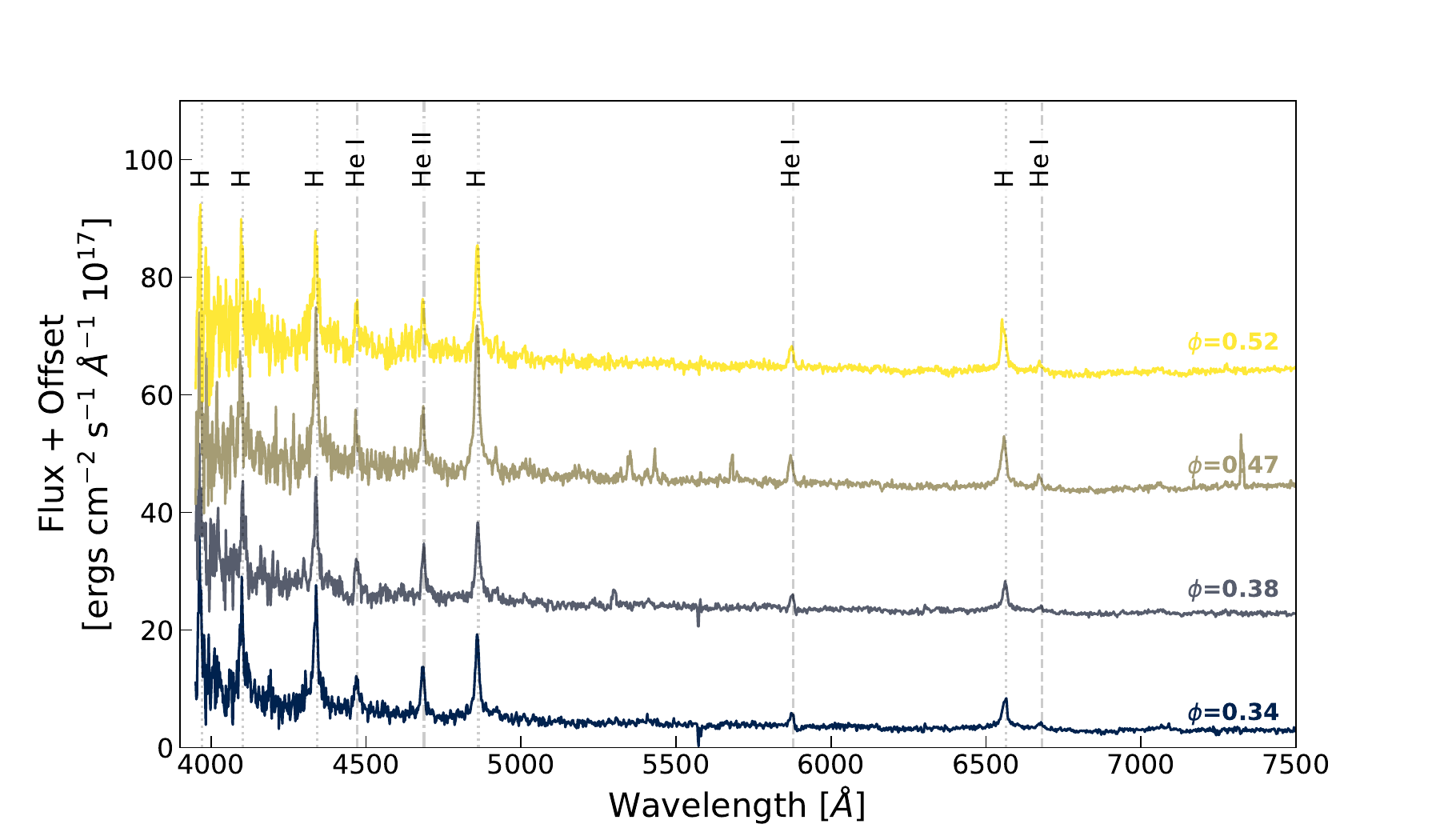}
\caption{\textit{LDSS3} spectra of \gaiashort. We show each of the $10$\,min spectra from both observing nights with an offset and plot the rest wavelengths for the Hydrogen Balmer series (dotted), Helium I (dashed), and Helium II emission lines (dot-dashed). The orbital phases $\phi$ are shown next to each corresponding spectrum.}
\label{fig: ldss3}
\end{figure}

\begin{figure}
\includegraphics[width=12.5cm]{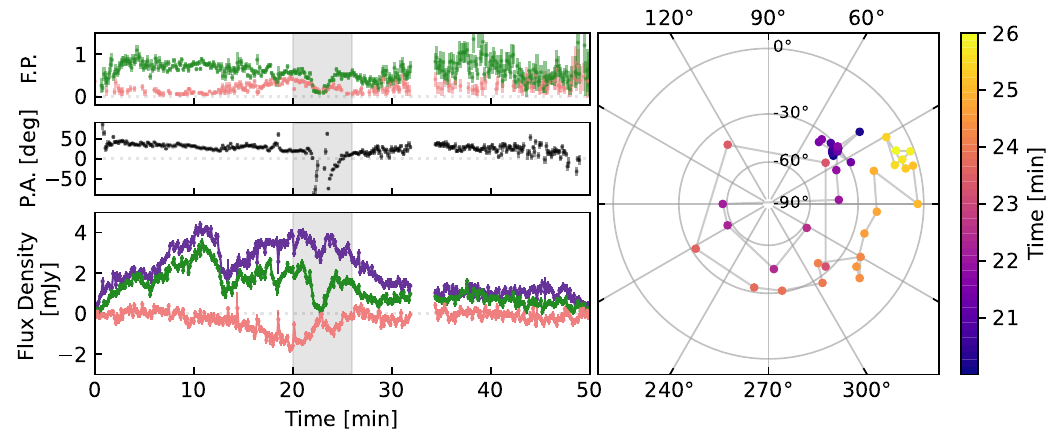}
\caption{Polarisation characteristics of a pulse detected in observation MKT Epoch~1. \textit{Left:} The lower panel show the Stokes \textit{I} (purple), Stokes \textit{V} (pink), and \textit{PI} (green) lightcurves. The middle panel shows the polarisation position angle and the upper panel shows the circular (pink) and linear (green) polarisation fractions. We use $1\sigma$  error bars in all panels. \textit{Right:} A stereographic projection of the polarisation state on the Poincar{\'e} sphere during the grey shaded region of the pulse, showing two complete revolutions about the Stokes \textit{V} axis (\SI{-90}{\degree}) before the polarisation state returns to the original position angle.}
\label{fig: MKT polarisation}
\end{figure}


\begin{figure}
\centering
\includegraphics[width=11.5cm]{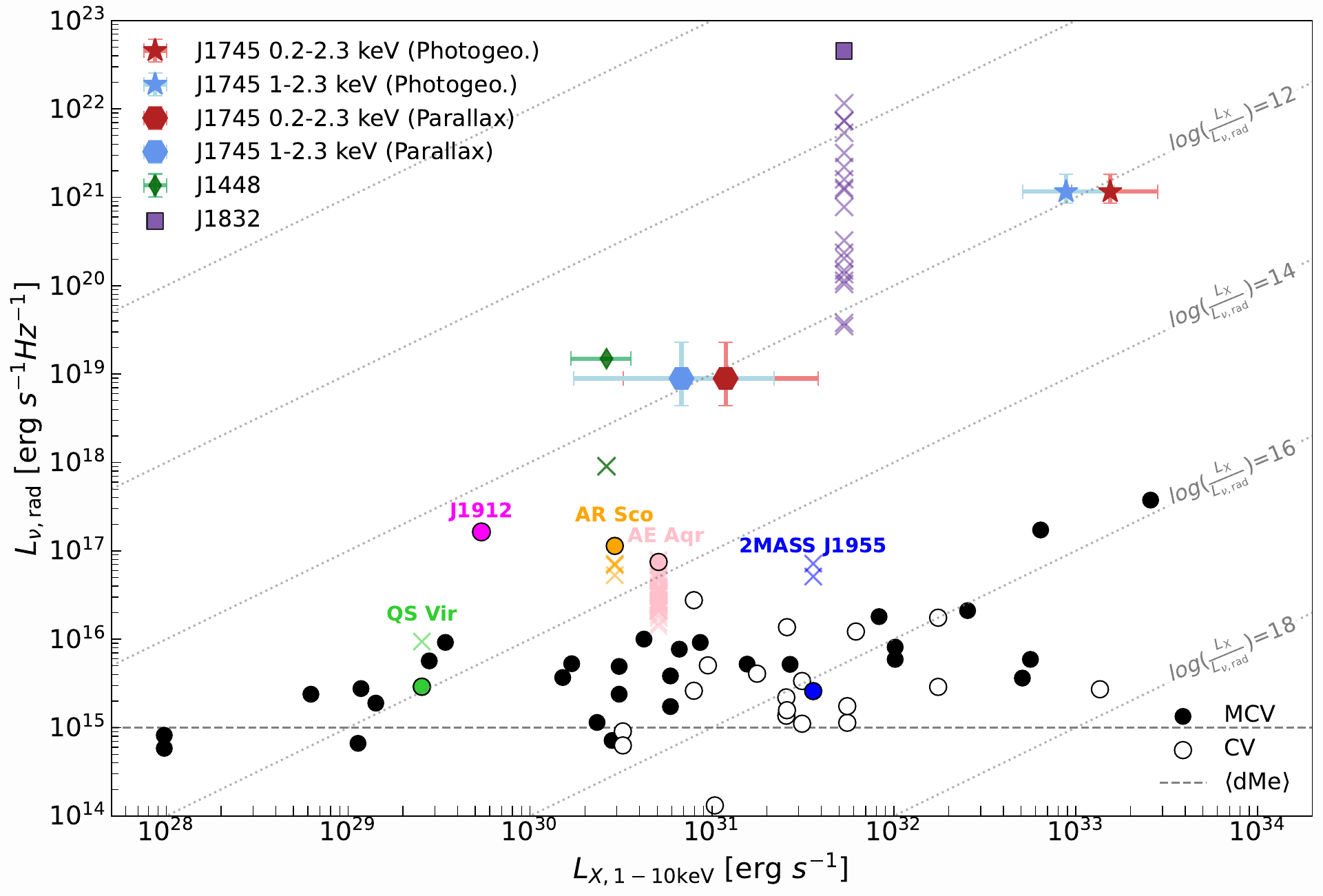}
\caption{
%
Radio and X-ray luminosities for CVs, MCVs, and LPTs with X-ray detections. The circles denote all CVs and confirmed white dwarf binaries \citep{barrett2020radio,ridder_cv_2023,j1912_pelisoli} with crosses showing coincident ASKAP measurements obtained from CASDA \citep{Huynh2020}. The horizontal dashed line denotes the mean flux for isolated M dwarfs. The radio luminosities for \askapshort, J1448 (diamond marker), and J1832 (square marker) are calculated with the maximum observed flux densities. We use the nominal \SI{1}{\kilo\parsec} distance for J1448 as indicative \citep{2025MNRAS.tmp.1214A}. For \askapshort\ we show both \textit{eROSITA} bands assuming the median Bailer-Jones photogeometric distance (star markers), and the \textit{Gaia} parallax distance (hexagon markers). We use the $1\sigma$ parallax uncertainties and the 16th and 84th percentiles of the Bailer-Jones distance posterior to calculate to error bars.}
\label{fig: rad vs xray}
\end{figure}

\clearpage


\begin{figure}
\centering
\includegraphics[width=11.5cm]{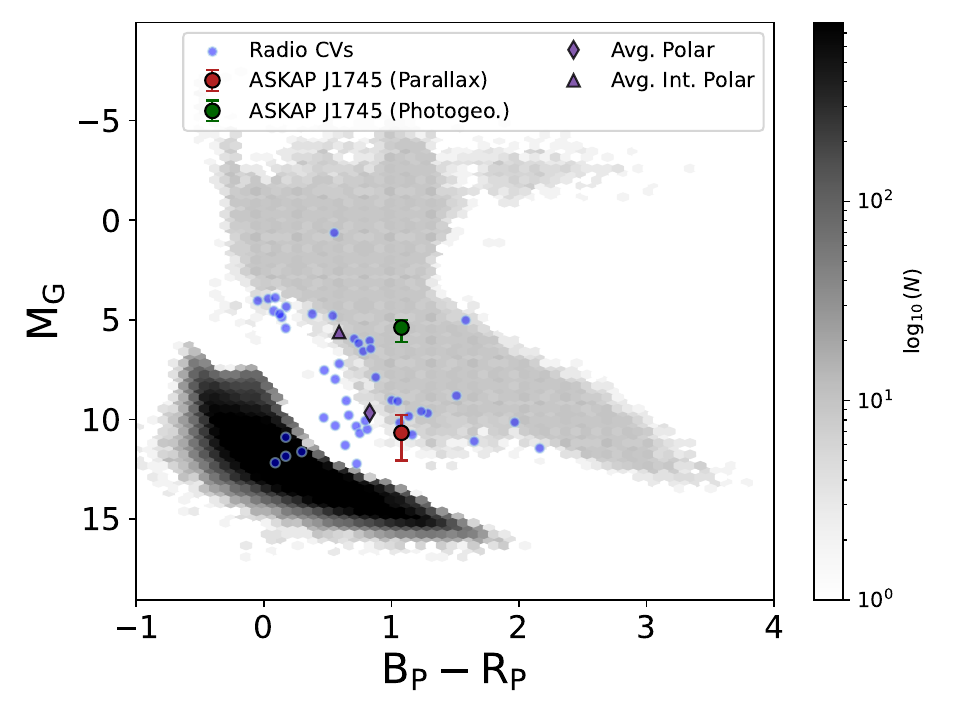}
\caption{\textit{Gaia} DR3 colour-magnitude diagram. We show radio-loud CVs with detected X-ray emission \citep[][and references therein]{ridder_cv_2023} in blue and \askapshort ; with the red marker denoting the value obtained from the \textit{Gaia} DR3 parallax distance and the green marker denoting the value calculated with the median Bailer-Jones photogeometric distance. We use the $1\sigma$ parallax uncertainties and the 16th and 84th percentiles of the Bailer-Jones distance posterior to calculate to error bars. 
We also show the average \textit{Gaia} centroids for polars and intermediate polars \citep{2020MNRAS.492L..40A} as a purple diamond and triangle, respectively. The background points show the \textit{Gaia} the white dwarf population (black) \citep{gentile_fusilio_gaia_edr3_wd_2021} as well as a sample of $10^7$ stars from \textit{Gaia} DR3 (grey) \citep{gaia_dr3}. 
}
\label{fig: gaia_cmd}
\end{figure}


\begin{figure}
\includegraphics[width=12.5cm]{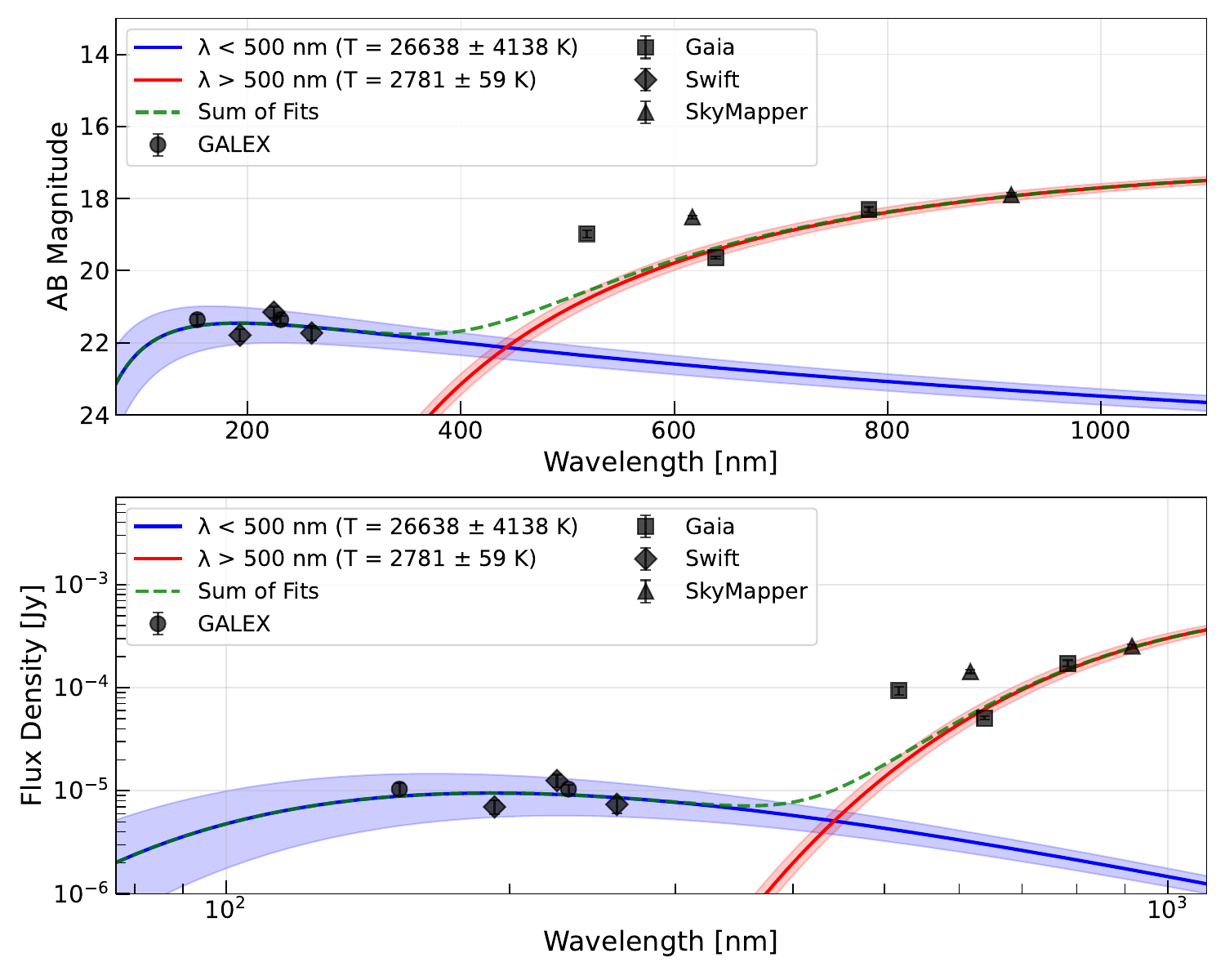}
\caption{Blackbody temperature fits to photometric data. We show AB magnitudes (Top) and flux density (Bottom), both with $1\sigma$ errors. We fit the shorter wavelengths with $\lambda<500$\,nm (blue) and longer wavelength with $\lambda>500$\,nm separately (red), shown with $1\sigma$ uncertainty regions. We also show the sum of the fits as a dashed (green) line.}
\label{fig: blackbody}
\end{figure}

\begin{figure}
\includegraphics[width=12.5cm]{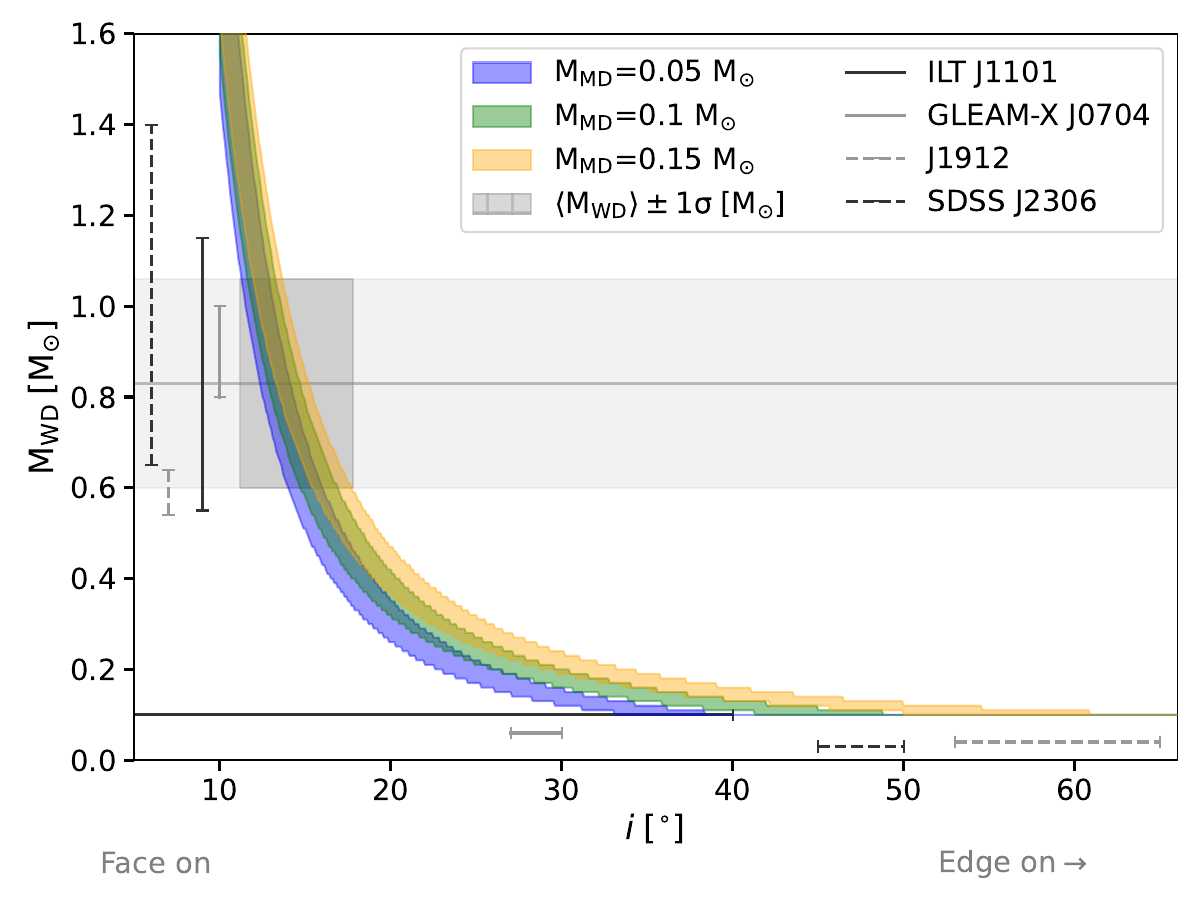}
\caption{Binary mass functions for companion estimates. The grey line shows the empirical mean mass white dwarf for white dwarfs in a cataclysmic variable, within a $\pm1\sigma$ region. The high-opacity subsection shows the range of inclinations ($11$\,deg--$18\deg$) that would correspond to this white dwarf mass range for three test companion masses ($0.05,0.1,0.15$\,$M_\odot$). We show the inclination and mass ranges for similar systems for comparison.}
\label{fig: binary mass}
\end{figure}

\begin{figure}
\includegraphics[width=11cm]{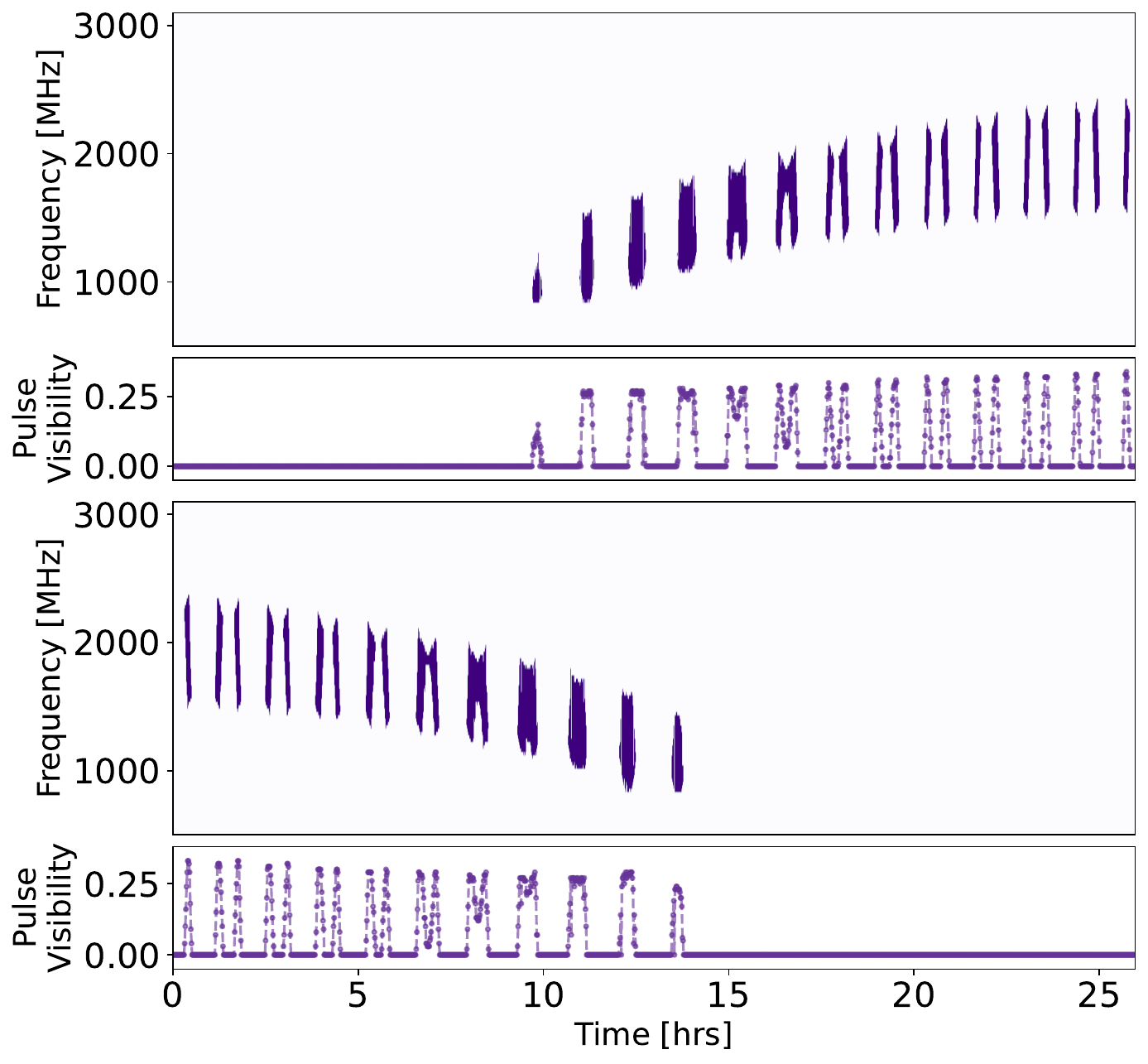}
\caption{
Example synthetic dynamic spectra generated from a geometric model representing \askapshort\ as a pair of interacting magnetic dipoles  in a near-synchronous binary. Each panel shows the simulated visibility of ECME as a function of frequency and time over several orbital cycles, where non-zero values indicate the presence of at least one viable emission site with the appropriate local field strength and orientation. The rising (Top) and falling (Bottom) of the pulse frequency corresponds to different phases of the beat period. Lower panels show the simulations as a time series of emission visibility. The time axis begins at inferior white dwarf conjunction.  The model reproduces several key features of radio pulses in our observations including: double-peaked pulses, variable spacing been pulse pairs, modulation of frequency cutoffs, and intermittency.}

\label{fig: DS sim}
\end{figure}

\section*{Declarations}

\begin{itemize}


\item \textbf{Data availability} The ASKAP data used in this paper can be accessed through the CSIRO ASKAP Science Data Archive \citep[\href{CASDA}{https://data.csiro.au/collections/domain/casdaObservation/search/};][]{Huynh2020} under project codes AS110, AS107, and AS113. The ATCA data used in this paper can be accessed through the Australia Telescope Online Archive (\href{ATOA}{https://atoa.atnf.csiro.au/query.jsp}) under project codes C3363, CX553, and C3587. The Murriyang Parkes data (Project ID: PX113) is also accessible via the ATOA website.
This manuscript makes use of data from MeerKAT (Project ID: SCI-20241101-KR-02) which can be obtained through the SARAO archive (\url{https://archive.sarao.ac.za}). 
Other auxiliary datasets can be made available upon request via email to the corresponding author. The data that support the findings of this study are available at Zenodo: \url{10.5281/zenodo.17365566} \cite{zenodo_ref}

\item \textbf{Code availability} Code to reproduce the figures and main results are available on Github:
\url{https://github.com/krose098/askap_j1745}.
Additional Python scripts used in the data reduction and analysis are available on request from K.R. 

\item \textbf{Acknowledgements} 
K.R. thanks the LSST-DA Data Science Fellowship Program, which is funded by LSST-DA, the Brinson Foundation, the WoodNext Foundation, and the Research Corporation for Science Advancement Foundation; their participation in the program has benefited this work. K.R. thanks W. Garrett Levine for their insightful suggestions. The authors would also like to thank Tim Bedding, Helen Johnston, Courtney Crawford, Scott Croom, and Patrick Woudt for their helpful discussions.

Part of this research was supported by the Australian Research Council Centre of Excellence for Gravitational Wave Discovery (OzGrav), project number CE230100016. M.C. acknowledges support of an Australian Research Council Discovery Early Career Research Award (project number DE220100819) funded by the Australian Government. D.L.K. was supported by NSF grants AST-1816492 and AST-2511757. G.R.S. is supported by NSERC RGPIN-2021-04001. D.H. acknowledges support from the Alfred P. Sloan Foundation, the National Aeronautics and Space Administration (80NSSC21K0652, 80NSSC22K0781), and the Australian Research Council (FT200100871). I.A. is supported by the National Science Foundation award AST 2505775, NASA grant 24-ADAP24- 0159, Scialog award SA-LSST-2024-102a and LSST2025-112b. M.G.P. recognises support from the Professor Harry Messel Research Fellowship in Physics Endowment, at the University of Sydney. N.R. and Y.L.W. are supported by the European Research Council via the ERC Consolidator grant ``MAGNESIA'' (No. 817661), the ERC Proof of Concept ``DeepSpacePULSE" (No. 101189496), and by the Spanish grant PID2023-153099NA-I00. Y. L. W. is supported by the China Scholarship Council (No. 202404910397).

The Australia Telescope Compact Array is part of the Australia Telescope National Facility (ATNF, \href{https://ror.org/05qajvd42}{Australia Telescope National Facility}) which is funded by the Australian Government for operation as a National Facility managed by CSIRO. We acknowledge the Gomeroi people as the Traditional Owners of the Observatory site.

The authors would like to thank Sarah Buchner and SARAO for scheduling our MeerKAT observations. The MeerKAT telescope is operated by the South African Radio Astronomy Observatory, which is a facility of the National Research Foundation, an agency of the Department of Science and Innovation. 

This scientific work uses data obtained from Inyarrimanha Ilgari Bundara, the CSIRO Murchison Radio-astronomy Observatory. We acknowledge the Wajarri Yamaji People as the Traditional Owners and Native Title Holders of the Observatory site. CSIRO’s ASKAP radio telescope is part of the ATNF \href{https://ror.org/05qajvd42}{Australia Telescope National Facility}. Operation of ASKAP is funded by the Australian Government with support from the National Collaborative Research Infrastructure Strategy. ASKAP uses the resources of the Pawsey Supercomputing Research Centre. Establishment of ASKAP, Inyarrimanha Ilgari Bundara, the CSIRO Murchison Radio-astronomy Observatory and the Pawsey Supercomputing Research Centre are initiatives of the Australian Government, with support from the Government of Western Australia and the Science and Industry Endowment Fund. The authors would like to thank Matthew Whiting, Minh Huynh, and the ASKAP observatory staff for the scheduling and processing of our ASKAP ToO observation.

Murriyang, CSIRO’s Parkes radio telescope, is part of the ATNF \href{https://ror.org/05qajvd42}{Australia Telescope National Facility} which is funded by the Australian Government for operation as a National Facility managed by CSIRO. We acknowledge the Wiradjuri people as the Traditional Owners of the Observatory site.

This research has made use of the VizieR catalogue access tool, CDS,
Strasbourg, France \citep{vizier}. The original description 
of the VizieR service was published in \cite{vizier2000}

This research has made use of data and/or software provided by the High Energy Astrophysics Science Archive Research Center (HEASARC), which is a service of the Astrophysics Science Division at NASA/GSFC. Our results are based on observations obtained at the Southern Astrophysical Research (SOAR) telescope, which is a joint project of the Minist\'{e}rio da Ci\^{e}ncia, Tecnologia e Inova\c{c}\~{o}es (MCTI/LNA) do Brasil, the US National Science Foundation’s NOIRLab, the University of North Carolina at Chapel Hill (UNC), and Michigan State University (MSU).

This work also uses data obtained with Einstein Probe, a space mission supported by Strategic Priority Program on Space Science of Chinese Academy of Sciences, in collaboration with ESA, MPE and CNES (Grant No. XDA15310000), the Strategic Priority Research Program of the Chinese Academy of Sciences (Grant No. XDB0550200), and the National Key R\&D Program of China (2022YFF0711500). The authors thank Jingwei Hu and Jun Yang for arranging the EP/FXT ToO observations. 

\item \textbf{Author contributions}
K.R. drafted the manuscript with suggestions and comments from all co-authors and is the PI of the ATCA and MeerKAT data. K.R. discovered \askapshort\ in the ASKAP data, as part of a project conceived by T.M.. K.R. also identified the \textit{Gaia} optical counterpart, undertook the spectral line analyses,  reduced and analysed the \textit{Swift} data, conducted the ATCA observations, and performed the multi-wavelength archival data search. J.P.P. developed the reduction pipeline for the ATCA and MeerKAT data, conducted the processing, generated the dynamic spectra, and produced the simulated dynamic spectra. T.M., L.N.D., D.L.K., and M.C. provided extensive discussions on the nature of this source and feedback on manuscript. D.H. helped with the early discussion and in obtaining the \textit{LDSS-3} observations. R.M. and F.Z. conducted the \textit{LDSS-3} observations, which were calibrated and processed by M.G.. I.A., B.N.B., and J.C. conducted the \textit{SOAR} observations which were calibrated and processed by I.A.. J.K.L. and Z.W. helped with ATCA observations.  Z.W. conducted the Murriyang observation and the pulse search of that data, and provided advice on the X-ray analysis. D.D. and E.L. provided helpful discussion on this source. G.H. and A.J.M.T provided extensive insights into the polarisation analysis and helpful discussion. A.Z. performed the ToA analysis. I.d.R. provided helpful discussions and feedback on the manuscript as well as help with analysis of the orbital dynamics. N.R. and Y.L.W. triggered and analysed the EP X-ray data and provided useful feedback on the manuscript. G.J.Y scheduled the EP/FXT ToO observations. K.S. performed the pulse search of the MeerKAT PTUSE data. Y.Q. and W.L. both provided helpful discussion on the theoretical aspects of the manuscript. M.G.P. supplied the code for generating the \textit{Gaia} sample and HR diagram. G.R.S. provided useful feedback on the MeerKAT proposal, extensive review of the manuscript, and first suggested the CV textbook. \citep{1995cvs..book.....W}.

\item \textbf{Conflict of interest/Competing interests} 
The authors declare no competing interests.

\end{itemize}

\bibliography{refs.bib}

@ARTICLE{atca_wilson_2011,
       author = {{Wilson}, Warwick E. and {Ferris}, R.~H. and {Axtens}, P. and {Brown}, A. and {Davis}, E. and {Hampson}, G. and {Leach}, M. and {Roberts}, P. and {Saunders}, S. and {Koribalski}, B.~S. and {Caswell}, J.~L. and {Lenc}, E. and {Stevens}, J. and {Voronkov}, M.~A. and {Wieringa}, M.~H. and {Brooks}, K. and {Edwards}, P.~G. and {Ekers}, R.~D. and {Emonts}, B. and {Hindson}, L. and {Johnston}, S. and {Maddison}, S.~T. and {Mahony}, E.~K. and {Malu}, S.~S. and {Massardi}, M. and {Mao}, M.~Y. and {McConnell}, D. and {Norris}, R.~P. and {Schnitzeler}, D. and {Subrahmanyan}, R. and {Urquhart}, J.~S. and {Thompson}, M.~A. and {Wark}, R.~M.},
        title = "{The Australia Telescope Compact Array Broad-band Backend: description and first results}",
      journal = {\mnras},
         year = 2011,
        month = sep,
       volume = {416},
       number = {2},
        pages = {832-856},
          doi = {10.1111/j.1365-2966.2011.19054.x},
archivePrefix = {arXiv},
       eprint = {1105.3532},
 primaryClass = {astro-ph.IM},
       adsurl = {https://ui.adsabs.harvard.edu/abs/2011MNRAS.416..832W}
}

@ARTICLE{racs_mid_duchesne_2024,
       author = {{Duchesne}, S.~W. and {Grundy}, J.~A. and {Heald}, George H. and {Lenc}, Emil and {Leung}, James K. and {McConnell}, David and {Murphy}, Tara and {Pritchard}, Joshua and {Rose}, Kovi and {Thomson}, Alec J.~M. and {Wang}, Yuanming and {Wang}, Ziteng and {Whiting}, Matthew T.},
        title = "{The Rapid ASKAP Continuum Survey V: Cataloguing the sky at 1 367.5 MHz and the second data release of RACS-mid}",
      journal = {\pasa},
         year = 2024,
        month = jan,
       volume = {41},
          eid = {e003},
        pages = {e003},
          doi = {10.1017/pasa.2023.60},
archivePrefix = {arXiv},
       eprint = {2311.12369},
 primaryClass = {astro-ph.GA},
       adsurl = {https://ui.adsabs.harvard.edu/abs/2024PASA...41....3D}
}

@ARTICLE{ar_sco_marsh,
       author = {{Marsh}, T.~R. and {G{\"a}nsicke}, B.~T. and {H{\"u}mmerich}, S. and {Hambsch}, F. -J. and {Bernhard}, K. and {Lloyd}, C. and {Breedt}, E. and {Stanway}, E.~R. and {Steeghs}, D.~T. and {Parsons}, S.~G. and {Toloza}, O. and {Schreiber}, M.~R. and {Jonker}, P.~G. and {van Roestel}, J. and {Kupfer}, T. and {Pala}, A.~F. and {Dhillon}, V.~S. and {Hardy}, L.~K. and {Littlefair}, S.~P. and {Aungwerojwit}, A. and {Arjyotha}, S. and {Koester}, D. and {Bochinski}, J.~J. and {Haswell}, C.~A. and {Frank}, P. and {Wheatley}, P.~J.},
        title = "{A radio-pulsing white dwarf binary star}",
      journal = {\nat},
         year = 2016,
        month = sep,
       volume = {537},
       number = {7620},
        pages = {374-377},
          doi = {10.1038/nature18620},
archivePrefix = {arXiv},
       eprint = {1607.08265},
 primaryClass = {astro-ph.SR},
       adsurl = {https://ui.adsabs.harvard.edu/abs/2016Natur.537..374M}
}

@ARTICLE{srsc,
       author = {{Driessen}, Laura Nicole and {Pritchard}, Joshua and {Murphy}, Tara and {Heald}, George and {Robrade}, Jan and {Das}, Barnali and {Duchesne}, Stefan William and {Kaplan}, David L. and {Lenc}, Emil and {Lynch}, Christene R. and {Mitchell-Bolton}, Jackson and {Pope}, Benjamin J.~S. and {Rose}, Kovi and {Stelzer}, Beate and {Wang}, Yuanming and {Zic}, Andrew},
        title = "{The Sydney Radio Star Catalogue: Properties of radio stars at megahertz to gigahertz frequencies}",
      journal = {\pasa},
         year = 2024,
        month = nov,
       volume = {41},
          eid = {e084},
        pages = {e084},
          doi = {10.1017/pasa.2024.72},
archivePrefix = {arXiv},
       eprint = {2404.07418},
 primaryClass = {astro-ph.SR},
       adsurl = {https://ui.adsabs.harvard.edu/abs/2024PASA...41...84D}
}

@ARTICLE{j1912_pelisoli,
       author = {{Pelisoli}, Ingrid and {Marsh}, T.~R. and {Buckley}, David A.~H. and {Heywood}, I. and {Potter}, Stephen. B. and {Schwope}, Axel and {Brink}, Jaco and {Standke}, Annie and {Woudt}, P.~A. and {Parsons}, S.~G. and {Green}, M.~J. and {Kepler}, S.~O. and {Munday}, James and {Romero}, A.~D. and {Breedt}, E. and {Brown}, A.~J. and {Dhillon}, V.~S. and {Dyer}, M.~J. and {Kerry}, P. and {Littlefair}, S.~P. and {Sahman}, D.~I. and {Wild}, J.~F.},
        title = "{A 5.3-min-period pulsing white dwarf in a binary detected from radio to X-rays}",
      journal = {\nastro},
         year = 2023,
        month = aug,
       volume = {7},
        pages = {931-942},
          doi = {10.1038/s41550-023-01995-x},
archivePrefix = {arXiv},
       eprint = {2306.09272},
 primaryClass = {astro-ph.SR},
       adsurl = {https://ui.adsabs.harvard.edu/abs/2023NatAs...7..931P}
}

@ARTICLE{circ_pol_ar_sco_stanway,
       author = {{Stanway}, E.~R. and {Marsh}, T.~R. and {Chote}, P. and {G{\"a}nsicke}, B.~T. and {Steeghs}, D. and {Wheatley}, P.~J.},
        title = "{VLA radio observations of AR Scorpii}",
      journal = {\aap},
         year = 2018,
        month = mar,
       volume = {611},
          eid = {A66},
        pages = {A66},
          doi = {10.1051/0004-6361/201732380},
archivePrefix = {arXiv},
       eprint = {1801.07258},
 primaryClass = {astro-ph.SR},
       adsurl = {https://ui.adsabs.harvard.edu/abs/2018A&A...611A..66S}
}

@ARTICLE{gentile_fusilio_gaia_edr3_wd_2021,
       author = {{Gentile Fusillo}, N.~P. and {Tremblay}, P. -E. and {Cukanovaite}, E. and {Vorontseva}, A. and {Lallement}, R. and {Hollands}, M. and {G{\"a}nsicke}, B.~T. and {Burdge}, K.~B. and {McCleery}, J. and {Jordan}, S.},
        title = "{A catalogue of white dwarfs in Gaia EDR3}",
      journal = {\mnras},
         year = 2021,
        month = dec,
       volume = {508},
       number = {3},
        pages = {3877-3896},
          doi = {10.1093/mnras/stab2672},
archivePrefix = {arXiv},
       eprint = {2106.07669},
 primaryClass = {astro-ph.SR},
       adsurl = {https://ui.adsabs.harvard.edu/abs/2021MNRAS.508.3877G}
}

@ARTICLE{ridder_cv_2023,
       author = {{Ridder}, M.~E. and {Heinke}, C.~O. and {Sivakoff}, G.~R. and {Hughes}, A.~K.},
        title = "{Radio detections of two unusual cataclysmic variables in the VLA Sky Survey}",
      journal = {\mnras},
         year = 2023,
        month = mar,
       volume = {519},
       number = {4},
        pages = {5922-5930},
          doi = {10.1093/mnras/stad038},
archivePrefix = {arXiv},
       eprint = {2303.01438},
 primaryClass = {astro-ph.HE},
       adsurl = {https://ui.adsabs.harvard.edu/abs/2023MNRAS.519.5922R}
}

@ARTICLE{gehrels_swift_2004,
       author = {{Gehrels}, N. and {Chincarini}, G. and {Giommi}, P. and {Mason}, K.~O. and {Nousek}, J.~A. and {Wells}, A.~A. and {White}, N.~E. and {Barthelmy}, S.~D. and {Burrows}, D.~N. and {Cominsky}, L.~R. and {Hurley}, K.~C. and {Marshall}, F.~E. and {M{\'e}sz{\'a}ros}, P. and {Roming}, P.~W.~A. and {Angelini}, L. and {Barbier}, L.~M. and {Belloni}, T. and {Campana}, S. and {Caraveo}, P.~A. and {Chester}, M.~M. and {Citterio}, O. and {Cline}, T.~L. and {Cropper}, M.~S. and {Cummings}, J.~R. and {Dean}, A.~J. and {Feigelson}, E.~D. and {Fenimore}, E.~E. and {Frail}, D.~A. and {Fruchter}, A.~S. and {Garmire}, G.~P. and {Gendreau}, K. and {Ghisellini}, G. and {Greiner}, J. and {Hill}, J.~E. and {Hunsberger}, S.~D. and {Krimm}, H.~A. and {Kulkarni}, S.~R. and {Kumar}, P. and {Lebrun}, F. and {Lloyd-Ronning}, N.~M. and {Markwardt}, C.~B. and {Mattson}, B.~J. and {Mushotzky}, R.~F. and {Norris}, J.~P. and {Osborne}, J. and {Paczynski}, B. and {Palmer}, D.~M. and {Park}, H. -S. and {Parsons}, A.~M. and {Paul}, J. and {Rees}, M.~J. and {Reynolds}, C.~S. and {Rhoads}, J.~E. and {Sasseen}, T.~P. and {Schaefer}, B.~E. and {Short}, A.~T. and {Smale}, A.~P. and {Smith}, I.~A. and {Stella}, L. and {Tagliaferri}, G. and {Takahashi}, T. and {Tashiro}, M. and {Townsley}, L.~K. and {Tueller}, J. and {Turner}, M.~J.~L. and {Vietri}, M. and {Voges}, W. and {Ward}, M.~J. and {Willingale}, R. and {Zerbi}, F.~M. and {Zhang}, W.~W.},
        title = "{The Swift Gamma-Ray Burst Mission}",
      journal = {\apj},
         year = 2004,
        month = aug,
       volume = {611},
       number = {2},
        pages = {1005-1020},
          doi = {10.1086/422091},
archivePrefix = {arXiv},
       eprint = {astro-ph/0405233},
 primaryClass = {astro-ph},
       adsurl = {https://ui.adsabs.harvard.edu/abs/2004ApJ...611.1005G}
}

@ARTICLE{Taghizadeh-Popp_sciserver_2020,
       author = {{Taghizadeh-Popp}, M. and {Kim}, J.~W. and {Lemson}, G. and {Medvedev}, D. and {Raddick}, M.~J. and {Szalay}, A.~S. and {Thakar}, A.~R. and {Booker}, J. and {Chhetri}, C. and {Dobos}, L. and {Rippin}, M.},
        title = "{SciServer: A science platform for astronomy and beyond}",
      journal = {\ac},
         year = 2020,
        month = oct,
       volume = {33},
          eid = {100412},
        pages = {100412},
          doi = {10.1016/j.ascom.2020.100412},
archivePrefix = {arXiv},
       eprint = {2001.08619},
 primaryClass = {astro-ph.IM},
       adsurl = {https://ui.adsabs.harvard.edu/abs/2020A&C....3300412T}
}

@ARTICLE{gaia_dr3,
       author = {{Gaia Collaboration} and others},
        title = "{Gaia Data Release 3. Summary of the content and survey properties}",
      journal = {\aap},
         year = 2023,
        month = jun,
       volume = {674},
          eid = {A1},
        pages = {A1},
          doi = {10.1051/0004-6361/202243940},
archivePrefix = {arXiv},
       eprint = {2208.00211},
 primaryClass = {astro-ph.GA},
       adsurl = {https://ui.adsabs.harvard.edu/abs/2023A&A...674A...1G}
}

@INPROCEEDINGS{clemens_goodman_2004,
       author = {{Clemens}, J. Christopher and {Crain}, J. Adam and {Anderson}, Robert},
        title = "{The Goodman spectrograph}",
    booktitle = {Ground-based Instrumentation for Astronomy},
         year = 2004,
       editor = {{Moorwood}, Alan F.~M. and {Iye}, Masanori},
       series = {Society of Photo-Optical Instrumentation Engineers (SPIE) Conference Series},
       volume = {5492},
        month = sep,
        pages = {331-340},
          doi = {10.1117/12.550069},
       adsurl = {https://ui.adsabs.harvard.edu/abs/2004SPIE.5492..331C}
}

@ARTICLE{parkes_uwl_hobbs_2020,
       author = {{Hobbs}, George and {Manchester}, Richard N. and {Dunning}, Alex and {Jameson}, Andrew and {Roberts}, Paul and {George}, Daniel and {Green}, J.~A. and {Tuthill}, John and {Toomey}, Lawrence and {Kaczmarek}, Jane F. and {Mader}, Stacy and {Marquarding}, Malte and {Ahmed}, Azeem and {Amy}, Shaun W. and {Bailes}, Matthew and {Beresford}, Ron and {Bhat}, N.~D.~R. and {Bock}, Douglas C. -J. and {Bourne}, Michael and {Bowen}, Mark and {Brothers}, Michael and {Cameron}, Andrew D. and {Carretti}, Ettore and {Carter}, Nick and {Castillo}, Santy and {Chekkala}, Raji and {Cheng}, Wan and {Chung}, Yoon and {Craig}, Daniel A. and {Dai}, Shi and {Dawson}, Joanne and {Dempsey}, James and {Doherty}, Paul and {Dong}, Bin and {Edwards}, Philip and {Ergesh}, Tuohutinuer and {Gao}, Xuyang and {Han}, JinLin and {Hayman}, Douglas and {Indermuehle}, Balthasar and {Jeganathan}, Kanapathippillai and {Johnston}, Simon and {Kanoniuk}, Henry and {Kesteven}, Michael and {Kramer}, Michael and {Leach}, Mark and {Mcintyre}, Vince and {Moss}, Vanessa and {Os{\l}owski}, Stefan and {Phillips}, Chris and {Pope}, Nathan and {Preisig}, Brett and {Price}, Daniel and {Reeves}, Ken and {Reilly}, Les and {Reynolds}, John and {Robishaw}, Tim and {Roush}, Peter and {Ruckley}, Tim and {Sadler}, Elaine and {Sarkissian}, John and {Severs}, Sean and {Shannon}, Ryan and {Smart}, Ken and {Smith}, Malcolm and {Smith}, Stephanie and {Sobey}, Charlotte and {Staveley-Smith}, Lister and {Tzioumis}, Anastasios and {van Straten}, Willem and {Wang}, Nina and {Wen}, Linqing and {Whiting}, Matthew},
        title = "{An ultra-wide bandwidth (704 to 4 032 MHz) receiver for the Parkes radio telescope}",
      journal = {\pasa},
         year = 2020,
        month = apr,
       volume = {37},
          eid = {e012},
        pages = {e012},
          doi = {10.1017/pasa.2020.2},
archivePrefix = {arXiv},
       eprint = {1911.00656},
 primaryClass = {astro-ph.IM},
       adsurl = {https://ui.adsabs.harvard.edu/abs/2020PASA...37...12H}
}

@ARTICLE{ymw16_dm,
       author = {{Yao}, J.~M. and {Manchester}, R.~N. and {Wang}, N.},
        title = "{A New Electron-density Model for Estimation of Pulsar and FRB Distances}",
      journal = {\apj},
         year = 2017,
        month = jan,
       volume = {835},
       number = {1},
          eid = {29},
        pages = {29},
          doi = {10.3847/1538-4357/835/1/29},
archivePrefix = {arXiv},
       eprint = {1610.09448},
 primaryClass = {astro-ph.GA},
       adsurl = {https://ui.adsabs.harvard.edu/abs/2017ApJ...835...29Y}
}

@ARTICLE{hotan_21,
       author = {{Hotan}, A.~W. and {Bunton}, J.~D. and {Chippendale}, A.~P. and {Whiting}, M. and {Tuthill}, J. and {Moss}, V.~A. and {McConnell}, D. and {Amy}, S.~W. and {Huynh}, M.~T. and {Allison}, J.~R. and {Anderson}, C.~S. and {Bannister}, K.~W. and {Bastholm}, E. and {Beresford}, R. and {Bock}, D.~C. -J. and {Bolton}, R. and {Chapman}, J.~M. and {Chow}, K. and {Collier}, J.~D. and {Cooray}, F.~R. and {Cornwell}, T.~J. and {Diamond}, P.~J. and {Edwards}, P.~G. and {Feain}, I.~J. and {Franzen}, T.~M.~O. and {George}, D. and {Gupta}, N. and {Hampson}, G.~A. and {Harvey-Smith}, L. and {Hayman}, D.~B. and {Heywood}, I. and {Jacka}, C. and {Jackson}, C.~A. and {Jackson}, S. and {Jeganathan}, K. and {Johnston}, S. and {Kesteven}, M. and {Kleiner}, D. and {Koribalski}, B.~S. and {Lee-Waddell}, K. and {Lenc}, E. and {Lensson}, E.~S. and {Mackay}, S. and {Mahony}, E.~K. and {McClure-Griffiths}, N.~M. and {McConigley}, R. and {Mirtschin}, P. and {Ng}, A.~K. and {Norris}, R.~P. and {Pearce}, S.~E. and {Phillips}, C. and {Pilawa}, M.~A. and {Raja}, W. and {Reynolds}, J.~E. and {Roberts}, P. and {Roxby}, D.~N. and {Sadler}, E.~M. and {Shields}, M. and {Schinckel}, A.~E.~T. and {Serra}, P. and {Shaw}, R.~D. and {Sweetnam}, T. and {Troup}, E.~R. and {Tzioumis}, A. and {Voronkov}, M.~A. and {Westmeier}, T.},
        title = "{Australian square kilometre array pathfinder: I. system description}",
      journal = {\pasa},
         year = 2021,
        month = mar,
       volume = {38},
          eid = {e009},
        pages = {e009},
          doi = {10.1017/pasa.2021.1},
archivePrefix = {arXiv},
       eprint = {2102.01870},
 primaryClass = {astro-ph.IM},
       adsurl = {https://ui.adsabs.harvard.edu/abs/2021PASA...38....9H}
}

@ARTICLE{bailer-jones_21,
       author = {{Bailer-Jones}, C.~A.~L. and {Rybizki}, J. and {Fouesneau}, M. and {Demleitner}, M. and {Andrae}, R.},
        title = "{Estimating Distances from Parallaxes. V. Geometric and Photogeometric Distances to 1.47 Billion Stars in Gaia Early Data Release 3}",
      journal = {\aj},
         year = 2021,
        month = mar,
       volume = {161},
       number = {3},
          eid = {147},
        pages = {147},
          doi = {10.3847/1538-3881/abd806},
archivePrefix = {arXiv},
       eprint = {2012.05220},
 primaryClass = {astro-ph.SR},
       adsurl = {https://ui.adsabs.harvard.edu/abs/2021AJ....161..147B}
}

@software{dstools,
  author       = {Joshua Pritchard},
  title        = {askap-vast/dstools: v2.0.0},
  month        = apr,
  year         = 2025,
  publisher    = {Zenodo},
  version      = {v2.0.0},
  doi          = {10.5281/zenodo.15232974},
  url          = {https://doi.org/10.5281/zenodo.15232974},
  swhid        = {swh:1:dir:768be72ee3fd49d0df27426778a6ef4ea87d223c
                   ;origin=https://doi.org/10.5281/zenodo.13626182;vi
                   sit=swh:1:snp:8db00c2f26d92872b3675ffd9c6f3fba79c3
                   eac1;anchor=swh:1:rel:5cae0f6e0c8762f81b826df80f54
                   76adf3b52db3;path=askap-vast-dstools-f8971fb
                  },
}

@ARTICLE{casa_2022,
       author = {{CASA Team} and {Bean}, Ben and {Bhatnagar}, Sanjay and {Castro}, Sandra and {Donovan Meyer}, Jennifer and {Emonts}, Bjorn and {Garcia}, Enrique and {Garwood}, Robert and {Golap}, Kumar and {Gonzalez Villalba}, Justo and {Harris}, Pamela and {Hayashi}, Yohei and {Hoskins}, Josh and {Hsieh}, Mingyu and {Jagannathan}, Preshanth and {Kawasaki}, Wataru and {Keimpema}, Aard and {Kettenis}, Mark and {Lopez}, Jorge and {Marvil}, Joshua and {Masters}, Joseph and {McNichols}, Andrew and {Mehringer}, David and {Miel}, Renaud and {Moellenbrock}, George and {Montesino}, Federico and {Nakazato}, Takeshi and {Ott}, Juergen and {Petry}, Dirk and {Pokorny}, Martin and {Raba}, Ryan and {Rau}, Urvashi and {Schiebel}, Darrell and {Schweighart}, Neal and {Sekhar}, Srikrishna and {Shimada}, Kazuhiko and {Small}, Des and {Steeb}, Jan-Willem and {Sugimoto}, Kanako and {Suoranta}, Ville and {Tsutsumi}, Takahiro and {van Bemmel}, Ilse M. and {Verkouter}, Marjolein and {Wells}, Akeem and {Xiong}, Wei and {Szomoru}, Arpad and {Griffith}, Morgan and {Glendenning}, Brian and {Kern}, Jeff},
        title = "{CASA, the Common Astronomy Software Applications for Radio Astronomy}",
      journal = {\pasp},
         year = 2022,
        month = nov,
       volume = {134},
       number = {1041},
          eid = {114501},
        pages = {114501},
          doi = {10.1088/1538-3873/ac9642},
archivePrefix = {arXiv},
       eprint = {2210.02276},
 primaryClass = {astro-ph.IM},
       adsurl = {https://ui.adsabs.harvard.edu/abs/2022PASP..134k4501C}
}

@ARTICLE{rose_2023,
       author = {{Rose}, Kovi and {Pritchard}, Joshua and {Murphy}, Tara and {Caleb}, Manisha and {Dobie}, Dougal and {Driessen}, Laura and {Duchesne}, Stefan W. and {Kaplan}, David L. and {Lenc}, Emil and {Wang}, Ziteng},
        title = "{Periodic Radio Emission from the T8 Dwarf WISE J062309.94-045624.6}",
      journal = {\apjl},
         year = 2023,
        month = jul,
       volume = {951},
       number = {2},
          eid = {L43},
        pages = {L43},
          doi = {10.3847/2041-8213/ace188},
archivePrefix = {arXiv},
       eprint = {2306.15219},
 primaryClass = {astro-ph.SR},
       adsurl = {https://ui.adsabs.harvard.edu/abs/2023ApJ...951L..43R}
}

@ARTICLE{pritchard_2021,
       author = {{Pritchard}, Joshua and {Murphy}, Tara and {Zic}, Andrew and {Lynch}, Christene and {Heald}, George and {Kaplan}, David L. and {Anderson}, Craig and {Banfield}, Julie and {Hale}, Catherine and {Hotan}, Aidan and {Lenc}, Emil and {Leung}, James K. and {McConnell}, David and {Moss}, Vanessa A. and {Raja}, Wasim and {Stewart}, Adam J. and {Whiting}, Matthew},
        title = "{A circular polarization survey for radio stars with the Australian SKA Pathfinder}",
      journal = {\mnras},
         year = 2021,
        month = apr,
       volume = {502},
       number = {4},
        pages = {5438-5454},
          doi = {10.1093/mnras/stab299},
archivePrefix = {arXiv},
       eprint = {2102.01801},
 primaryClass = {astro-ph.SR},
       adsurl = {https://ui.adsabs.harvard.edu/abs/2021MNRAS.502.5438P}
}

@ARTICLE{astropy:2022,
       author = {{Astropy Collaboration} and {Price-Whelan}, Adrian M. and {Lim}, Pey Lian and {Earl}, Nicholas and {Starkman}, Nathaniel and {Bradley}, Larry and {Shupe}, David L. and {Patil}, Aarya A. and {Corrales}, Lia and {Brasseur}, C.~E. and {N{"o}the}, Maximilian and {Donath}, Axel and {Tollerud}, Erik and {Morris}, Brett M. and {Ginsburg}, Adam and {Vaher}, Eero and {Weaver}, Benjamin A. and {Tocknell}, James and {Jamieson}, William and {van Kerkwijk}, Marten H. and {Robitaille}, Thomas P. and {Merry}, Bruce and {Bachetti}, Matteo and {G{"u}nther}, H. Moritz and {Aldcroft}, Thomas L. and {Alvarado-Montes}, Jaime A. and {Archibald}, Anne M. and {B{'o}di}, Attila and {Bapat}, Shreyas and {Barentsen}, Geert and {Baz{'a}n}, Juanjo and {Biswas}, Manish and {Boquien}, M{'e}d{'e}ric and {Burke}, D.~J. and {Cara}, Daria and {Cara}, Mihai and {Conroy}, Kyle E. and {Conseil}, Simon and {Craig}, Matthew W. and {Cross}, Robert M. and {Cruz}, Kelle L. and {D'Eugenio}, Francesco and {Dencheva}, Nadia and {Devillepoix}, Hadrien A.~R. and {Dietrich}, J{"o}rg P. and {Eigenbrot}, Arthur Davis and {Erben}, Thomas and {Ferreira}, Leonardo and {Foreman-Mackey}, Daniel and {Fox}, Ryan and {Freij}, Nabil and {Garg}, Suyog and {Geda}, Robel and {Glattly}, Lauren and {Gondhalekar}, Yash and {Gordon}, Karl D. and {Grant}, David and {Greenfield}, Perry and {Groener}, Austen M. and {Guest}, Steve and {Gurovich}, Sebastian and {Handberg}, Rasmus and {Hart}, Akeem and {Hatfield-Dodds}, Zac and {Homeier}, Derek and {Hosseinzadeh}, Griffin and {Jenness}, Tim and {Jones}, Craig K. and {Joseph}, Prajwel and {Kalmbach}, J. Bryce and {Karamehmetoglu}, Emir and {Ka{l}uszy{'n}ski}, Miko{l}aj and {Kelley}, Michael S.~P. and {Kern}, Nicholas and {Kerzendorf}, Wolfgang E. and {Koch}, Eric W. and {Kulumani}, Shankar and {Lee}, Antony and {Ly}, Chun and {Ma}, Zhiyuan and {MacBride}, Conor and {Maljaars}, Jakob M. and {Muna}, Demitri and {Murphy}, N.~A. and {Norman}, Henrik and {O'Steen}, Richard and {Oman}, Kyle A. and {Pacifici}, Camilla and {Pascual}, Sergio and {Pascual-Granado}, J. and {Patil}, Rohit R. and {Perren}, Gabriel I. and {Pickering}, Timothy E. and {Rastogi}, Tanuj and {Roulston}, Benjamin R. and {Ryan}, Daniel F. and {Rykoff}, Eli S. and {Sabater}, Jose and {Sakurikar}, Parikshit and {Salgado}, Jes{'u}s and {Sanghi}, Aniket and {Saunders}, Nicholas and {Savchenko}, Volodymyr and {Schwardt}, Ludwig and {Seifert-Eckert}, Michael and {Shih}, Albert Y. and {Jain}, Anany Shrey and {Shukla}, Gyanendra and {Sick}, Jonathan and {Simpson}, Chris and {Singanamalla}, Sudheesh and {Singer}, Leo P. and {Singhal}, Jaladh and {Sinha}, Manodeep and {Sip{H{o}}cz}, Brigitta M. and {Spitler}, Lee R. and {Stansby}, David and {Streicher}, Ole and {{{S}}umak}, Jani and {Swinbank}, John D. and {Taranu}, Dan S. and {Tewary}, Nikita and {Tremblay}, Grant R. and {Val-Borro}, Miguel de and {Van Kooten}, Samuel J. and {Vasovi{'c}}, Zlatan and {Verma}, Shresth and {de Miranda Cardoso}, Jos{'e} Vin{'i}cius and {Williams}, Peter K.~G. and {Wilson}, Tom J. and {Winkel}, Benjamin and {Wood-Vasey}, W.~M. and {Xue}, Rui and {Yoachim}, Peter and {Zhang}, Chen and {Zonca}, Andrea and {Astropy Project Contributors}},
        title = "{The Astropy Project: Sustaining and Growing a Community-oriented Open-source Project and the Latest Major Release (v5.0) of the Core Package}",
      journal = {\apj},
         year = 2022,
        month = aug,
       volume = {935},
       number = {2},
          eid = {167},
        pages = {167},
          doi = {10.3847/1538-4357/ac7c74},
archivePrefix = {arXiv},
       eprint = {2206.14220},
 primaryClass = {astro-ph.IM},
       adsurl = {https://ui.adsabs.harvard.edu/abs/2022ApJ...935..167A}
}

@ARTICLE{thejoker_2017,
       author = {{Price-Whelan}, Adrian M. and {Hogg}, David W. and {Foreman-Mackey}, Daniel and {Rix}, Hans-Walter},
        title = "{The Joker: A Custom Monte Carlo Sampler for Binary-star and Exoplanet Radial Velocity Data}",
      journal = {\apj},
         year = 2017,
        month = mar,
       volume = {837},
       number = {1},
          eid = {20},
        pages = {20},
          doi = {10.3847/1538-4357/aa5e50},
archivePrefix = {arXiv},
       eprint = {1610.07602},
 primaryClass = {astro-ph.SR},
       adsurl = {https://ui.adsabs.harvard.edu/abs/2017ApJ...837...20P}
}

@ARTICLE{galex_bianchi,
       author = {{Bianchi}, Luciana and {Shiao}, Bernie and {Thilker}, David},
        title = "{Revised Catalog of GALEX Ultraviolet Sources. I. The All-Sky Survey: GUVcat\_AIS}",
      journal = {\apjs},
         year = 2017,
        month = jun,
       volume = {230},
       number = {2},
          eid = {24},
        pages = {24},
          doi = {10.3847/1538-4365/aa7053},
archivePrefix = {arXiv},
       eprint = {1704.05903},
 primaryClass = {astro-ph.GA},
       adsurl = {https://ui.adsabs.harvard.edu/abs/2017ApJS..230...24B}
}

@software{rmtools_purcell_2020,
       author = {{Purcell}, C.~R. and {Van Eck}, C.~L. and {West}, J. and {Sun}, X.~H. and {Gaensler}, B.~M.},
        title = "{RM-Tools: Rotation measure (RM) synthesis and Stokes QU-fitting}",
 howpublished = {Astrophysics Source Code Library, record ascl:2005.003},
         year = 2020,
        month = may,
          eid = {ascl:2005.003},
       adsurl = {https://ui.adsabs.harvard.edu/abs/2020ascl.soft05003P}
}

@ARTICLE{nifty_garrison_2024,
       author = {{Garrison}, Lehman H. and {Foreman-Mackey}, Dan and {Shih}, Yu-hsuan and {Barnett}, Alex},
        title = "{NIFTY-LS: Fast and Accurate Lomb{\textendash}Scargle Periodograms Using a Non-uniform FFT}",
      journal = {\rnaas},
         year = 2024,
        month = oct,
       volume = {8},
       number = {10},
          eid = {250},
        pages = {250},
          doi = {10.3847/2515-5172/ad82cd},
archivePrefix = {arXiv},
       eprint = {2409.08090},
 primaryClass = {astro-ph.IM},
       adsurl = {https://ui.adsabs.harvard.edu/abs/2024RNAAS...8..250G}
}

@ARTICLE{ne2001_dm,
       author = {{Cordes}, J.~M. and {Lazio}, T.~J.~W.},
        title = "{NE2001.I. A New Model for the Galactic Distribution of Free Electrons and its Fluctuations}",
      journal = {arXiv e-prints},
         year = 2002,
        month = jul,
          eid = {astro-ph/0207156},
        pages = {astro-ph/0207156},
          doi = {10.48550/arXiv.astro-ph/0207156},
archivePrefix = {arXiv},
       eprint = {astro-ph/0207156},
 primaryClass = {astro-ph},
       adsurl = {https://ui.adsabs.harvard.edu/abs/2002astro.ph..7156C}
}

@ARTICLE{alma_2009,
       author = {{Wootten}, Alwyn and {Thompson}, A. Richard},
        title = "{The Atacama Large Millimeter/Submillimeter Array}",
      journal = {IEEE Proceedings},
         year = 2009,
        month = aug,
       volume = {97},
       number = {8},
        pages = {1463-1471},
          doi = {10.1109/JPROC.2009.2020572},
archivePrefix = {arXiv},
       eprint = {0904.3739},
 primaryClass = {astro-ph.IM},
       adsurl = {https://ui.adsabs.harvard.edu/abs/2009IEEEP..97.1463W}
}

@ARTICLE{gaia_mission_2016,
       author = {{Gaia Collaboration} and {Prusti}, T. and {de Bruijne}, J.~H.~J. and {Brown}, A.~G.~A. and {Vallenari}, A. and {Babusiaux}, C. and {Bailer-Jones}, C.~A.~L. and {Bastian}, U. and {Biermann}, M. and {Evans}, D.~W. and {Eyer}, L. and {Jansen}, F. and {Jordi}, C. and {Klioner}, S.~A. and {Lammers}, U. and {Lindegren}, L. and {Luri}, X. and {Mignard}, F. and {Milligan}, D.~J. and {Panem}, C. and {Poinsignon}, V. and {Pourbaix}, D. and {Randich}, S. and {Sarri}, G. and {Sartoretti}, P. and {Siddiqui}, H.~I. and {Soubiran}, C. and {Valette}, V. and {van Leeuwen}, F. and {Walton}, N.~A. and {Aerts}, C. and {Arenou}, F. and {Cropper}, M. and {Drimmel}, R. and {H{\o}g}, E. and {Katz}, D. and {Lattanzi}, M.~G. and {O'Mullane}, W. and {Grebel}, E.~K. and {Holland}, A.~D. and {Huc}, C. and {Passot}, X. and {Bramante}, L. and {Cacciari}, C. and {Casta{\~n}eda}, J. and {Chaoul}, L. and {Cheek}, N. and {De Angeli}, F. and {Fabricius}, C. and {Guerra}, R. and {Hern{\'a}ndez}, J. and {Jean-Antoine-Piccolo}, A. and {Masana}, E. and {Messineo}, R. and {Mowlavi}, N. and {Nienartowicz}, K. and {Ord{\'o}{\~n}ez-Blanco}, D. and {Panuzzo}, P. and {Portell}, J. and {Richards}, P.~J. and {Riello}, M. and {Seabroke}, G.~M. and {Tanga}, P. and {Th{\'e}venin}, F. and {Torra}, J. and {Els}, S.~G. and {Gracia-Abril}, G. and {Comoretto}, G. and {Garcia-Reinaldos}, M. and {Lock}, T. and {Mercier}, E. and {Altmann}, M. and {Andrae}, R. and {Astraatmadja}, T.~L. and {Bellas-Velidis}, I. and {Benson}, K. and {Berthier}, J. and {Blomme}, R. and {Busso}, G. and {Carry}, B. and {Cellino}, A. and {Clementini}, G. and {Cowell}, S. and {Creevey}, O. and {Cuypers}, J. and {Davidson}, M. and {De Ridder}, J. and {de Torres}, A. and {Delchambre}, L. and {Dell'Oro}, A. and {Ducourant}, C. and {Fr{\'e}mat}, Y. and {Garc{\'\i}a-Torres}, M. and {Gosset}, E. and {Halbwachs}, J. -L. and {Hambly}, N.~C. and {Harrison}, D.~L. and {Hauser}, M. and {Hestroffer}, D. and {Hodgkin}, S.~T. and {Huckle}, H.~E. and {Hutton}, A. and {Jasniewicz}, G. and {Jordan}, S. and {Kontizas}, M. and {Korn}, A.~J. and {Lanzafame}, A.~C. and {Manteiga}, M. and {Moitinho}, A. and {Muinonen}, K. and {Osinde}, J. and {Pancino}, E. and {Pauwels}, T. and {Petit}, J. -M. and {Recio-Blanco}, A. and {Robin}, A.~C. and {Sarro}, L.~M. and {Siopis}, C. and {Smith}, M. and {Smith}, K.~W. and {Sozzetti}, A. and {Thuillot}, W. and {van Reeven}, W. and {Viala}, Y. and {Abbas}, U. and {Abreu Aramburu}, A. and {Accart}, S. and {Aguado}, J.~J. and {Allan}, P.~M. and {Allasia}, W. and {Altavilla}, G. and {{\'A}lvarez}, M.~A. and {Alves}, J. and {Anderson}, R.~I. and {Andrei}, A.~H. and {Anglada Varela}, E. and {Antiche}, E. and {Antoja}, T. and {Ant{\'o}n}, S. and {Arcay}, B. and {Atzei}, A. and {Ayache}, L. and {Bach}, N. and {Baker}, S.~G. and {Balaguer-N{\'u}{\~n}ez}, L. and {Barache}, C. and {Barata}, C. and {Barbier}, A. and {Barblan}, F. and {Baroni}, M. and {Barrado y Navascu{\'e}s}, D. and {Barros}, M. and {Barstow}, M.~A. and {Becciani}, U. and {Bellazzini}, M. and {Bellei}, G. and {Bello Garc{\'\i}a}, A. and {Belokurov}, V. and {Bendjoya}, P. and {Berihuete}, A. and {Bianchi}, L. and {Bienaym{\'e}}, O. and {Billebaud}, F. and {Blagorodnova}, N. and {Blanco-Cuaresma}, S. and {Boch}, T. and {Bombrun}, A. and {Borrachero}, R. and {Bouquillon}, S. and {Bourda}, G. and {Bouy}, H. and {Bragaglia}, A. and {Breddels}, M.~A. and {Brouillet}, N. and {Br{\"u}semeister}, T. and {Bucciarelli}, B. and {Budnik}, F. and {Burgess}, P. and {Burgon}, R. and {Burlacu}, A. and {Busonero}, D. and {Buzzi}, R. and {Caffau}, E. and {Cambras}, J. and {Campbell}, H. and {Cancelliere}, R. and {Cantat-Gaudin}, T. and {Carlucci}, T. and {Carrasco}, J.~M. and {Castellani}, M. and {Charlot}, P. and {Charnas}, J. and {Charvet}, P. and {Chassat}, F. and {Chiavassa}, A. and {Clotet}, M. and {Cocozza}, G. and {Collins}, R.~S. and {Collins}, P. and {Costigan}, G.},
        title = "{The Gaia mission}",
      journal = {\aap},
         year = 2016,
        month = nov,
       volume = {595},
          eid = {A1},
        pages = {A1},
          doi = {10.1051/0004-6361/201629272},
archivePrefix = {arXiv},
       eprint = {1609.04153},
 primaryClass = {astro-ph.IM},
       adsurl = {https://ui.adsabs.harvard.edu/abs/2016A&A...595A...1G}
}

@ARTICLE{Lindegren2012,
       author = {{Lindegren}, L. and {Lammers}, U. and {Hobbs}, D. and {O'Mullane}, W. and {Bastian}, U. and {Hern{\'a}ndez}, J.},
        title = "{The astrometric core solution for the Gaia mission. Overview of models, algorithms, and software implementation}",
      journal = {\aap},
         year = 2012,
        month = feb,
       volume = {538},
          eid = {A78},
        pages = {A78},
          doi = {10.1051/0004-6361/201117905},
archivePrefix = {arXiv},
       eprint = {1112.4139},
 primaryClass = {astro-ph.IM},
       adsurl = {https://ui.adsabs.harvard.edu/abs/2012A&A...538A..78L}
}

@ARTICLE{Luri2018_gaia_dr2_parallax,
       author = {{Luri}, X. and {Brown}, A.~G.~A. and {Sarro}, L.~M. and {Arenou}, F. and {Bailer-Jones}, C.~A.~L. and {Castro-Ginard}, A. and {de Bruijne}, J. and {Prusti}, T. and {Babusiaux}, C. and {Delgado}, H.~E.},
        title = "{Gaia Data Release 2. Using Gaia parallaxes}",
      journal = {\aap},
         year = 2018,
        month = aug,
       volume = {616},
          eid = {A9},
        pages = {A9},
          doi = {10.1051/0004-6361/201832964},
archivePrefix = {arXiv},
       eprint = {1804.09376},
 primaryClass = {astro-ph.IM},
       adsurl = {https://ui.adsabs.harvard.edu/abs/2018A&A...616A...9L}
}

@ARTICLE{Lindegren2021,
       author = {{Lindegren}, L. and {Klioner}, S.~A. and {Hern{\'a}ndez}, J. and {Bombrun}, A. and {Ramos-Lerate}, M. and {Steidelm{\"u}ller}, H. and {Bastian}, U. and {Biermann}, M. and {de Torres}, A. and {Gerlach}, E. and {Geyer}, R. and {Hilger}, T. and {Hobbs}, D. and {Lammers}, U. and {McMillan}, P.~J. and {Stephenson}, C.~A. and {Casta{\~n}eda}, J. and {Davidson}, M. and {Fabricius}, C. and {Gracia-Abril}, G. and {Portell}, J. and {Rowell}, N. and {Teyssier}, D. and {Torra}, F. and {Bartolom{\'e}}, S. and {Clotet}, M. and {Garralda}, N. and {Gonz{\'a}lez-Vidal}, J.~J. and {Torra}, J. and {Abbas}, U. and {Altmann}, M. and {Anglada Varela}, E. and {Balaguer-N{\'u}{\~n}ez}, L. and {Balog}, Z. and {Barache}, C. and {Becciani}, U. and {Bernet}, M. and {Bertone}, S. and {Bianchi}, L. and {Bouquillon}, S. and {Brown}, A.~G.~A. and {Bucciarelli}, B. and {Busonero}, D. and {Butkevich}, A.~G. and {Buzzi}, R. and {Cancelliere}, R. and {Carlucci}, T. and {Charlot}, P. and {Cioni}, M.-R.~L. and {Crosta}, M. and {Crowley}, C. and {del Peloso}, E.~F. and {del Pozo}, E. and {Drimmel}, R. and {Esquej}, P. and {Fienga}, A. and {Fraile}, E. and {Gai}, M. and {Garcia-Reinaldos}, M. and {Guerra}, R. and {Hambly}, N.~C. and {Hauser}, M. and {Jan{\ss}en}, K. and {Jordan}, S. and {Kostrzewa-Rutkowska}, Z. and {Lattanzi}, M.~G. and {Liao}, S. and {Licata}, E. and {Lister}, T.~A. and {L{\"o}ffler}, W. and {Marchant}, J.~M. and {Masip}, A. and {Mignard}, F. and {Mints}, A. and {Molina}, D. and {Mora}, A. and {Morbidelli}, R. and {Murphy}, C.~P. and {Pagani}, C. and {Panuzzo}, P. and {Pe{\~n}alosa Esteller}, X. and {Poggio}, E. and {Re Fiorentin}, P. and {Riva}, A. and {Sagrist{\`a} Sell{\'e}s}, A. and {Sanchez Gimenez}, V. and {Sarasso}, M. and {Sciacca}, E. and {Siddiqui}, H.~I. and {Smart}, R.~L. and {Souami}, D. and {Spagna}, A. and {Steele}, I.~A. and {Taris}, F. and {Utrilla}, E. and {van Reeven}, W. and {Vecchiato}, A.},
        title = "{Gaia Early Data Release 3. The astrometric solution}",
      journal = {\aap},
         year = 2021,
        month = may,
       volume = {649},
          eid = {A2},
        pages = {A2},
          doi = {10.1051/0004-6361/202039709},
archivePrefix = {arXiv},
       eprint = {2012.03380},
 primaryClass = {astro-ph.IM},
       adsurl = {https://ui.adsabs.harvard.edu/abs/2021A&A...649A...2L}
}

@ARTICLE{erosita_merloni_24,
       author = {{Merloni}, A. and {Lamer}, G. and {Liu}, T. and {Ramos-Ceja}, M.~E. and {Brunner}, H. and {Bulbul}, E. and {Dennerl}, K. and {Doroshenko}, V. and {Freyberg}, M.~J. and {Friedrich}, S. and {Gatuzz}, E. and {Georgakakis}, A. and {Haberl}, F. and {Igo}, Z. and {Kreykenbohm}, I. and {Liu}, A. and {Maitra}, C. and {Malyali}, A. and {Mayer}, M.~G.~F. and {Nandra}, K. and {Predehl}, P. and {Robrade}, J. and {Salvato}, M. and {Sanders}, J.~S. and {Stewart}, I. and {Tub{\'\i}n-Arenas}, D. and {Weber}, P. and {Wilms}, J. and {Arcodia}, R. and {Artis}, E. and {Aschersleben}, J. and {Avakyan}, A. and {Aydar}, C. and {Bahar}, Y.~E. and {Balzer}, F. and {Becker}, W. and {Berger}, K. and {Boller}, T. and {Bornemann}, W. and {Br{\"u}ggen}, M. and {Brusa}, M. and {Buchner}, J. and {Burwitz}, V. and {Camilloni}, F. and {Clerc}, N. and {Comparat}, J. and {Coutinho}, D. and {Czesla}, S. and {Dannhauer}, S.~M. and {Dauner}, L. and {Dauser}, T. and {Dietl}, J. and {Dolag}, K. and {Dwelly}, T. and {Egg}, K. and {Ehl}, E. and {Freund}, S. and {Friedrich}, P. and {Gaida}, R. and {Garrel}, C. and {Ghirardini}, V. and {Gokus}, A. and {Gr{\"u}nwald}, G. and {Grandis}, S. and {Grotova}, I. and {Gruen}, D. and {Gueguen}, A. and {H{\"a}mmerich}, S. and {Hamaus}, N. and {Hasinger}, G. and {Haubner}, K. and {Homan}, D. and {Ider Chitham}, J. and {Joseph}, W.~M. and {Joyce}, A. and {K{\"o}nig}, O. and {Kaltenbrunner}, D.~M. and {Khokhriakova}, A. and {Kink}, W. and {Kirsch}, C. and {Kluge}, M. and {Knies}, J. and {Krippendorf}, S. and {Krumpe}, M. and {Kurpas}, J. and {Li}, P. and {Liu}, Z. and {Locatelli}, N. and {Lorenz}, M. and {M{\"u}ller}, S. and {Magaudda}, E. and {Mannes}, C. and {McCall}, H. and {Meidinger}, N. and {Michailidis}, M. and {Migkas}, K. and {Mu{\~n}oz-Giraldo}, D. and {Musiimenta}, B. and {Nguyen-Dang}, N.~T. and {Ni}, Q. and {Olechowska}, A. and {Ota}, N. and {Pacaud}, F. and {Pasini}, T. and {Perinati}, E. and {Pires}, A.~M. and {Pommranz}, C. and {Ponti}, G. and {Poppenhaeger}, K. and {P{\"u}hlhofer}, G. and {Rau}, A. and {Reh}, M. and {Reiprich}, T.~H. and {Roster}, W. and {Saeedi}, S. and {Santangelo}, A. and {Sasaki}, M. and {Schmitt}, J. and {Schneider}, P.~C. and {Schrabback}, T. and {Schuster}, N. and {Schwope}, A. and {Seppi}, R. and {Serim}, M.~M. and {Shreeram}, S. and {Sokolova-Lapa}, E. and {Starck}, H. and {Stelzer}, B. and {Stierhof}, J. and {Suleimanov}, V. and {Tenzer}, C. and {Traulsen}, I. and {Tr{\"u}mper}, J. and {Tsuge}, K. and {Urrutia}, T. and {Veronica}, A. and {Waddell}, S.~G.~H. and {Willer}, R. and {Wolf}, J. and {Yeung}, M.~C.~H. and {Zainab}, A. and {Zangrandi}, F. and {Zhang}, X. and {Zhang}, Y. and {Zheng}, X.},
        title = "{The SRG/eROSITA all-sky survey. First X-ray catalogues and data release of the western Galactic hemisphere}",
      journal = {\aap},
         year = 2024,
        month = feb,
       volume = {682},
          eid = {A34},
        pages = {A34},
          doi = {10.1051/0004-6361/202347165},
archivePrefix = {arXiv},
       eprint = {2401.17274},
 primaryClass = {astro-ph.HE},
       adsurl = {https://ui.adsabs.harvard.edu/abs/2024A&A...682A..34M}
}

@ARTICLE{Qu&Zhang2025,
       author = {{Qu}, Yuanhong and {Zhang}, Bing},
        title = "{Magnetic Interactions in White Dwarf Binaries as Mechanism for Long-period Radio Transients}",
      journal = {\apj},
         year = 2025,
        month = mar,
       volume = {981},
       number = {1},
          eid = {34},
        pages = {34},
          doi = {10.3847/1538-4357/adb1b5},
archivePrefix = {arXiv},
       eprint = {2409.05978},
 primaryClass = {astro-ph.HE},
       adsurl = {https://ui.adsabs.harvard.edu/abs/2025ApJ...981...34Q}
}

@ARTICLE{guver_ozel_nh_2009,
       author = {{G{\"u}ver}, Tolga and {{\"O}zel}, Feryal},
        title = "{The relation between optical extinction and hydrogen column density in the Galaxy}",
      journal = {\mnras},
         year = 2009,
        month = dec,
       volume = {400},
       number = {4},
        pages = {2050-2053},
          doi = {10.1111/j.1365-2966.2009.15598.x},
archivePrefix = {arXiv},
       eprint = {0903.2057},
 primaryClass = {astro-ph.GA},
       adsurl = {https://ui.adsabs.harvard.edu/abs/2009MNRAS.400.2050G}
}

@ARTICLE{molecfit_1,
       author = {{Smette}, A. and {Sana}, H. and {Noll}, S. and {Horst}, H. and {Kausch}, W. and {Kimeswenger}, S. and {Barden}, M. and {Szyszka}, C. and {Jones}, A.~M. and {Gallenne}, A. and {Vinther}, J. and {Ballester}, P. and {Taylor}, J.},
        title = "{Molecfit: A general tool for telluric absorption correction. I. Method and application to ESO instruments}",
      journal = {\aap},
         year = 2015,
        month = apr,
       volume = {576},
          eid = {A77},
        pages = {A77},
          doi = {10.1051/0004-6361/201423932},
archivePrefix = {arXiv},
       eprint = {1501.07239},
 primaryClass = {astro-ph.IM},
       adsurl = {https://ui.adsabs.harvard.edu/abs/2015A&A...576A..77S}
}

@ARTICLE{molecfit_2,
       author = {{Kausch}, W. and {Noll}, S. and {Smette}, A. and {Kimeswenger}, S. and {Barden}, M. and {Szyszka}, C. and {Jones}, A.~M. and {Sana}, H. and {Horst}, H. and {Kerber}, F.},
        title = "{Molecfit: A general tool for telluric absorption correction. II. Quantitative evaluation on ESO-VLT/X-Shooterspectra}",
      journal = {\aap},
         year = 2015,
        month = apr,
       volume = {576},
          eid = {A78},
        pages = {A78},
          doi = {10.1051/0004-6361/201423909},
archivePrefix = {arXiv},
       eprint = {1501.07265},
 primaryClass = {astro-ph.IM},
       adsurl = {https://ui.adsabs.harvard.edu/abs/2015A&A...576A..78K}
}

@ARTICLE{2017ApJ...845..157N,
       author = {{Newman}, Andrew B. and {Smith}, Russell J. and {Conroy}, Charlie and {Villaume}, Alexa and {van Dokkum}, Pieter},
        title = "{The Initial Mass Function in the Nearest Strong Lenses from SNELLS: Assessing the Consistency of Lensing, Dynamical, and Spectroscopic Constraints}",
      journal = {\apj},
         year = 2017,
        month = aug,
       volume = {845},
       number = {2},
          eid = {157},
        pages = {157},
          doi = {10.3847/1538-4357/aa816d},
archivePrefix = {arXiv},
       eprint = {1612.00065},
 primaryClass = {astro-ph.GA},
       adsurl = {https://ui.adsabs.harvard.edu/abs/2017ApJ...845..157N}
}

@ARTICLE{2022ApJ...932..103G,
       author = {{Gu}, Meng and {Greene}, Jenny E. and {Newman}, Andrew B. and {Kreisch}, Christina and {Quenneville}, Matthew E. and {Ma}, Chung-Pei and {Blakeslee}, John P.},
        title = "{The MASSIVE Survey. XVI. The Stellar Initial Mass Function in the Center of MASSIVE Early-type Galaxies}",
      journal = {\apj},
         year = 2022,
        month = jun,
       volume = {932},
       number = {2},
          eid = {103},
        pages = {103},
          doi = {10.3847/1538-4357/ac69ea},
archivePrefix = {arXiv},
       eprint = {2110.11985},
 primaryClass = {astro-ph.GA},
       adsurl = {https://ui.adsabs.harvard.edu/abs/2022ApJ...932..103G}
}

@software{2012ascl.soft07005V,
       author = {{van Dokkum}, Pieter G. and {Bloom}, J. and {Tewes}, Malte},
        title = "{L.A.Cosmic: Laplacian Cosmic Ray Identification}",
 howpublished = {Astrophysics Source Code Library, record ascl:1207.005},
         year = 2012,
        month = jul,
          eid = {ascl:1207.005},
       adsurl = {https://ui.adsabs.harvard.edu/abs/2012ascl.soft07005V}
}

@ARTICLE{marcote_2017_arsco,
       author = {{Marcote}, B. and {Marsh}, T.~R. and {Stanway}, E.~R. and {Paragi}, Z. and {Blanchard}, J.~M.},
        title = "{Towards the origin of the radio emission in AR Scorpii, the first radio-pulsing white dwarf binary}",
      journal = {\aap},
         year = 2017,
        month = may,
       volume = {601},
          eid = {L7},
        pages = {L7},
          doi = {10.1051/0004-6361/201730948},
archivePrefix = {arXiv},
       eprint = {1705.00600},
 primaryClass = {astro-ph.HE},
       adsurl = {https://ui.adsabs.harvard.edu/abs/2017A&A...601L...7M}
}

@ARTICLE{iris_ilt_j1101,
       author = {{de Ruiter}, I. and {Rajwade}, K.~M. and {Bassa}, C.~G. and {Rowlinson}, A. and {Wijers}, R.~A.~M.~J. and {Kilpatrick}, C.~D. and {Stefansson}, G. and {Callingham}, J.~R. and {Hessels}, J.~W.~T. and {Clarke}, T.~E. and {Peters}, W. and {Wijnands}, R.~A.~D. and {Shimwell}, T.~W. and {ter Veen}, S. and {Morello}, V. and {Zeimann}, G.~R. and {Mahadevan}, S.},
        title = "{Sporadic radio pulses from a white dwarf binary at the orbital period}",
      journal = {\nastro},
         year = 2025,
        month = mar,
        volume = {9},
        pages = {672-684},
          doi = {10.1038/s41550-025-02491-0},
archivePrefix = {arXiv},
       eprint = {2408.11536},
 primaryClass = {astro-ph.HE},
       adsurl = {https://ui.adsabs.harvard.edu/abs/2025NatAs.tmp...65D}
}

@ARTICLE{2025Natur.642..583W,
       author = {{Wang}, Ziteng and {Rea}, Nanda and {Bao}, Tong and {Kaplan}, David L. and {Lenc}, Emil and {Wadiasingh}, Zorawar and {Hare}, Jeremy and {Zic}, Andrew and {Anumarlapudi}, Akash and {Bera}, Apurba and {Beniamini}, Paz and {Cooper}, A.~J. and {Clarke}, Tracy E. and {Deller}, Adam T. and {Dawson}, J.~R. and {Glowacki}, Marcin and {Hurley-Walker}, Natasha and {McSweeney}, S.~J. and {Polisensky}, Emil J. and {Peters}, Wendy M. and {Younes}, George and {Bannister}, Keith W. and {Caleb}, Manisha and {Dage}, Kristen C. and {James}, Clancy W. and {Kasliwal}, Mansi M. and {Karambelkar}, Viraj and {Lower}, Marcus E. and {Mori}, Kaya and {Ocker}, Stella Koch and {P{\'e}rez-Torres}, Miguel and {Qiu}, Hao and {Rose}, Kovi and {Shannon}, Ryan M. and {Taub}, Rhianna and {Wang}, Fayin and {Wang}, Yuanming and {Zhao}, Zhenyin and {Bhat}, N.~D. Ramesh and {Dobie}, Dougal and {Driessen}, Laura N. and {Murphy}, Tara and {Jaini}, Akhil and {Deng}, Xinping and {Jahns-Schindler}, Joscha N. and {Lee}, Y.~W. Joshua and {Pritchard}, Joshua and {Tuthill}, John and {Thyagarajan}, Nithyanandan},
        title = "{Detection of X-ray emission from a bright long-period radio transient}",
      journal = {\nat},
         year = 2025,
        month = jun,
       volume = {642},
       number = {8068},
        pages = {583-586},
          doi = {10.1038/s41586-025-09077-w},
archivePrefix = {arXiv},
       eprint = {2411.16606},
 primaryClass = {astro-ph.HE},
       adsurl = {https://ui.adsabs.harvard.edu/abs/2025Natur.642..583W}
}

@ARTICLE{2025A&A...695L...8R,
       author = {{Rodriguez}, Antonio C.},
        title = "{Spectroscopic detection of a 2.9-hour orbit in a long-period radio transient}",
      journal = {\aap},
         year = 2025,
        month = mar,
       volume = {695},
          eid = {L8},
        pages = {L8},
          doi = {10.1051/0004-6361/202553684},
archivePrefix = {arXiv},
       eprint = {2501.03315},
 primaryClass = {astro-ph.SR},
       adsurl = {https://ui.adsabs.harvard.edu/abs/2025A&A...695L...8R}
}

@BOOK{1995cvs..book.....W,
       author = {{Warner}, Brian},
        title = "{Cataclysmic variable stars}",
         year = 1995,
        publisher = {Cambridge University Press},
       volume = {28},
       adsurl = {https://ui.adsabs.harvard.edu/abs/1995cvs..book.....W}
}

@ARTICLE{2011A&A...536A..42Z,
       author = {{Zorotovic}, M. and {Schreiber}, M.~R. and {G{\"a}nsicke}, B.~T.},
        title = "{Post common envelope binaries from SDSS. XI. The white dwarf mass distributions of CVs and pre-CVs}",
      journal = {\aap},
         year = 2011,
        month = dec,
       volume = {536},
          eid = {A42},
        pages = {A42},
          doi = {10.1051/0004-6361/201116626},
archivePrefix = {arXiv},
       eprint = {1108.4600},
 primaryClass = {astro-ph.SR},
       adsurl = {https://ui.adsabs.harvard.edu/abs/2011A&A...536A..42Z}
}

@ARTICLE{2018ApJS..237...25K,
       author = {{Kao}, Melodie M. and {Hallinan}, Gregg and {Pineda}, J. Sebastian and {Stevenson}, David and {Burgasser}, Adam},
        title = "{The Strongest Magnetic Fields on the Coolest Brown Dwarfs}",
      journal = {\apjs},
         year = 2018,
        month = aug,
       volume = {237},
       number = {2},
          eid = {25},
        pages = {25},
          doi = {10.3847/1538-4365/aac2d5},
archivePrefix = {arXiv},
       eprint = {1808.02485},
 primaryClass = {astro-ph.SR},
       adsurl = {https://ui.adsabs.harvard.edu/abs/2018ApJS..237...25K}
}

@ARTICLE{2017ApJ...846...75P,
       author = {{Pineda}, J. Sebastian and {Hallinan}, Gregg and {Kao}, Melodie M.},
        title = "{A Panchromatic View of Brown Dwarf Aurorae}",
      journal = {\apj},
         year = 2017,
        month = sep,
       volume = {846},
       number = {1},
          eid = {75},
        pages = {75},
          doi = {10.3847/1538-4357/aa8596},
archivePrefix = {arXiv},
       eprint = {1708.02942},
 primaryClass = {astro-ph.SR},
       adsurl = {https://ui.adsabs.harvard.edu/abs/2017ApJ...846...75P}
}

@ARTICLE{castrosegura2025,
        author = {{Castro Segura}, N. and {Pelisoli}, I. and {G{\"a}nsicke}, B.~T. and {Coppejans}, D.~L. and {Steeghs}, D. and {Aungwerojwit}, A. and {Inight}, K. and {Romero}, A. and {Sahu}, A. and {Dhillon}, V.~S. and {Munday}, J. and {Parsons}, S.~G. and {Kennedy}, M.~R. and {Green}, M.~J. and {Brown}, A.~J. and {Dyer}, M.~J. and {Pike}, E. and {Garbutt}, J.~A. and {Jarvis}, D. and {Kerry}, P. and {Littlefair}, S.~P. and {McCormac}, J. and {Sahman}, D.~I. and {Buckley}, D.~A.~H.},
        title = "{A sibling of AR Scorpii: SDSS J230641.47+244055.8 and the observational blueprint of white dwarf pulsars}",
      journal = {\mnras},
         year = 2025,
        month = nov,
       volume = {543},
       number = {3},
        pages = {2116-2129},
          doi = {10.1093/mnras/staf1511},
archivePrefix = {arXiv},
       eprint = {2506.20455},
 primaryClass = {astro-ph.SR},
       adsurl = {https://ui.adsabs.harvard.edu/abs/2025MNRAS.543.2116C}
}

@ARTICLE{2024ApJ...976L..21H,
       author = {{Hurley-Walker}, N. and {McSweeney}, S.~J. and {Bahramian}, A. and {Rea}, N. and {Horv{\'a}th}, C. and {Buchner}, S. and {Williams}, A. and {Meyers}, B.~W. and {Strader}, Jay and {Aydi}, Elias and {Urquhart}, Ryan and {Chomiuk}, Laura and {Galvin}, T.~J. and {Coti Zelati}, F. and {Bailes}, Matthew},
        title = "{A 2.9 hr Periodic Radio Transient with an Optical Counterpart}",
      journal = {\apjl},
         year = 2024,
        month = dec,
       volume = {976},
       number = {2},
          eid = {L21},
        pages = {L21},
          doi = {10.3847/2041-8213/ad890e},
archivePrefix = {arXiv},
       eprint = {2408.15757},
 primaryClass = {astro-ph.SR},
       adsurl = {https://ui.adsabs.harvard.edu/abs/2024ApJ...976L..21H}
}

@ARTICLE{2002JGRA..107.1081I,
       author = {{Imai}, Kazumasa and {Riihimaa}, Jorma J. and {Reyes}, Francisco and {Carr}, Thomas D.},
        title = "{Measurement of Jupiter's decametric radio source parameters by the modulation lane method}",
      journal = {\jgrsp},
         year = 2002,
        month = jun,
       volume = {107},
       number = {A6},
          eid = {1081},
        pages = {1081},
          doi = {10.1029/2001JA007555},
       adsurl = {https://ui.adsabs.harvard.edu/abs/2002JGRA..107.1081I}
}

@ARTICLE{2017PASP..129f2001M,
       author = {{Mukai}, K.},
        title = "{X-Ray Emissions from Accreting White Dwarfs: A Review}",
      journal = {\pasp},
         year = 2017,
        month = jun,
       volume = {129},
       number = {976},
        pages = {062001},
          doi = {10.1088/1538-3873/aa6736},
archivePrefix = {arXiv},
       eprint = {1703.06171},
 primaryClass = {astro-ph.HE},
       adsurl = {https://ui.adsabs.harvard.edu/abs/2017PASP..129f2001M}
}

@ARTICLE{2024AJ....167..186S,
       author = {{Szkody}, Paula and {van Roestel}, Jan and {Mason}, Paul A. and {Littlefield}, Colin and {Rich}, R. Michael and {Bellm}, Eric C. and {Romanov}, Filipp D. and {Healy}, Brian F. and {Jegou du Laz}, Theophile and {Laher}, Russ R. and {Rusholme}, Ben},
        title = "{Spectroscopic Follow-up on Potential Magnetic Cataclysmic Variables}",
      journal = {\aj},
         year = 2024,
        month = may,
       volume = {167},
       number = {5},
          eid = {186},
        pages = {186},
          doi = {10.3847/1538-3881/ad2fcd},
       adsurl = {https://ui.adsabs.harvard.edu/abs/2024AJ....167..186S}
}

@ARTICLE{2025MNRAS.540..821P,
       author = {{Pelisoli}, Ingrid and {Marsh}, T.~R. and {Tovmassian}, G. and {Amaral}, L.~A. and {Aungwerojwit}, Amornrat and {Green}, M.~J. and {Ashley}, R.~P. and {Buckley}, David A.~H. and {G{\"a}nsicke}, B.~T. and {Hambsch}, F. -J. and {Inight}, K. and {Potter}, S.~B. and {Brown}, A.~J. and {Castro Segura}, N. and {Dhillon}, V.~S. and {Dyer}, M.~J. and {Garbutt}, J.~A. and {Jarvis}, D. and {Kennedy}, M.~R. and {Kepler}, S.~O. and {Kerry}, P. and {Littlefair}, S.~P. and {McCormac}, J. and {Munday}, J. and {Parsons}, S.~G. and {Pike}, E. and {Sahman}, D.~I.},
        title = "{A targeted search for binary white dwarf pulsars using Gaia and WISE}",
      journal = {\mnras},
         year = 2025,
        month = jun,
       volume = {540},
       number = {1},
        pages = {821-836},
          doi = {10.1093/mnras/staf761},
archivePrefix = {arXiv},
       eprint = {2505.04693},
 primaryClass = {astro-ph.SR},
       adsurl = {https://ui.adsabs.harvard.edu/abs/2025MNRAS.540..821P}
}

@ARTICLE{2006MNRAS.373..484K,
       author = {{Knigge}, Christian},
        title = "{The donor stars of cataclysmic variables}",
      journal = {\mnras},
         year = 2006,
        month = dec,
       volume = {373},
       number = {2},
        pages = {484-502},
          doi = {10.1111/j.1365-2966.2006.11096.x},
archivePrefix = {arXiv},
       eprint = {astro-ph/0609671},
 primaryClass = {astro-ph},
       adsurl = {https://ui.adsabs.harvard.edu/abs/2006MNRAS.373..484K}
}

@ARTICLE{2023MNRAS.521.4190K,
       author = {{Koljonen}, K.~I.~I. and {Long}, K.~S. and {Matthews}, J.~H. and {Knigge}, C.},
        title = "{The origin of optical emission lines in the soft state of X-ray binary outbursts: the case of MAXI J1820+070}",
      journal = {\mnras},
         year = 2023,
        month = may,
       volume = {521},
       number = {3},
        pages = {4190-4206},
          doi = {10.1093/mnras/stad809},
archivePrefix = {arXiv},
       eprint = {2303.09242},
 primaryClass = {astro-ph.HE},
       adsurl = {https://ui.adsabs.harvard.edu/abs/2023MNRAS.521.4190K}
}

@ARTICLE{2021JApA...42...83S,
       author = {{Singh}, K.~P. and {Girish}, V. and {Tiwari}, J. and {Barrett}, P.~E. and {Buckley}, D.~A.~H. and {Potter}, S.~B. and {Schlegel}, E. and {Rana}, V. and {Stewart}, G.},
        title = "{Observations of AR Sco with Chandra and AstroSat soft X-ray telescope}",
      journal = {Journal of Astrophysics and Astronomy},
         year = 2021,
        month = oct,
       volume = {42},
       number = {2},
          eid = {83},
        pages = {83},
          doi = {10.1007/s12036-021-09756-w},
archivePrefix = {arXiv},
       eprint = {2104.08662},
 primaryClass = {astro-ph.HE},
       adsurl = {https://ui.adsabs.harvard.edu/abs/2021JApA...42...83S}
}

@ARTICLE{2023A&A...674L...9S,
       author = {{Schwope}, A. and {Marsh}, T.~R. and {Standke}, A. and {Pelisoli}, I. and {Potter}, S. and {Buckley}, D. and {Munday}, J. and {Dhillon}, V.},
        title = "{X-ray properties of the white dwarf pulsar eRASSU J191213.9{\ensuremath{-}}441044}",
      journal = {\aap},
         year = 2023,
        month = jun,
       volume = {674},
          eid = {L9},
        pages = {L9},
          doi = {10.1051/0004-6361/202346589},
archivePrefix = {arXiv},
       eprint = {2306.09732},
 primaryClass = {astro-ph.HE},
       adsurl = {https://ui.adsabs.harvard.edu/abs/2023A&A...674L...9S}
}

@ARTICLE{2012PASA...29..214G,
       author = {{George}, Samuel J. and {Stil}, Jeroen M. and {Keller}, Ben W.},
        title = "{Detection Thresholds and Bias Correction in Polarized Intensity}",
      journal = {\pasa},
         year = 2012,
        month = oct,
       volume = {29},
       number = {3},
        pages = {214-220},
          doi = {10.1071/AS11027},
archivePrefix = {arXiv},
       eprint = {1106.5362},
 primaryClass = {astro-ph.IM},
       adsurl = {https://ui.adsabs.harvard.edu/abs/2012PASA...29..214G}
}

@ARTICLE{2025A&A...698A.120G,
       author = {{Gupta}, N. and {Kerp}, J. and {Balashev}, S.~A. and {Morelli}, A.~P.~M. and {Combes}, F. and {Krogager}, J. -K. and {Momjian}, E. and {Borgaonkar}, D. and {Deka}, P.~P. and {Emig}, K.~L. and {Jose}, J. and {J{\'o}zsa}, G.~I.~G. and {Kl{\"o}ckner}, H. -R. and {Moodley}, K. and {Muller}, S. and {Noterdaeme}, P. and {Petitjean}, P. and {Wagenveld}, J.~D.},
        title = "{The MeerKAT Absorption Line Survey (MALS) data release 3: Cold atomic gas associated with the Milky Way}",
      journal = {\aap},
         year = 2025,
        month = jun,
       volume = {698},
          eid = {A120},
        pages = {A120},
          doi = {10.1051/0004-6361/202452407},
archivePrefix = {arXiv},
       eprint = {2504.00097},
 primaryClass = {astro-ph.GA},
       adsurl = {https://ui.adsabs.harvard.edu/abs/2025A&A...698A.120G}
}

@ARTICLE{RVM_1969,
       author = {{Radhakrishnan}, V. and {Cooke}, D.~J.},
        title = "{Magnetic Poles and the Polarization Structure of Pulsar Radiation}",
      journal = {\aplett},
         year = 1969,
        month = jan,
       volume = {3},
        pages = {225},
       adsurl = {https://ui.adsabs.harvard.edu/abs/1969ApL.....3..225R}
}

@inproceedings{Huynh2020,
 adsurl = {https://ui.adsabs.harvard.edu/abs/2020ASPC..522..263H},
 author = {{Huynh}, Minh and {Dempsey}, James and {Whiting}, Matthew T. and {Ophel}, Margaret},
 booktitle = {Astronomical Data Analysis Software and Systems XXVII},
 editor = {{Ballester}, Pascal and {Ibsen}, Jorge and {Solar}, Mauricio and {Shortridge}, Keith},
 month = {April},
 pages = {263},
 series = {Astronomical Society of the Pacific Conference Series},
 title = {{The CSIRO ASKAP Science Data Archive}},
 volume = {522},
 year = {2020}
}

@ARTICLE{barrett2020radio,
       author = {{Barrett}, Paul and {Dieck}, Christopher and {Beasley}, Anthony J. and {Mason}, Paul A. and {Singh}, Kulinder P.},
        title = "{Radio observations of magnetic cataclysmic variables}",
      journal = {Advances in Space Research},
         year = 2020,
        month = sep,
       volume = {66},
       number = {5},
        pages = {1226-1234},
          doi = {10.1016/j.asr.2020.04.007},
archivePrefix = {arXiv},
       eprint = {2004.11418},
 primaryClass = {astro-ph.SR},
       adsurl = {https://ui.adsabs.harvard.edu/abs/2020AdSpR..66.1226B}
}

@article{coppejans2016dwarf,
  title={Dwarf nova-type cataclysmic variable stars are significant radio emitters},
  author={Coppejans, Deanne L and K{\"o}rding, Elmar G and Miller-Jones, James CA and Rupen, Michael P and Sivakoff, Gregory R and Knigge, Christian and Groot, Paul J and Woudt, Patrick A and Waagen, Elizabeth O and Templeton, Matthew},
  journal={\mnras},
  volume={463},
  number={2},
  pages={2229--2241},
  year={2016},
  publisher={Oxford University Press}
}

@article{coppejans2015novalike,
  title={Novalike cataclysmic variables are significant radio emitters},
  author={Coppejans, Deanne L and K{\"o}rding, Elmar G and Miller-Jones, James CA and Rupen, Michael P and Knigge, Christian and Sivakoff, Gregory R and Groot, Paul J},
  journal={\mnras},
  volume={451},
  number={4},
  pages={3801--3813},
  year={2015},
  publisher={Oxford University Press}
}

@ARTICLE{2025MNRAS.tmp.1214A,
       author = {{Anumarlapudi}, Akash and {Kaplan}, David L. and {Rea}, Nanda and {Erasmus}, Nicolas and {Kelson}, Daniel and {Ocker}, Stella Koch and {Lenc}, Emil and {Dobie}, Dougal and {Hurley-Walker}, Natasha and {Sivakoff}, Gregory and {Buckley}, David A.~H. and {Murphy}, Tara and {Pritchard}, Joshua and {Driessen}, Laura and {Rose}, Kovi and {Zic}, Andrew},
        title = "{ASKAP J144834-685644: a newly discovered long period radio transient detected from radio to X-rays}",
      journal = {\mnras},
         year = 2025,
        month = jul,
          volume = {542},
       number = {2},
        pages = {1208-1232},
          doi = {10.1093/mnras/staf1227},
archivePrefix = {arXiv},
       eprint = {2507.13453},
 primaryClass = {astro-ph.HE},
       adsurl = {https://ui.adsabs.harvard.edu/abs/2025MNRAS.tmp.1214A}
}

@ARTICLE{2023MNRAS.524.4867I,
       author = {{Inight}, Keith and {G{\"a}nsicke}, Boris T. and {Breedt}, Elm{\'e} and {Israel}, Henry T. and {Littlefair}, Stuart P. and {Manser}, Christopher J. and {Marsh}, Tom R. and {Mulvany}, Tim and {Pala}, Anna Francesca and {Thorstensen}, John R.},
        title = "{A catalogue of cataclysmic variables from 20 yr of the Sloan Digital Sky Survey with new classifications, periods, trends, and oddities}",
      journal = {\mnras},
         year = 2023,
        month = oct,
       volume = {524},
       number = {4},
        pages = {4867-4898},
          doi = {10.1093/mnras/stad2018},
archivePrefix = {arXiv},
       eprint = {2304.06749},
 primaryClass = {astro-ph.SR},
       adsurl = {https://ui.adsabs.harvard.edu/abs/2023MNRAS.524.4867I}
}

@inproceedings{2016mks..confE...1J.GRS,
       author = "{Jonas}, J. and {MeerKAT Team}",
        title = "{The MeerKAT Radio Telescope}",
    booktitle = {Proceedings of MeerKAT Science: On the Pathway to the SKA {\textemdash} PoS(MeerKAT2016)},
         year = 2018,
       editor = {{Taylor}, R. and {Camilo}, F. and Leeuw, L. and {Moodley}, K.},
       series = {Proceedings of Science},
       volume = {277},
        month = Feb,
        pages = {001},
          doi = {10.22323/1.277.0001"},
       adsurl = {https://ui.adsabs.harvard.edu/abs/2016mks..confE...1J}
}

@ARTICLE{2012MNRAS.422..379B,
       author = {{Barsdell}, B.~R. and {Bailes}, M. and {Barnes}, D.~G. and {Fluke}, C.~J.},
        title = "{Accelerating incoherent dedispersion}",
      journal = {\mnras},
         year = 2012,
        month = may,
       volume = {422},
       number = {1},
        pages = {379-392},
          doi = {10.1111/j.1365-2966.2012.20622.x},
archivePrefix = {arXiv},
       eprint = {1201.5380},
 primaryClass = {astro-ph.IM},
       adsurl = {https://ui.adsabs.harvard.edu/abs/2012MNRAS.422..379B}
}

@ARTICLE{2006A&ARv..13..229T,
       author = {{Treumann}, Rudolf A.},
        title = "{The electron-cyclotron maser for astrophysical application}",
      journal = {\aapr},
         year = 2006,
        month = aug,
       volume = {13},
       number = {4},
        pages = {229-315},
          doi = {10.1007/s00159-006-0001-y},
       adsurl = {https://ui.adsabs.harvard.edu/abs/2006A&ARv..13..229T}
}

@ARTICLE{2008ApJ...684..644H,
       author = {{Hallinan}, G. and {Antonova}, A. and {Doyle}, J.~G. and {Bourke}, S. and {Lane}, C. and {Golden}, A.},
        title = "{Confirmation of the Electron Cyclotron Maser Instability as the Dominant Source of Radio Emission from Very Low Mass Stars and Brown Dwarfs}",
      journal = {\apj},
         year = 2008,
        month = sep,
       volume = {684},
       number = {1},
        pages = {644-653},
          doi = {10.1086/590360},
archivePrefix = {arXiv},
       eprint = {0805.4010},
 primaryClass = {astro-ph},
       adsurl = {https://ui.adsabs.harvard.edu/abs/2008ApJ...684..644H}
}

@inproceedings{Sault1995,
       author = {{Sault}, R.~J. and {Teuben}, P.~J. and {Wright}, M.~C.~H.},
        title = "{A Retrospective View of MIRIAD}",
    booktitle = {Astronomical Data Analysis Software and Systems IV},
         year = 1995,
       editor = {{Shaw}, R.~A. and {Payne}, H.~E. and {Hayes}, J.~J.~E.},
       series = {Astronomical Society of the Pacific Conference Series},
       volume = {77},
        month = jan,
        pages = {433},
          doi = {10.48550/arXiv.astro-ph/0612759},
archivePrefix = {arXiv},
       eprint = {astro-ph/0612759},
 primaryClass = {astro-ph},
       adsurl = {https://ui.adsabs.harvard.edu/abs/1995ASPC...77..433S}
}

@article{Offringa2014,
       author = {{Offringa}, A.~R. and {McKinley}, B. and {Hurley-Walker}, N. and {Briggs}, F.~H. and {Wayth}, R.~B. and {Kaplan}, D.~L. and {Bell}, M.~E. and {Feng}, L. and {Neben}, A.~R. and {Hughes}, J.~D. and {Rhee}, J. and {Murphy}, T. and {Bhat}, N.~D.~R. and {Bernardi}, G. and {Bowman}, J.~D. and {Cappallo}, R.~J. and {Corey}, B.~E. and {Deshpande}, A.~A. and {Emrich}, D. and {Ewall-Wice}, A. and {Gaensler}, B.~M. and {Goeke}, R. and {Greenhill}, L.~J. and {Hazelton}, B.~J. and {Hindson}, L. and {Johnston-Hollitt}, M. and {Jacobs}, D.~C. and {Kasper}, J.~C. and {Kratzenberg}, E. and {Lenc}, E. and {Lonsdale}, C.~J. and {Lynch}, M.~J. and {McWhirter}, S.~R. and {Mitchell}, D.~A. and {Morales}, M.~F. and {Morgan}, E. and {Kudryavtseva}, N. and {Oberoi}, D. and {Ord}, S.~M. and {Pindor}, B. and {Procopio}, P. and {Prabu}, T. and {Riding}, J. and {Roshi}, D.~A. and {Shankar}, N. Udaya and {Srivani}, K.~S. and {Subrahmanyan}, R. and {Tingay}, S.~J. and {Waterson}, M. and {Webster}, R.~L. and {Whitney}, A.~R. and {Williams}, A. and {Williams}, C.~L.},
        title = "{WSCLEAN: an implementation of a fast, generic wide-field imager for radio astronomy}",
      journal = {\mnras},
         year = 2014,
        month = oct,
       volume = {444},
       number = {1},
        pages = {606-619},
          doi = {10.1093/mnras/stu1368},
archivePrefix = {arXiv},
       eprint = {1407.1943},
 primaryClass = {astro-ph.IM},
       adsurl = {https://ui.adsabs.harvard.edu/abs/2014MNRAS.444..606O}
}

@article{Brentjens2005,
  author =	 {{Brentjens}, M.~A. and {de Bruyn}, A.~G.},
  title =	 "{Faraday rotation measure synthesis}",
  journal =	 {\aap},
  year =	 2005,
  month =	 oct,
  volume =	 441,
  number =	 3,
  pages =	 {1217-1228},
  doi =		 {10.1051/0004-6361:20052990},
  archivePrefix ={arXiv},
  eprint =	 {astro-ph/0507349},
  primaryClass = {astro-ph},
  adsurl =
                  {https://ui.adsabs.harvard.edu/abs/2005A&A...441.1217B}
}

@ARTICLE{2018PASA...35...10W,
       author = {{Wolf}, Christian and {Onken}, Christopher A. and {Luvaul}, Lance C. and {Schmidt}, Brian P. and {Bessell}, Michael S. and {Chang}, Seo-Won and {Da Costa}, Gary S. and {Mackey}, Dougal and {Martin-Jones}, Tony and {Murphy}, Simon J. and {Preston}, Tim and {Scalzo}, Richard A. and {Shao}, Li and {Smillie}, Jon and {Tisserand}, Patrick and {White}, Marc C. and {Yuan}, Fang},
        title = "{SkyMapper Southern Survey: First Data Release (DR1)}",
      journal = {\pasa},
         year = 2018,
        month = feb,
       volume = {35},
          eid = {e010},
        pages = {e010},
          doi = {10.1017/pasa.2018.5},
archivePrefix = {arXiv},
       eprint = {1801.07834},
 primaryClass = {astro-ph.IM},
       adsurl = {https://ui.adsabs.harvard.edu/abs/2018PASA...35...10W}
}

@misc{vizier,
  doi = {10.26093/CDS/VIZIER},
  url = {https://vizier.cds.unistra.fr},
  author = {Ochsenbein,  Francois},
  title = {The VizieR database of astronomical catalogues},
  publisher = {CDS,  Centre de DonnÃ©es astronomiques de Strasbourg},
  year = {1996},
  copyright = {Refer to CDS usage}
}

@ARTICLE{vizier2000,
       author = {{Ochsenbein}, F. and {Bauer}, P. and {Marcout}, J.},
        title = "{The VizieR database of astronomical catalogues}",
      journal = {\aaps},
         year = 2000,
        month = apr,
       volume = {143},
        pages = {23-32},
          doi = {10.1051/aas:2000169},
archivePrefix = {arXiv},
       eprint = {astro-ph/0002122},
 primaryClass = {astro-ph},
       adsurl = {https://ui.adsabs.harvard.edu/abs/2000A&AS..143...23O}
}

@ARTICLE{2013ApJS..208....9P,
       author = {{Pecaut}, Mark J. and {Mamajek}, Eric E.},
        title = "{Intrinsic Colors, Temperatures, and Bolometric Corrections of Pre-main-sequence Stars}",
      journal = {\apjs},
         year = 2013,
        month = sep,
       volume = {208},
       number = {1},
          eid = {9},
        pages = {9},
          doi = {10.1088/0067-0049/208/1/9},
archivePrefix = {arXiv},
       eprint = {1307.2657},
 primaryClass = {astro-ph.SR},
       adsurl = {https://ui.adsabs.harvard.edu/abs/2013ApJS..208....9P}
}

@ARTICLE{pritchard_2024,
       author = {{Pritchard}, Joshua and {Murphy}, Tara and {Heald}, George and {Wheatland}, Michael S. and {Kaplan}, David L. and {Lenc}, Emil and {O'Brien}, Andrew and {Wang}, Ziteng},
        title = "{Multi-epoch sampling of the radio star population with the Australian SKA Pathfinder}",
      journal = {\mnras},
         year = 2024,
        month = apr,
       volume = {529},
       number = {2},
        pages = {1258-1270},
          doi = {10.1093/mnras/stae127},
archivePrefix = {arXiv},
       eprint = {2312.11031},
 primaryClass = {astro-ph.SR},
       adsurl = {https://ui.adsabs.harvard.edu/abs/2024MNRAS.529.1258P}
}

@techreport{Perley2022,
  author       = {{Perley}, Rick A. and {Greisen}, Eric and {Hugo}, Ben},
  title        = {Enabling MeerKAT Polarimetric Imaging in AIPS},
  institution  = {EVLA Memo 219, National Radio Astronomy Observatory},
  year         = {2022},
  month        = sep,
  day          = {9},
}

@ARTICLE{2012AR&T....9..237V,
       author = {{van Straten}, Willem and {Demorest}, Paul and {Oslowski}, Stefan},
        title = "{Pulsar Data Analysis with PSRCHIVE}",
      journal = {Astronomical Research and Technology},
         year = 2012,
        month = jul,
       volume = {9},
       number = {3},
        pages = {237-256},
          doi = {10.48550/arXiv.1205.6276},
archivePrefix = {arXiv},
       eprint = {1205.6276},
 primaryClass = {astro-ph.IM},
       adsurl = {https://ui.adsabs.harvard.edu/abs/2012AR&T....9..237V}
}

@article{Aggarwal2020Fetch,
    author = {Agarwal, Devansh and Aggarwal, Kshitij and Burke-Spolaor, Sarah and Lorimer, Duncan R and Garver-Daniels, Nathaniel},
    title = {FETCH: A deep-learning based classifier for fast transient classification},
    journal = {\mnras},
    volume = {497},
    number = {2},
    pages = {1661-1674},
    year = {2020},
    month = {06},
    issn = {0035-8711},
    doi = {10.1093/mnras/staa1856},
    url = {https://doi.org/10.1093/mnras/staa1856},
    eprint = {https://academic.oup.com/mnras/article-pdf/497/2/1661/33562918/staa1856.pdf},
}

@article{10.1093/mnras/sty3328,
    author = {Morello, V and Barr, E D and Cooper, S and Bailes, M and Bates, S and Bhat, N D R and Burgay, M and Burke-Spolaor, S and Cameron, A D and Champion, D J and Eatough, R P and Flynn, C M L and Jameson, A and Johnston, S and Keith, M J and Keane, E F and Kramer, M and Levin, L and Ng, C and Petroff, E and Possenti, A and Stappers, B W and van Straten, W and Tiburzi, C},
    title = {The High Time Resolution Universe survey – XIV. Discovery of 23 pulsars through GPU-accelerated reprocessing},
    journal = {\mnras},
    volume = {483},
    number = {3},
    pages = {3673-3685},
    year = {2018},
    month = {12},
    issn = {0035-8711},
    doi = {10.1093/mnras/sty3328},
    url = {https://doi.org/10.1093/mnras/sty3328},
    eprint = {https://academic.oup.com/mnras/article-pdf/483/3/3673/27299713/sty3328.pdf},
}

@ARTICLE{2020JOSS....5.2308P,
       author = {{Prochaska}, J. and {Hennawi}, Joseph and {Westfall}, Kyle and {Cooke}, Ryan and {Wang}, Feige and {Hsyu}, Tiffany and {Davies}, Frederick and {Farina}, Emanuele and {Pelliccia}, Debora},
        title = "{PypeIt: The Python Spectroscopic Data Reduction Pipeline}",
      journal = {\joss},
         year = 2020,
        month = dec,
       volume = {5},
       number = {56},
          eid = {2308},
        pages = {2308},
          doi = {10.21105/joss.02308},
archivePrefix = {arXiv},
       eprint = {2005.06505},
 primaryClass = {astro-ph.IM},
       adsurl = {https://ui.adsabs.harvard.edu/abs/2020JOSS....5.2308P}
}

@ARTICLE{ep_paper,
       author = {{Yuan}, Weimin and {Dai}, Lixin and {Feng}, Hua and {Jin}, Chichuan and {Jonker}, Peter and {Kuulkers}, Erik and {Liu}, Yuan and {Nandra}, Kirpal and {O'Brien}, Paul and {Piro}, Luigi and {Rau}, Arne and {Rea}, Nanda and {Sanders}, Jeremy and {Tao}, Lian and {Wang}, Junfeng and {Wu}, Xuefeng and {Zhang}, Bing and {Zhang}, Shuangnan and {Ai}, Shunke and {Buchner}, Johannes and {Bulbul}, Esra and {Chen}, Hechao and {Chen}, Minghua and {Chen}, Yong and {Chen}, Yu-Peng and {Coleiro}, Alexis and {Zelati}, Francesco Coti and {Dai}, Zigao and {Fan}, Xilong and {Fan}, Zhou and {Friedrich}, Susanne and {Gao}, He and {Ge}, Chong and {Ge}, Mingyu and {Geng}, Jinjun and {Ghirlanda}, Giancarlo and {Gianfagna}, Giulia and {Gou}, Lijun and {Guillot}, S{\'e}bastien and {Hou}, Xian and {Hu}, Jingwei and {Huang}, Yongfeng and {Ji}, Long and {Jia}, Shumei and {Komossa}, S. and {Kong}, Albert K.~H. and {Lan}, Lin and {Li}, An and {Li}, Ang and {Li}, Chengkui and {Li}, Dongyue and {Li}, Jian and {Li}, Zhaosheng and {Ling}, Zhixing and {Liu}, Ang and {Liu}, Jinzhong and {Liu}, Liangduan and {Liu}, Zhu and {Luo}, Jiawei and {Ma}, Ruican and {Maggi}, Pierre and {Maitra}, Chandreyee and {Marino}, Alessio and {Ng}, Stephen Chi-Yung and {Pan}, Haiwu and {Rukdee}, Surangkhana and {Soria}, Roberto and {Sun}, Hui and {Tam}, Pak-Hin Thomas and {Thakur}, Aishwarya Linesh and {Tian}, Hui and {Troja}, Eleonora and {Wang}, Wei and {Wang}, Xiangyu and {Wang}, Yanan and {Wei}, Junjie and {Wen}, Sixiang and {Wu}, Jianfeng and {Wu}, Ting and {Xiao}, Di and {Xu}, Dong and {Xu}, Renxin and {Xu}, Yanjun and {Xu}, Yu and {Yang}, Haonan and {You}, Bei and {Yu}, Heng and {Yu}, Yunwei and {Zhang}, Binbin and {Zhang}, Chen and {Zhang}, Guobao and {Zhang}, Liang and {Zhang}, Wenda and {Zhang}, Yu and {Zhou}, Ping and {Zou}, Zecheng},
        title = "{Science objectives of the Einstein Probe mission}",
      journal = {Science China Physics, Mechanics, and Astronomy},
         year = 2025,
        month = mar,
       volume = {68},
       number = {3},
          eid = {239501},
        pages = {239501},
          doi = {10.1007/s11433-024-2600-3},
archivePrefix = {arXiv},
       eprint = {2501.07362},
 primaryClass = {astro-ph.HE},
       adsurl = {https://ui.adsabs.harvard.edu/abs/2025SCPMA..6839501Y}
}

@ARTICLE{2020SciPy-NMeth,
  author  = {Virtanen, Pauli and Gommers, Ralf and Oliphant, Travis E. and
            Haberland, Matt and Reddy, Tyler and Cournapeau, David and
            Burovski, Evgeni and Peterson, Pearu and Weckesser, Warren and
            Bright, Jonathan and {van der Walt}, St{\'e}fan J. and
            Brett, Matthew and Wilson, Joshua and Millman, K. Jarrod and
            Mayorov, Nikolay and Nelson, Andrew R. J. and Jones, Eric and
            Kern, Robert and Larson, Eric and Carey, C J and
            Polat, {\.I}lhan and Feng, Yu and Moore, Eric W. and
            {VanderPlas}, Jake and Laxalde, Denis and Perktold, Josef and
            Cimrman, Robert and Henriksen, Ian and Quintero, E. A. and
            Harris, Charles R. and Archibald, Anne M. and
            Ribeiro, Ant{\^o}nio H. and Pedregosa, Fabian and
            {van Mulbregt}, Paul and {SciPy 1.0 Contributors}},
  title   = {{{SciPy} 1.0: Fundamental Algorithms for Scientific
            Computing in Python}},
  journal = {\nmeth},
  year    = {2020},
  volume  = {17},
  pages   = {261--272},
  adsurl  = {https://rdcu.be/b08Wh},
  doi     = {10.1038/s41592-019-0686-2},
}

@ARTICLE{2007ApJS..173..682M,
       author = {{Morrissey}, Patrick and {Conrow}, Tim and {Barlow}, Tom A. and {Small}, Todd and {Seibert}, Mark and {Wyder}, Ted K. and {Budav{\'a}ri}, Tam{\'a}s and {Arnouts}, Stephane and {Friedman}, Peter G. and {Forster}, Karl and {Martin}, D. Christopher and {Neff}, Susan G. and {Schiminovich}, David and {Bianchi}, Luciana and {Donas}, Jos{\'e} and {Heckman}, Timothy M. and {Lee}, Young-Wook and {Madore}, Barry F. and {Milliard}, Bruno and {Rich}, R. Michael and {Szalay}, Alex S. and {Welsh}, Barry Y. and {Yi}, Sukyoung K.},
        title = "{The Calibration and Data Products of GALEX}",
      journal = {\apjs},
         year = 2007,
        month = dec,
       volume = {173},
       number = {2},
        pages = {682-697},
          doi = {10.1086/520512},
archivePrefix = {arXiv},
       eprint = {0706.0755},
 primaryClass = {astro-ph},
       adsurl = {https://ui.adsabs.harvard.edu/abs/2007ApJS..173..682M}
}

@ARTICLE{2007A&A...474..951S,
       author = {{Sing}, D.~K. and {Green}, E.~M. and {Howell}, S.~B. and {Holberg}, J.~B. and {Lopez-Morales}, M. and {Shaw}, J.~S. and {Schmidt}, G.~D.},
        title = "{Discovery of a bright eclipsing cataclysmic variable}",
      journal = {\aap},
         year = 2007,
        month = nov,
       volume = {474},
       number = {3},
        pages = {951-960},
          doi = {10.1051/0004-6361:20078026},
archivePrefix = {arXiv},
       eprint = {0708.0725},
 primaryClass = {astro-ph},
       adsurl = {https://ui.adsabs.harvard.edu/abs/2007A&A...474..951S}
}

@ARTICLE{2007A&A...473..511S,
       author = {{Schwarz}, R. and {Schwope}, A.~D. and {Staude}, A. and {Rau}, A. and {Hasinger}, G. and {Urrutia}, T. and {Motch}, C.},
        title = "{Paloma (RX J0524+42): the missing link in magnetic CV evolution?}",
      journal = {\aap},
         year = 2007,
        month = oct,
       volume = {473},
       number = {2},
        pages = {511-521},
          doi = {10.1051/0004-6361:20077684},
       adsurl = {https://ui.adsabs.harvard.edu/abs/2007A&A...473..511S}
}

@ARTICLE{2023ApJ...943L..24L,
       author = {{Littlefield}, Colin and {Mason}, Paul A. and {Garnavich}, Peter and {Szkody}, Paula and {Thorstensen}, John and {Scaringi}, Simone and {I{\l}kiewicz}, Krystian and {Kennedy}, Mark R. and {Wells}, Natalie},
        title = "{SDSS J134441.83+204408.3: A Highly Asynchronous Short-period Magnetic Cataclysmic Variable with a 56 MG Field Strength}",
      journal = {\apjl},
         year = 2023,
        month = feb,
       volume = {943},
       number = {2},
          eid = {L24},
        pages = {L24},
          doi = {10.3847/2041-8213/acaf04},
archivePrefix = {arXiv},
       eprint = {2301.05723},
 primaryClass = {astro-ph.SR},
       adsurl = {https://ui.adsabs.harvard.edu/abs/2023ApJ...943L..24L}
}

@ARTICLE{2004ApJ...614..349N,
       author = {{Norton}, A.~J. and {Wynn}, G.~A. and {Somerscales}, R.~V.},
        title = "{The Spin Periods and Magnetic Moments of White Dwarfs in Magnetic Cataclysmic Variables}",
      journal = {\apj},
         year = 2004,
        month = oct,
       volume = {614},
       number = {1},
        pages = {349-357},
          doi = {10.1086/423333},
archivePrefix = {arXiv},
       eprint = {astro-ph/0406363},
 primaryClass = {astro-ph},
       adsurl = {https://ui.adsabs.harvard.edu/abs/2004ApJ...614..349N}
}

@ARTICLE{1982ApJ...259..844M,
       author = {{Melrose}, D.~B. and {Dulk}, G.~A.},
        title = "{Electron-cyclotron masers as the source of certain solar and stellar radio bursts.}",
      journal = {\apj},
         year = 1982,
        month = aug,
       volume = {259},
        pages = {844-858},
          doi = {10.1086/160219},
       adsurl = {https://ui.adsabs.harvard.edu/abs/1982ApJ...259..844M}
}

@article{sullivan2019pyvista,
  doi = {10.21105/joss.01450},
  url = {https://doi.org/10.21105/joss.01450},
  year = {2019},
  month = {May},
  publisher = {The Open Journal},
  volume = {4},
  number = {37},
  pages = {1450},
  author = {Bane Sullivan and Alexander Kaszynski},
  title = {{PyVista}: {3D} plotting and mesh analysis through a streamlined interface for the {Visualization Toolkit} ({VTK})},
  journal = {Journal of Open Source Software}
}

@ARTICLE{2026NatAs.tmp...27H,
       author = {{Horv{\'a}th}, Csan{\'a}d and {Rea}, Nanda and {Hurley-Walker}, Natasha and {McSweeney}, Samuel J. and {Perley}, Richard A. and {Lenc}, Emil},
        title = "{A binary model of long-period radio transients and white dwarf pulsars}",
      journal = {\nastro},
         year = 2026,
        month = jan,
          doi = {10.1038/s41550-025-02760-y},
archivePrefix = {arXiv},
       eprint = {2507.15352},
 primaryClass = {astro-ph.HE},
       adsurl = {https://ui.adsabs.harvard.edu/abs/2026NatAs.tmp...27H}
}

@INPROCEEDINGS{2008SPIE.7014E..0AO,
       author = {{Osip}, David J. and {Floyd}, David and {Covarrubias}, Ricardo},
        title = "{Instrumentation at the Magellan Telescopes 2008}",
    booktitle = {Ground-based and Airborne Instrumentation for Astronomy II},
         year = 2008,
       editor = {{McLean}, Ian S. and {Casali}, Mark M.},
       series = {Society of Photo-Optical Instrumentation Engineers (SPIE) Conference Series},
       volume = {7014},
        month = jul,
          eid = {70140A},
        pages = {70140A},
          doi = {10.1117/12.790011},
       adsurl = {https://ui.adsabs.harvard.edu/abs/2008SPIE.7014E..0AO}
}

@ARTICLE{2015PASP..127..994B,
       author = {{Bailer-Jones}, Coryn A.~L.},
        title = "{Estimating Distances from Parallaxes}",
      journal = {\pasp},
         year = 2015,
        month = oct,
       volume = {127},
       number = {956},
        pages = {994},
          doi = {10.1086/683116},
archivePrefix = {arXiv},
       eprint = {1507.02105},
 primaryClass = {astro-ph.IM},
       adsurl = {https://ui.adsabs.harvard.edu/abs/2015PASP..127..994B}
}

@ARTICLE{2020MNRAS.492L..40A,
       author = {{Abril}, Javier and {Schmidtobreick}, Linda and {Ederoclite}, Alessandro and {L{\'o}pez-Sanjuan}, Carlos},
        title = "{Disentangling cataclysmic variables in Gaia's HR diagram}",
      journal = {\mnras},
         year = 2020,
        month = feb,
       volume = {492},
       number = {1},
        pages = {L40-L44},
          doi = {10.1093/mnrasl/slz181},
archivePrefix = {arXiv},
       eprint = {1912.01531},
 primaryClass = {astro-ph.SR},
       adsurl = {https://ui.adsabs.harvard.edu/abs/2020MNRAS.492L..40A}
}

@ARTICLE{2026ApJ...997..124Y,
       author = {{Yang}, Yuan-Pei},
        title = "{Magnetic White Dwarf--M Dwarf Binaries in Pre-mCV Phase as Special Population of Long-period Radio Transients}",
      journal = {\apj},
         year = 2026,
        month = jan,
       volume = {997},
       number = {1},
          eid = {124},
        pages = {124},
          doi = {10.3847/1538-4357/ae2864},
archivePrefix = {arXiv},
       eprint = {2509.09224},
 primaryClass = {astro-ph.HE},
       adsurl = {https://ui.adsabs.harvard.edu/abs/2026ApJ...997..124Y}
}

@ARTICLE{2021ApJ...911...45L,
       author = {{Luo}, Jing and {Ransom}, Scott and {Demorest}, Paul and {Ray}, Paul S. and {Archibald}, Anne and {Kerr}, Matthew and {Jennings}, Ross J. and {Bachetti}, Matteo and {van Haasteren}, Rutger and {Champagne}, Chloe A. and {Colen}, Jonathan and {Phillips}, Camryn and {Zimmerman}, Josef and {Stovall}, Kevin and {Lam}, Michael T. and {Jenet}, Fredrick A.},
        title = "{PINT: A Modern Software Package for Pulsar Timing}",
      journal = {\apj},
         year = 2021,
        month = apr,
       volume = {911},
       number = {1},
          eid = {45},
        pages = {45},
          doi = {10.3847/1538-4357/abe62f},
archivePrefix = {arXiv},
       eprint = {2012.00074},
 primaryClass = {astro-ph.IM},
       adsurl = {https://ui.adsabs.harvard.edu/abs/2021ApJ...911...45L}
}

@ARTICLE{gansicke2009,
       author = {{G{\"a}nsicke}, B.~T. and {Dillon}, M. and {Southworth}, J. and {Thorstensen}, J.~R. and {Rodr{\'\i}guez-Gil}, P. and {Aungwerojwit}, A. and {Marsh}, T.~R. and {Szkody}, P. and {Barros}, S.~C.~C. and {Casares}, J. and {de Martino}, D. and {Groot}, P.~J. and {Hakala}, P. and {Kolb}, U. and {Littlefair}, S.~P. and {Mart{\'\i}nez-Pais}, I.~G. and {Nelemans}, G. and {Schreiber}, M.~R.},
        title = "{SDSS unveils a population of intrinsically faint cataclysmic variables at the minimum orbital period}",
      journal = {\mnras},
         year = 2009,
        month = aug,
       volume = {397},
       number = {4},
        pages = {2170-2188},
          doi = {10.1111/j.1365-2966.2009.15126.x},
archivePrefix = {arXiv},
       eprint = {0905.3476},
 primaryClass = {astro-ph.SR},
       adsurl = {https://ui.adsabs.harvard.edu/abs/2009MNRAS.397.2170G}
}

@ARTICLE{rodriguez2025,
       author = {{Rodriguez}, Antonio C. and {El-Badry}, Kareem and {Hakala}, Pasi and {Rodr{\'\i}guez-Gil}, Pablo and {Bao}, Tong and {Galiullin}, Ilkham and {Kurlander}, Jacob A. and {Law}, Casey J. and {Pelisoli}, Ingrid and {Schreiber}, Matthias R. and {Burdge}, Kevin and {Caiazzo}, Ilaria and {Roestel}, Jan van and {Szkody}, Paula and {Drake}, Andrew J. and {Buckley}, David A.~H. and {Potter}, Stephen B. and {Gaensicke}, Boris and {Mori}, Kaya and {Bellm}, Eric C. and {Kulkarni}, Shrinivas R. and {Prince}, Thomas A. and {Graham}, Matthew and {Kasliwal}, Mansi M. and {Rose}, Sam and {Sharma}, Yashvi and {Ahumada}, Tom{\'a}s and {Anand}, Shreya and {Viitanen}, Akke and {Wold}, Avery and {Chen}, Tracy X. and {Riddle}, Reed and {Smith}, Roger},
        title = "{A Link Between White Dwarf Pulsars and Polars: Multiwavelength Observations of the 9.36-minute Period Variable Gaia22ayj}",
      journal = {\pasp},
         year = 2025,
        month = feb,
       volume = {137},
       number = {2},
          eid = {024202},
        pages = {024202},
          doi = {10.1088/1538-3873/adb0f1},
archivePrefix = {arXiv},
       eprint = {2501.01490},
 primaryClass = {astro-ph.SR},
       adsurl = {https://ui.adsabs.harvard.edu/abs/2025PASP..137b4202R}
}

@ARTICLE{pala2017,
       author = {{Pala}, A.~F. and {G{\"a}nsicke}, B.~T. and {Townsley}, D. and {Boyd}, D. and {Cook}, M.~J. and {De Martino}, D. and {Godon}, P. and {Haislip}, J.~B. and {Henden}, A.~A. and {Hubeny}, I. and {Ivarsen}, K.~M. and {Kafka}, S. and {Knigge}, C. and {LaCluyze}, A.~P. and {Long}, K.~S. and {Marsh}, T.~R. and {Monard}, B. and {Moore}, J.~P. and {Myers}, G. and {Nelson}, P. and {Nogami}, D. and {Oksanen}, A. and {Pickard}, R. and {Poyner}, G. and {Reichart}, D.~E. and {Rodriguez Perez}, D. and {Schreiber}, M.~R. and {Shears}, J. and {Sion}, E.~M. and {Stubbings}, R. and {Szkody}, P. and {Zorotovic}, M.},
        title = "{Effective temperatures of cataclysmic-variable white dwarfs as a probe of their evolution}",
      journal = {\mnras},
         year = 2017,
        month = apr,
       volume = {466},
       number = {3},
        pages = {2855-2878},
          doi = {10.1093/mnras/stw3293},
archivePrefix = {arXiv},
       eprint = {1701.02738},
 primaryClass = {astro-ph.SR},
       adsurl = {https://ui.adsabs.harvard.edu/abs/2017MNRAS.466.2855P}
}

@ARTICLE{pala2022,
       author = {{Pala}, A.~F. and {G{\"a}nsicke}, B.~T. and {Belloni}, D. and {Parsons}, S.~G. and {Marsh}, T.~R. and {Schreiber}, M.~R. and {Breedt}, E. and {Knigge}, C. and {Sion}, E.~M. and {Szkody}, P. and {Townsley}, D. and {Bildsten}, L. and {Boyd}, D. and {Cook}, M.~J. and {De Martino}, D. and {Godon}, P. and {Kafka}, S. and {Kouprianov}, V. and {Long}, K.~S. and {Monard}, B. and {Myers}, G. and {Nelson}, P. and {Nogami}, D. and {Oksanen}, A. and {Pickard}, R. and {Poyner}, G. and {Reichart}, D.~E. and {Rodriguez Perez}, D. and {Shears}, J. and {Stubbings}, R. and {Toloza}, O.},
        title = "{Constraining the evolution of cataclysmic variables via the masses and accretion rates of their underlying white dwarfs}",
      journal = {\mnras},
         year = 2022,
        month = mar,
       volume = {510},
       number = {4},
        pages = {6110-6132},
          doi = {10.1093/mnras/stab3449},
archivePrefix = {arXiv},
       eprint = {2111.13706},
 primaryClass = {astro-ph.SR},
       adsurl = {https://ui.adsabs.harvard.edu/abs/2022MNRAS.510.6110P}
}

@ARTICLE{dulk1985,
       author = {{Dulk}, G.~A.},
        title = "{Radio emission from the sun and stars.}",
      journal = {\araa},
         year = 1985,
        month = jan,
       volume = {23},
        pages = {169-224},
          doi = {10.1146/annurev.aa.23.090185.001125},
       adsurl = {https://ui.adsabs.harvard.edu/abs/1985ARA&A..23..169D}
}

@ARTICLE{2014MNRAS.437.3004L,
       author = {{Lentati}, L. and {Alexander}, P. and {Hobson}, M.~P. and {Feroz}, F. and {van Haasteren}, R. and {Lee}, K.~J. and {Shannon}, R.~M.},
        title = "{TEMPONEST: a Bayesian approach to pulsar timing analysis}",
      journal = {\mnras},
         year = 2014,
        month = jan,
       volume = {437},
       number = {3},
        pages = {3004-3023},
          doi = {10.1093/mnras/stt2122},
archivePrefix = {arXiv},
       eprint = {1310.2120},
 primaryClass = {astro-ph.IM},
       adsurl = {https://ui.adsabs.harvard.edu/abs/2014MNRAS.437.3004L}
}

@ARTICLE{sokoloff1998,
       author = {{Sokoloff}, D.~D. and {Bykov}, A.~A. and {Shukurov}, A. and {Berkhuijsen}, E.~M. and {Beck}, R. and {Poezd}, A.~D.},
        title = "{Depolarization and Faraday effects in galaxies}",
      journal = {\mnras},
         year = 1998,
        month = aug,
       volume = {299},
       number = {1},
        pages = {189-206},
          doi = {10.1046/j.1365-8711.1998.01782.x},
       adsurl = {https://ui.adsabs.harvard.edu/abs/1998MNRAS.299..189S}
}

@ARTICLE{blandhawthorn2016,
       author = {{Bland-Hawthorn}, Joss and {Gerhard}, Ortwin},
        title = "{The Galaxy in Context: Structural, Kinematic, and Integrated Properties}",
      journal = {\araa},
         year = 2016,
        month = sep,
       volume = {54},
        pages = {529-596},
          doi = {10.1146/annurev-astro-081915-023441},
archivePrefix = {arXiv},
       eprint = {1602.07702},
 primaryClass = {astro-ph.GA},
       adsurl = {https://ui.adsabs.harvard.edu/abs/2016ARA&A..54..529B}
}

@ARTICLE{arsco_gyro2025,
       author = {{Barrett}, Paul E. and {Gurwell}, Mark A.},
        title = "{Submillimeter Observations of the White Dwarf Pulsar AR Sco}",
      journal = {\apj},
         year = 2025,
        month = jun,
       volume = {986},
       number = {1},
          eid = {78},
        pages = {78},
          doi = {10.3847/1538-4357/add725},
archivePrefix = {arXiv},
       eprint = {2505.06468},
 primaryClass = {astro-ph.SR},
       adsurl = {https://ui.adsabs.harvard.edu/abs/2025ApJ...986...78B}
}

@dataset{zenodo_ref,
  author       = {Rose, Kovi},
  title        = {Periodic Radio and X-ray Emission from an
                   Accreting White Dwarf Binary
                  },
  year         = 2026,
  publisher    = {Zenodo},
  version      = {0.0.1},
  doi          = {10.5281/zenodo.17365566},
  url          = {https://doi.org/10.5281/zenodo.17365566},
}

\clearpage

\section*{Supplementary Information}
\label{sec: supplementary information}

\subsection*{Gaia Astrometric Uncertainty}
\label{subsec:gaia}

The astrometric solution for \askapshort\ in  \textit{Gaia} DR3 \citep{gaia_dr3} indicates significant unreliability in the derived parallax distance, proper motion, and space velocity. The measured parallax of $\varpi = 1.75 \pm 0.91$\,mas corresponds to a signal-to-noise ratio of $\sim2$, resulting in a simple distance inversion ($d = 1/\varpi$) that is unreliable \citep{2015PASP..127..994B, Luri2018_gaia_dr2_parallax}. Furthermore, the \textit{Gaia} astrometric quality indicators for give an \texttt{astrometric\_excess\_noise} value of $3.84$\,mas (typically of order $0.1$--$1.0$\,mas), indicating substantial excess scatter in the along-scan residuals that cannot be accounted for by the measurement uncertainties alone \citep{Lindegren2012, Lindegren2021}. This is consistent with a photocentre wobble induced by an unresolved binary companion, which we have demonstrated is the case for \gaiashort. Similarly, the \texttt{astrometric\_sigma5d\_max} value of 1.43\,mas  exceeds the recommended threshold of $\sim$1\,mas for good astrometric solutions \citep{Lindegren2021}. Together, these indicators imply that the reported proper motion measurements ($\mu_{\alpha} = 2.98 \pm 0.85$\,mas\,yr$^{-1}$, $\mu_{\delta} = 3.77 \pm 0.75$\,mas\,yr$^{-1}$) have underestimated uncertainties, and any derived tangential velocity is unreliable.

For these reasons we do not adopt the naive parallax inversion distance, and instead adopt a conservative range for \askapshort. As a lower bound we adopt $d_\mathrm{min} = 1/(\varpi + \sigma_\varpi)$\,pc, corresponding to the $1\sigma$ parallax limit, and as an upper bound we adopt the 84th percentile of the  ``Bailer-Jones" photogeometric distance posterior \citep{bailer-jones_21}. This method uses a direction-dependent prior on the Milky Way stellar distance distribution and incorporates \textit{Gaia} photometry alongside the parallax. As noted in the Methods, it is considered more reliable for \textit{Gaia} sources with fractional parallax uncertainties in the range $0.1\leq\sigma_{\varpi}/\varpi\leq1$ \cite{bailer-jones_21}, as is the case for \gaiashort. 

Therefore, combining the parallax and photogeometric distances, we adopt the  distance range \SIrange{0.4}{9.1}{\kilo\parsec}. This range is intended to conservatively constrain the true distance of \askapshort\ given the unreliable astrometric solution and the binary nature of the system.

\subsection*{MALS}
\label{subsec:mals}

Each MALS observation has a \SI{56}{\minute} integration time from three combined \SI{1120}{\second} scans. The MALS \href{catalogue}{https://mals.iucaa.in/release/malsdr1v3/search} provides $15$ \SI{100}{\arcsec} cutout images, each covering a \SI{60}{\mega\hertz} spectral window within the band. We find eight nominal detections in $15$ of the sub-band cutouts between \SIrange{900}{1400}{\mega\hertz}, with non-detections between \SIrange{1250}{1350}{\mega\hertz} and above \SI{\sim1500}{\mega\hertz}. Six of these detections, are above the $3\sigma$ threshold, with a median root mean square (RMS) of background noise $\sigma=$\SI{0.21}{\mjpb}. We fit these detections, with fluxes ranging from \SIrange{0.62}{1.82}{\mjpb}, to a simple $F(\nu)\sim\nu^{\alpha}$ power-law and obtain a spectral index of $\alpha=2.68\pm0.29$.
Further analysis of the archival MALS data was not undertaken as it would not have added to this work.

\subsection*{LDSS-3 Full Reduction}
\label{subsec:ldss3 full reduction}

Wavelength calibration was performed using He, Ne, and Ar arc lamp exposures. Initial line identification was carried out on the central spectrum, followed by a correction for spectral curvature across the spatial direction. A two-dimensional master flat was generated in each night from dome flat exposures. We then modelled the lamp spectrum and divided that out to isolate the pixel-to-pixel sensitivity variations and the non-uniformity in the slit. Cosmic rays are identified and removed using the L.A.Cosmic algorithm \citep{2012ascl.soft07005V}. We also acquire standard star observations, specifically of LTT~3864, to perform flux calibration.  For each science exposure, a sky model is constructed with an iterative two-dimensional b-spline function, with a break points spacing of $1.0$ pixel in the wavelength direction, and a 5th order polynomial in the spatial direction.  The resulting sky-subtracted frames, along with their inverse variance arrays, are rectified onto a uniform wavelength and spatial grid. Spectra are then linearly resampled onto a common grid spanning \SIrange{3000}{10500}{\angstrom}. Science targets are extracted using a 10-pixel wide aperture. The spatial scale is \SI{0.189}{\arcsec} per pixel. For standard stars, a circular extraction aperture with a $2$ pixel radius is used. Telluric features are modelled and removed using \texttt{MOLECFIT} \citep{molecfit_1,molecfit_2}, assuming H$_2$O and O$_2$ as the dominant molecular contributors. The absorption bands used in the telluric correction span \SIrange{6820}{6970}{\angstrom}, \SIrange{7210}{7330}{\angstrom}, \SIrange{7590}{7690}{\angstrom}, \SIrange{8170}{8360}{\angstrom}, \SIrange{8170}{8360}{\angstrom}, and \SIrange{9100}{9400}{\angstrom}.

\subsection*{Additional Swift Details}
\label{subsec: additional swift details}

The first \textit{Swift} observation (\href{ID 00016563001}{https://heasarc.gsfc.nasa.gov/FTP/swift/data/obs/2024_03//00016563001/}) was allocated \SI{1.5}{\kilo\second} but only \SI{0.24}{\kilo\second} was observed (starting at 2024-03-12 23:07 UTC) due to \textit{Swift} downtime.

We used the \texttt{uvotdetect} tool from \texttt{HEASoft} to identify sources in the UVOT images.  We input the sky image and exposure map for each of the UVOT bands and set the standard detection threshold of $3\sigma$, where $\sigma$ is defined as a standard deviation of the noise. This produced a catalogue of sources and photometry for each of the three images.\\

We filtered the \texttt{uvotdetect}-generated source catalogues for sources within \SI{5}{\arcsec} of \askapshort's position. In the first observation (2024-03-12 23:07 UTC) we did not detect any coincident $\geq3\sigma$ sources; likely due to the short exposure time.

In  Supplementary Table 2 we convert the \textit{Swift} magnitudes to the AB system using the additive corrections factors $1.51,1.69,1.73$ for the UVOT filters UVW1, UVM2, and UVW2, respectively; following the \href{AB zero point}{https://swift.gsfc.nasa.gov/caldb/docs/uvot/uvot_caldb_AB_10wa.pdf} \textit{Swift} UVOT calibration document.

For the PIMMS calculation we assumed a photon power-law index of $2$ and Galactic Hydrogen column density of n$_{\rm{H}}=1.4\times10^{21}$\,cm$^{-2}$, using the standard scaling relation \cite{guver_ozel_nh_2009} with an estimated maximum extinction of $A_V=0.6$\,mag.

\subsection*{Additional Line Fitting}
\label{subsec: additional line fitting}

We used the sum of a linear function and a Lorentzian to fit each spectral peak within a \SI{\pm 100}{\angstrom} range of the vacuum rest wavelength $\lambda_{\rm{rest}}$ of the respective emission line, using the \texttt{astropy} \citep{astropy:2022} implementation of the Levenberg-Marquardt algorithm and least squares statistic \texttt{LevMarLSQFitter}. 

We calculated the equivalent widths by summing the values of the fitted Lorentzian within \SI{\pm 50}{\angstrom} of $\lambda_{\rm{rest}}$ and normalising by the average of the linear continuum fit across that range. 

\subsection*{Plasma Frequency}
\label{subsec:plasma frequency}

A condition for the production of ECME at a frequency $\nu_{\rm{B}}$ is that $\nu_{\rm{B}}>\nu_{\rm{p}}$, where $\nu_{\rm{p}}$ is the plasma frequency. Since $\nu_{\rm{p}}\approx9\times10^3 n_{\rm{e}}^{1/2}$ \citep{dulk1985}, the \SI{2.7}{\giga\hertz} upper frequency cutoff of the ECME implies an electron density upper limit of $n_{\rm{e}}<9\times10^{10}$\,cm$^{-3}$.

Another characteristic of ECME is that $\nu_{\rm{B}}\approx2.8\times10^6 B$ \citep{dulk1985}, where $B$ is the magnetic field strength. Using the same \SI{2.7}{\giga\hertz} upper frequency cutoff gives us the lower limit of $B>$\SI{0.96}{\kilo\gauss} in the base of the emission region, which is consistent with the magnetic field strength expected in the interaction region.

\subsection*{Blackbody Fitting}
\label{subsec:blackbody}

We use the blackbody function

\begin{equation}
    \label{eq: blackbody}    
    B(\lambda)=\frac{2hc^2/\lambda^5}{\exp(hc/(\lambda k_B T)-1)},
\end{equation}

\noindent where $h$ is the Planck constant, $c$ the speed of light, $k_B$ the Boltzmann constant, and $T$ the temperature and assess the fit with a reduced $\chi^2$ statistic

\begin{equation}
    \label{eq: chi-squared}    
    \chi^2=\frac{1}{N-2}\sum\left(\frac{F_{\rm{obs}}-F_{\rm{fit}}}{\Delta F_{\rm{obs}}}\right)^2,
\end{equation}

\noindent where $F_{\rm{obs}}$ and $F_{\rm{fit}}$ is the observed and fitted flux, respectively, $N$ is the number of measurements, and $\Delta F_{\rm{obs}}$ are the uncertainties on the observed flux.

We use the \texttt{curve\_fit} method from the \texttt{scipy.optimize} library \citep{2020SciPy-NMeth}.

We apply an extinction correction of $A_V=0.6$\,mag to the short wavelength ($\lambda\leq500$\,nm) fluxes to obtain the AB magnitudes listed in Supplementary Table 1. The flux scaling was calculated with the \texttt{\href{extinction}{https://extinction.readthedocs.io/}} Python package

\subsection*{Pulse Sub-Structure Analysis}
\label{subsec: pulse sub-structure analysis}

We carried out a periodicity search of the Murriyang data with \texttt{pulsar\_miner} -- an automated pulsar searching pipeline based on \texttt{PRESTO}. For computational efficiency we performed an initial test for periodic pulses followed by a more detailed search for single pulses. In the first case, we searched for pulses in the first \SI{15}{\minute}. We used a dispersion measure (DM) range of up to \SI{300}{\parsec\,\cm^{-3}} -- double the plausible Galactic DMs along line-of-sight based on the YMW16 electron density model \citep{ymw16_dm} -- and periods ranging from \SI{1}{\milli\second} to \SI{10}{\second}. We performed acceleration searches in the Fourier domain using the \texttt{accelsearch} routine, with a maximum allowed Doppler-induced Fourier bin drift $z_{\rm{max}}=\pm200$. We found no convincing pulsar-like candidates in our search with a threshold of $8$.

From the PTUSE search, we obtained $\sim3\times10^4$ initial candidates. These were subsequently reduced to $\sim5\times10^3$ candidates using the machine-learning classifier \texttt{FETCH} \citep{Aggarwal2020Fetch}, which evaluates candidates based on their time–frequency morphology. The candidates were further processed with the RFI removal tool \texttt{\href{clfd}{https://github.com/v-morello/clfd?tab=readme-ov-file}} \citep{10.1093/mnras/sty3328}. We then inspected the remaining events manually and verified with the \texttt{pdmp} tool from \texttt{PSRCHIVE} \citep{2012AR&T....9..237V} that all of them correspond to zero-DM signals, consistent with terrestrial radio-frequency interference. We therefore conclude that no astrophysical short duration bursts are present in these data.

\subsection*{Time of Arrival Analysis}
\label{subsec: Time of Arrival analysis}

We used a triangle-shaped pulse model, with variable height, phase width, and centroid to measure times of arrival (ToAs) from the ATCA and ASKAP observations, adopting the centroid as the derived ToA. For observations exhibiting a double-peaked profile, we use a double-triangle model, and arbitrarily adopt the flux-weighted centroid as the ToA. We used time-integrated, folded pulse profiles per observation to measure the ToAs, which we produced using a weighted average for each observation, weighted by the single-pulse signal-to-noise ratio. This improved the ToA precision while mitigating against decreases in flux density observed during some observations.

We fit the timing model using PINT \citep{2021ApJ...911...45L}. We first performed an initial fit to the ToAs with a simple timing model where only the spin frequency $\nu$ was fit as a free parameter. After establishing a stable solution for the spin period alone, we then allowed the spin frequency time derivative $\dot{\nu}$, along with so-called $\mathrm{EFAC}$ and $\mathrm{EQUAD}$ parameters, which we used to augment the ATCA ToA uncertainties, to vary. $\mathrm{EFAC}$ and $\mathrm{EQUAD}$ are parameters used in pulsar timing analyses \citep[e.g.,][]{2014MNRAS.437.3004L} that adjust the ToA uncertainties ($\sigma_\mathrm{ToA}$) multiplicatively and in quadrature as $\sigma_{\mathrm{ToA}} \rightarrow \mathrm{EFAC} \sqrt{\sigma_\mathrm{ToA}^2 + \mathrm{EQUAD}^2}$.

We find $\nu = 0.000206529 ^{+6\times 10^{-9}}_{-5\times 10^{-9}}$, $\dot{\nu} < 5.7\times 10^{-16}$ at 95\% confidence, $\mathrm{EFAC} = 0.8^{+0.7}_{-0.2}$, and $\log_{10} (\mathrm{EQUAD}/\mathrm{s}) = -2^{+3}_{-2}$, with a period epoch of MJD 60350.9997.

\subsection*{Simulated Dynamic Spectra Details}

Synthetic dynamic spectra were generated with a geometric model
implemented in \textsc{PyVista} \citep{sullivan2019pyvista},
representing the system as two magnetic dipoles of moments
\SI{7e32}{\gauss\centi\metre^3} (white dwarf) and \SI{2e33}{\gauss\centi\metre^3} (M dwarf) separated by \SI{0.61}{\rsol}. These correspond to surface polar field strengths of \SI{4}{\mega\gauss} and \SI{1.2}{\kilo\gauss} for stellar radii of \SI{0.01}{\rsol} (white dwarf) and \SI{0.2}{\rsol} (M dwarf),
respectively. The binary was evolved in \SI{1}{\minute} timesteps over a circular orbit with period \SI{1.35}{\hour}, adopting a synchronised M dwarf rotation period and an asynchronous white dwarf rotation period of \SI{1.33}{\hour}. The spin axes of both stars and the M dwarf magnetic axis were aligned with the orbital axis, while the white dwarf magnetic axis was inclined by \SI{10}{\degree}. The system inclination was set to \SI{15}{\degree}.

We computed the three-dimensional magnetic vector field at each
timestep and traced field lines along a flux tube initially connecting the M dwarf’s northern magnetic hemisphere to the white dwarf’s southern magnetic pole. Voxels within the flux tube were tested for viable ECME generation and visibility, requiring the local magnetic field vector to lie within $\pm\SI{5}{\degree}$ of perpendicular to the observer’s line of sight and to fall within a prescribed acceleration region extending from roughly \SIrange{70}{75}{\percent} along the tube length toward the white dwarf. This region was chosen to be consistent with the presence of relativistic electrons implied by the observed elliptical
polarisation of the radio pulses. The resulting synthetic dynamic
spectrum assigns each frequency–time bin a value of $1$ when at least one visible emission site exists with a local field strength matching that ECME frequency, and $0$ otherwise --- indicating where ECME generation and visibility are viable rather than the emitted intensity.

Several model parameters -- such as the magnetic field strengths, 
white dwarf rotation period and resulting beat cycle, the inclination of the white dwarf magnetic axis, orbital inclination, and emission beaming angle -- cannot be uniquely constrained in these simulations. We therefore adopted values consistent with the observed orbital geometry and within plausible physical ranges for magnetic cataclysmic variables. This model explores only the geometric aspects of the emission process; additional effects due to gravitational influence, plasma flows and supply, and variations in the electron distribution are expected to further modify the observed emission properties, and may explain departures from oursimulated dynamic spectra.

\subsection*{Orbital Phase Delay}
\label{subsec:phase delay}

To obtain the X-ray orbital phase we fit a sinusoid to the phase-folded \textit{Einstein Probe}-FXT data, with all three observations binned to \SI{50}{\second} time resolution; shown in Fig. 2.
We obtain the peak orbital phase $\phi_{\rm{X}}=0.89\pm0.19$ at the maximum of the fitted curve, and take the uncertainty to be the square root of the phase term from the \texttt{curve\_fit} covariance matrix \citep{2020SciPy-NMeth}. \\ 
To calculate the orbital phase delays we take the difference in phases and use the quadrature sum of their respective uncertainties as the delay uncertainty:
\begin{itemize}
    \item[] $\text{X-ray vs. MKT: } \Delta\phi= 0.89-0.8 \pm \sqrt{0.19^2+0.1^2}=0.09\pm0.21$
    \item[] $\text{X-ray vs. ATCA/ASKAP: } \Delta\phi= 0.89-0.31 \pm \sqrt{0.19^2+0.03^2}=0.58\pm0.19$
    \item[] $\text{X-ray vs. RV: } \Delta\phi= 0.89-0.50 \pm \sqrt{0.19^2+0.32^2}=0.39\pm0.37$
\end{itemize}
We see that the phase delay between the MeerKAT pulses and the X-ray peak is negligible. Similarly the phase delay between the X-ray and radial velocity posterior is not significant. However the ATCA and ASKAP pulses are anti-phase with respect to the X-ray emission, with a nearly $3\sigma$ significance.

\setcounter{figure}{0}
\captionsetup[figure]{name={\bf Supplementary Figure}}


\begin{figure}
\centering
\includegraphics[width=11.5cm]{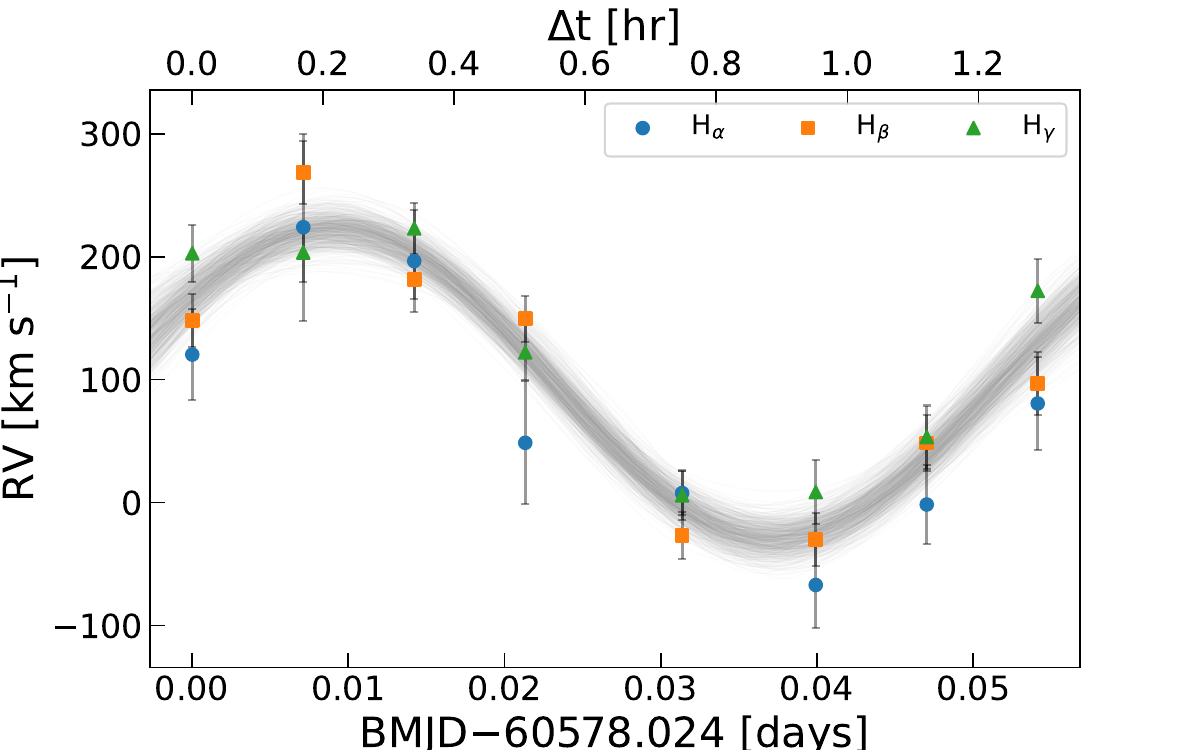}
\caption{Radial velocity measurements from \textit{SOAR} observations with $1\sigma$ error bars around the fitted values. We show posterior sample curves from \tj\ in grey. We use circle, square, and triangle markers for the $\rm{H}_{\alpha}$, $\rm{H}_{\beta}$, and $\rm{H}_{\gamma}$ radial velocity measurements, respectively.}
\label{fig: RVs}
\end{figure}


\begin{figure}
\centering
\includegraphics[width=11.5cm]{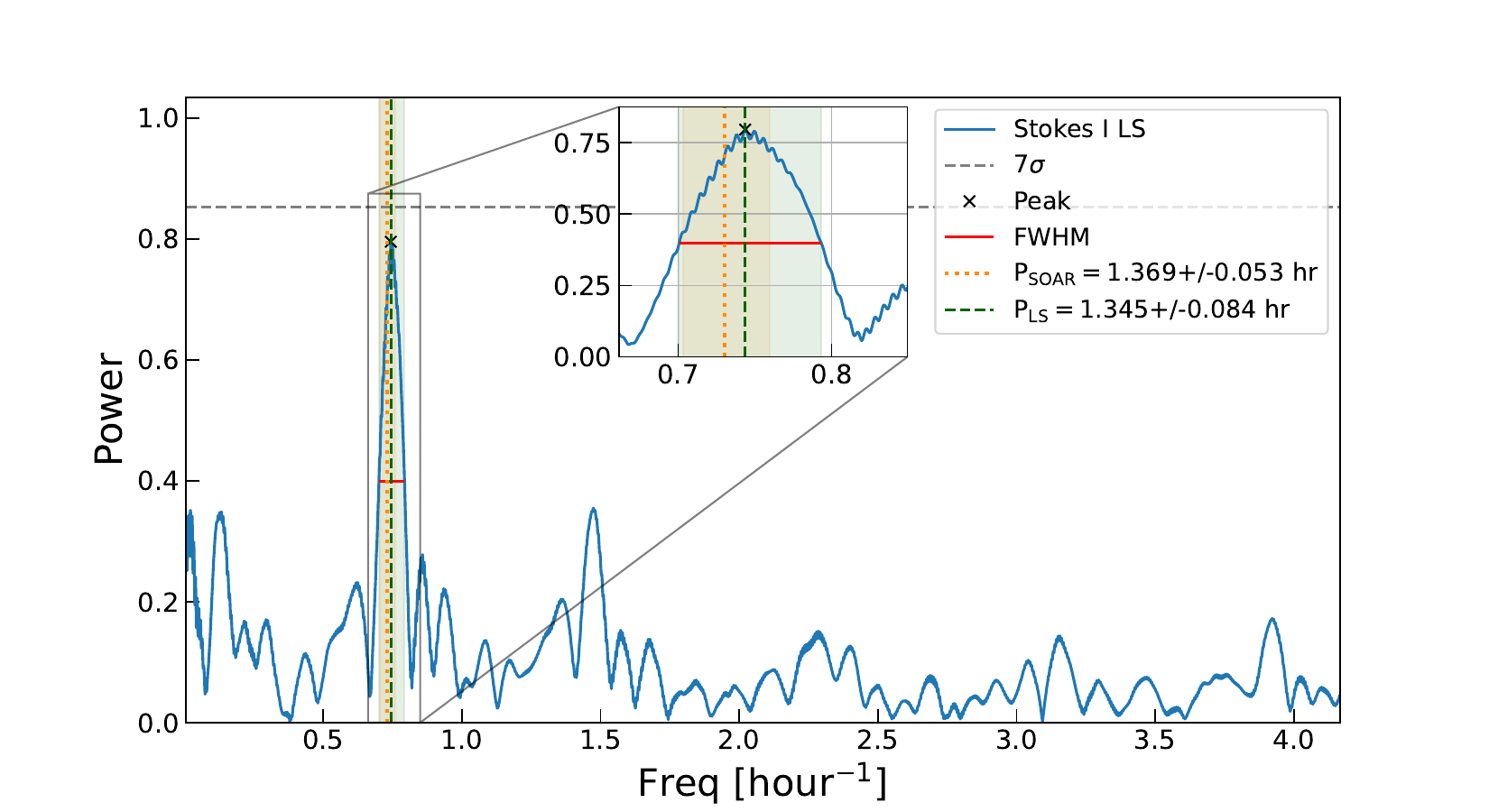}
\includegraphics[width=11.5cm]{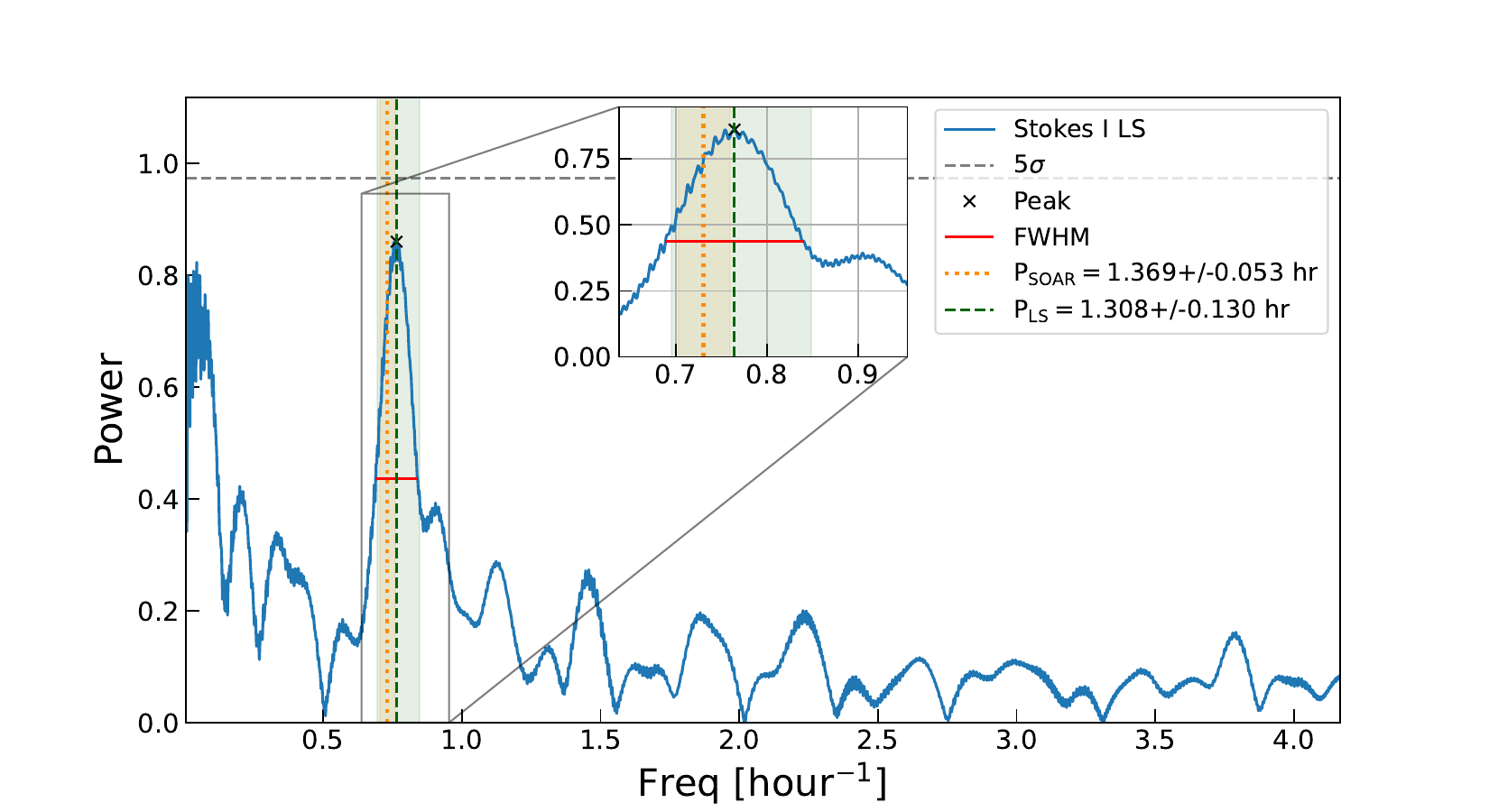}
        \caption{Lomb-Scargle Periodograms of \askapshort\ radio observations. We show combined ATCA Epochs~6--8 (Top) and MKT Epochs~1--3 (Bottom). The dashed green line shows the Lomb-Scargle period and the dotted orange line shows the period extracted from the \textit{SOAR} radial velocity fitting.}
\label{fig: Radio Periodograms}
\end{figure}

\setcounter{figure}{2}
\captionsetup[figure]{name={\bf Supplementary Figure}}

\begin{figure}
\centering
\includegraphics[width=11.5cm]{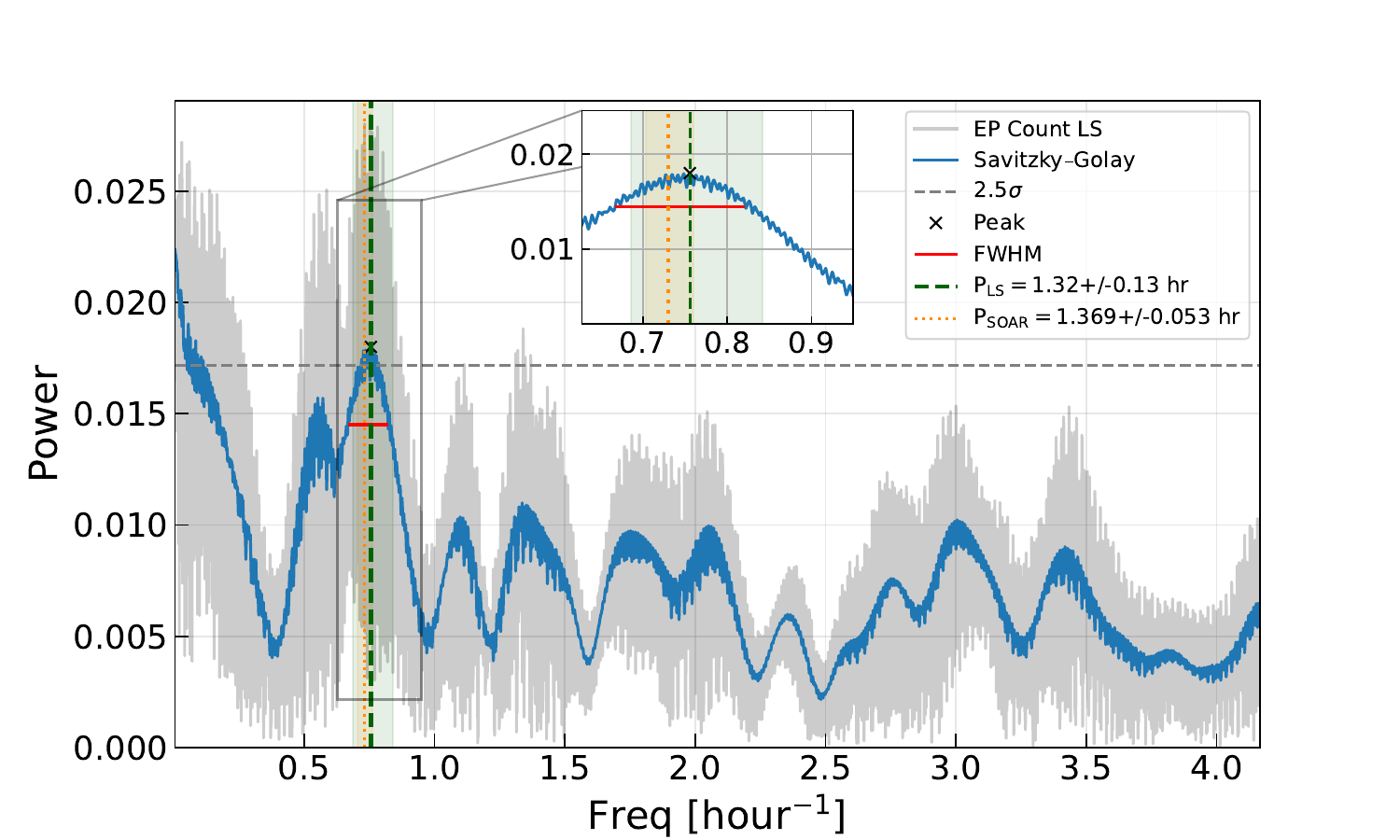}
\caption{Smoothed Lomb-Scargle Periodogram of combined \textit{Einstein Probe} observations of \askapshort\ binned with \SI{200}{\second} time resolution. The dashed green line shows the Lomb-Scargle period and the dotted orange line shows the period extracted from the \textit{SOAR} radial velocity fitting.}
\label{fig: ep_period}
\end{figure}

\section*{Supplementary Tables}
\label{sec: supplementary tables}

\setcounter{table}{0}
\captionsetup[table]{name={\bf Supplementary Table}}

\begin{table}[!ht]
    \centering
    \caption{AB magnitude photometry for \askapshort\ from archival UV and optical as well as UV bands from the \textit{Swift} observation (2024-05-16 17:43 UTC). We note the central wavelength $\lambda_{\rm{cent}}$ the instrumental filter name.}
    \begin{tabular}{cccc}
    \toprule
  
        $\lambda_{\rm{cent}}$ [nm] & AB Magnitude [mag] & Instrument & Filter \\ \\
        \hline
        152.8 & $21.36 \pm 0.149$ & GALEX & FUV \\
        192.8 & $21.792 \pm 0.163$ & Swift & UVW2 \\ 
        224.6 & $21.153 \pm 0.154$ & Swift & UVM2 \\ 
        231.0 & $21.36 \pm 0.105$ & GALEX & NUV \\ 
        260.0 & $21.732 \pm 0.201$ & Swift & UVW1 \\ 
        518.26 & $18.973 \pm 0.102$ & Gaia & G\_BP \\
        617.0 & $18.511 \pm 0.042$ & SkyMapper & r\_psf \\ 
        639.07 & $19.64 \pm 0.039$ & Gaia & G \\ 
        782.51 & $18.311 \pm 0.077$ & Gaia & G\_RP \\ 
        916 & $17.895 \pm 0.047$ & SkyMapper & z\_psf \\ 
        \bottomrule

    \end{tabular}
    \label{tab: photometry}

\end{table}

\begin{table}
\caption{Photometric rates, magnitudes, and fluxes for the different \textit{Swift} UVOT bands in the second observation (2024-05-16 17:43 UTC). We also calculate the angular separation between the positions of \askapshort\ in RACS-mid and each of the sources identified with \texttt{uvotdetect}.}
\centering
    \setlength{\tabcolsep}{3.2pt}
    \begin{tabular}{ccccc}
    \toprule
    Filter & Rate [count/s] & Vega Magnitude [mag] & Flux [erg s$^{-1}$ cm$^{-2}$] & Sep. [\arcsec] \\
    \\
    \hline
    UVW1    & $0.299 \pm 0.045$ & $18.46 \pm 0.16$
    & $4.3 \pm 0.6 \times10^{-13}$  & $0.79$\\
    UVM2    & $0.229 \pm 0.033$ & $18.17 \pm 0.15$
    & $5.6 \pm 0.8 \times10^{-13}$   & $0.54$\\
    UVW2    & $0.168 \pm 0.031$ & $18.93 \pm 0.20$
    & $2.8 \pm 0.5 \times10^{-13}$   & $0.32$\\
    \bottomrule
    \end{tabular}
    \label{tab: Swift UVOT Observation 2}
 \end{table}

 \begin{table}[htbp]
\centering
\caption{Fitted Values for SOAR spectra}
\label{tab:soar fits}
\begin{tabular}{|c|c|c|c|c|c|c|c|c|}
\hline
\textbf{Phase} $\phi$ & \textbf{$0.06$} & \textbf{$0.24$} & \textbf{$0.39$} & \textbf{$0.52$} & \textbf{$0.65$} & \textbf{$0.69$} & \textbf{$0.81$} & \textbf{$0.94$} \\
\hline
\hline
\multicolumn{9}{|c|}{\textbf{H$\alpha$ Line}} \\
\hline
RV [km/s] & 120.4 & 224.1 & 196.5 & 48.6 & 7.7 & -67.1 & -1.6 & 80.6 \\
err [km/s] & 37.0 & 76.1 & 41.6 & 50.0 & 18.1 & 34.8 & 32.4 & 37.7 \\
EQW [\AA] & 16.6 & 19.6 & 20.0 & 21.5 & 17.1 & 21.5 & 20.7 & 20.0 \\
FWHM [\AA] & 21.7 & 40.4 & 29.0 & 32.9 & 15.6 & 23.7 & 21.0 & 22.5 \\
\hline
\hline
\multicolumn{9}{|c|}{\textbf{H$\beta$ Line}} \\
\hline
RV [km/s] & 148.3 & 268.6 & 181.7 & 149.5 & -26.7 & -30.0 & 48.5 & 96.9 \\
err [km/s] & 21.4 & 26.0 & 16.3 & 18.7 & 19.1 & 21.9 & 22.9 & 25.8 \\
EQW [\AA] & 30.3 & 28.2 & 30.1 & 34.1 & 28.8 & 28.9 & 31.9 & 28.6 \\
FWHM [\AA] & 16.6 & 19.9 & 16.2 & 18.5 & 18.1 & 16.9 & 18.9 & 18.8 \\
\hline
\hline
\multicolumn{9}{|c|}{\textbf{H$\gamma$ Line}} \\
\hline
RV [km/s] & 202.8 & 203.5 & 223.1 & 122.2 & 5.8 & 8.5 & 53.2 & 172.2 \\
err [km/s] & 23.2 & 24.1 & 21.0 & 22.3 & 20.2 & 26.2 & 25.7 & 25.8 \\
EQW [\AA] & 33.3 & 31.6 & 36.5 & 33.6 & 28.7 & 30.0 & 27.4 & 32.6 \\
FWHM [\AA] & 18.1 & 19.0 & 20.1 & 20.1 & 17.5 & 18.1 & 17.3 & 18.7 \\
\hline
\hline
\multicolumn{9}{|c|}{\textbf{He II Line}} \\
\hline
RV [km/s] & 134.5 & 179.2 & 169.3 & 22.4 & -143.0 & -74.0 & 37.5 & 212.1 \\
err [km/s] & 58.9 & 32.7 & 33.6 & 43.9 & 33.0 & 41.8 & 29.8 & 48.1 \\
EQW [\AA] & 13.6 & 11.7 & 12.1 & 14.4 & 12.1 & 12.2 & 10.3 & 13.4 \\
FWHM [\AA] & 17.2 & 11.9 & 12.8 & 16.7 & 13.4 & 13.2 & 9.9 & 15.5 \\
\hline
\hline
\multicolumn{9}{|c|}{\textbf{Line Ratios}} \\
\hline
H$\alpha$/H$\beta$ & 0.549 & 0.696 & 0.664 & 0.629 & 0.593 & 0.744 & 0.648 & 0.698 \\
HeII/H$\beta$ & 0.450 & 0.415 & 0.402 & 0.423 & 0.418 & 0.420 & 0.324 & 0.468 \\
\hline
\end{tabular}
\end{table}

\begin{table}[htbp]
\centering
\caption{Fitted Values for LDSS-3 spectra}
\label{tab:LDSS-3 fits}
\begin{tabular}{|c|c|c|c|c|}
\hline
\textbf{Phase $\phi$} & \textbf{$0.34$} & \textbf{$0.38$} & \textbf{$0.47$} & \textbf{$0.52$} \\
\hline
\hline
\multicolumn{5}{|c|}{\textbf{H$\alpha$ Line}} \\
\hline
RV [km/s] & -81.2 & -54.1 & -251.8 & -329.9 \\
err [km/s] & 240.9 & 232.8 & 154.6 & 156.9 \\
EQW [\AA] & 21.1 & 17.3 & 27.9 & 28.5 \\
FWHM [\AA] & 19.2 & 16.5 & 19.9 & 19.1 \\
\hline
\hline
\multicolumn{5}{|c|}{\textbf{H$\beta$ Line}} \\
\hline
RV [km/s] & -148.0 & -32.1 & -188.3 & -125.5 \\
err [km/s] & 38.3 & 42.8 & 23.5 & 31.4 \\
EQW [\AA] & 28.7 & 26.7 & 47.6 & 31.9 \\
FWHM [\AA] & 14.2 & 14.7 & 16.5 & 15.2 \\
\hline
\hline
\multicolumn{5}{|c|}{\textbf{H$\gamma$ Line}} \\
\hline
RV [km/s] & -136.3 & -145.7 & -129.7 & -187.3 \\
err [km/s] & 14.8 & 13.0 & 12.6 & 19.1 \\
EQW [\AA] & 32.0 & 22.0 & 27.1 & 29.0 \\
FWHM [\AA] & 15.8 & 12.1 & 14.4 & 19.6 \\
\hline
\hline
\multicolumn{5}{|c|}{\textbf{H$\delta$ Line}} \\
\hline
RV [km/s] & -156.3 & -0.9 & -431.3 & -230.7 \\
err [km/s] & 6.4 & 6.0 & 6.7 & 6.4 \\
EQW [\AA] & 14.7 & 9.3 & 13.6 & 9.8 \\
FWHM [\AA] & 10.7 & 7.8 & 11.0 & 8.4 \\
\hline
\hline
\multicolumn{5}{|c|}{\textbf{He II Line}} \\
\hline
RV [km/s] & -159.2 & 27.7 & -178.9 & -137.6 \\
err [km/s] & 53.0 & 43.8 & 50.8 & 41.0 \\
EQW [\AA] & 16.3 & 13.4 & 13.8 & 8.4 \\
FWHM [\AA] & 13.4 & 10.9 & 14.6 & 8.9 \\
\hline
\hline
\multicolumn{5}{|c|}{\textbf{Line Ratios}} \\
\hline
H$\alpha$/H$\beta$ & 0.737 & 0.646 & 0.587 & 0.894 \\
HeII/H$\beta$ & 0.568 & 0.502 & 0.291 & 0.262 \\
\hline
\end{tabular}
\end{table}

\clearpage

\end{document}